\documentclass[11pt]{article}
\usepackage[utf8]{inputenc}
\usepackage{amsfonts,amsmath,amssymb,amsthm,graphicx,float,hyperref,comment,color}

\usepackage{algcompatible}
\usepackage[ruled,vlined]{algorithm2e}
\usepackage[title]{appendix}
\usepackage{lscape}
\usepackage{tikzscale}

\usepackage{framed}

\usepackage{authblk}

\usepackage[lmargin=0.6in,rmargin=1.2in,tmargin=1.0in,bmargin=1.0in]{geometry}

\usepackage{siunitx}

\usepackage{enumitem}

\newtheorem{protocol}{Protocol}

\newcommand{\lgl}{\langle}
\newcommand{\rgl}{\rangle}
\newcommand{\by}{\boldsymbol{y}}
\newcommand{\balpha}{\boldsymbol{\alpha}}

\newcommand{\psib}{\boldsymbol{\psi}}

\newcommand{\phib}{\boldsymbol{\phi}}

\newcommand{\tb}{{\mathbf t}}

\newcommand{\zb}{{\mathbf z}}

\newcommand{\ub}{{\mathbf u}}
\newcommand{\vb}{{\mathbf v}}

\newcommand{\pzc}{\textcolor{black}}

\usepackage{tikz}
\usetikzlibrary{shapes.geometric, arrows}
\usetikzlibrary[calc]
\usetikzlibrary{positioning}
\usetikzlibrary{fit}
\usetikzlibrary{arrows.meta}
\tikzset{block_device_small/.style={draw, thick, text width=1.5cm, minimum height=1.0cm, align=left,fill=cyan},   
}

\tikzset{block_device/.style={draw, thick, text width=4.5cm, minimum height=1.5cm, fill={cyan},  align=center}}
\tikzset{block_laptop/.style={draw, thick, text width=4cm, minimum height=1.5cm, fill={pink},  align=center}}
\tikzstyle{arrow} = [thick,-Stealth]
\tikzstyle{line} = [thick,-]
\tikzstyle{square-node} = [rectangle, rounded corners, minimum width=3cm, minimum height=1cm,text centered, draw=black, , fill=red!30]

\tikzset{
    *|/.style={
        to path={
            (perpendicular cs: horizontal line through={(\tikztostart)},
                                 vertical line through={(\tikztotarget)})
            -- (\tikztotarget) \tikztonodes
        }
    }
}

\usepackage{sansmath}

\usepackage{todonotes}

\usepackage{soul}

\usepackage{fancyhdr,graphicx,lastpage}
\title{Mathematical approaches for characterization, control, calibration and validation of a quantum computing device\footnote{This work was performed under the auspices of the U.S. Department of Energy by Lawrence Livermore
National Laboratory under Contract DE-AC52-07NA27344. This is contribution LLNL-TR-843608.}}
\author{Peng, Zhichao\thanks{{\tt pengzhic@msu.edu}, Mathematics Department, Michigan State University},
  Daniel Appel\"o\thanks{{\tt appeloda@msu.edu}, Mathematics Department, Michigan State University},
  N.~Anders Petersson\thanks{{\tt petersson1@llnl.gov}, Center for Applied Scientific Computing, Lawrence Livermore National
    Laboratory},
  Fortino Garcia\thanks{{\tt fortino.garcia@cims.nyu.edu}, Courant Institute of Mathematical Sciences, New York University}, and
  Yujin Cho\thanks{{\tt cho25@llnl.gov}, Quantum Coherent Device Physics, Lawrence Livermore National Laboratory}}
\date{\today}

\begin{document}

\maketitle

\begin{abstract}

Quantum computing has received significant amounts of interest from many different research communities over the last few years. Although there are many introductory texts that focus on the algorithmic parts of quantum computing, there is a dearth of publications that describe the modeling, calibration and operation of current quantum computing devices. One aim of this report is to fill that void by providing a case study that walks through the entire procedure from the characterization and optimal control of a qudit device at Lawrence Livermore National Laboratory (LLNL) to the validation of the results. A goal of the report is to provide an introduction for students and researchers, especially computational mathematicians, who are interested in but new to quantum computing. Both experimental and mathematical aspects of this procedure are discussed. We present a description of the LLNL QuDIT testbed, the mathematical models that are used to describe it, and the numerical methods that are used to to design optimal controls. We also present experimental and computational methods that can be used to characterize a quantum device. Finally, an experimental validation of an optimized control pulse is presented, which relies on the accuracy of the characterization and the optimal control methodologies. 

\end{abstract}

\clearpage
\tableofcontents
\clearpage


\section{Introduction}
During the middle of the last century, the hardware realization of a digital computer sparked an
intense interest from mathematicians who realized that the new computing power could be used for
solving problems in many areas of science and engineering. Of course, in order to reliably use these new
computing devices,  the effects of round-off errors in linear algebra and discretizations of
differential equations had to first be understood. Much of the efforts in developing such models and
theories can be traced back to the {\em The Institute for Numerical Analysis} (INA) at UCLA, which
arguably is the birthplace of the field of numerical analysis\footnote{In fact, given that the
founder of the computer science department at Stanford University, George Forsythe, was one of the
early members of INA, some numerical analysts claim that INA is even the birthplace of the field of
computer science}  \cite{todd1990prehistory,hestenes1991nbs}. 

Fast forward to today, quantum computing holds the promise of becoming the next technology to transform computing and scientific discovery. However, unlike the early days of classical computing, mathematicians have yet to play a major role in the developments of quantum computing. The number of (computational) mathematicians that engage with quantum computing will no
doubt grow as the community becomes more familiar with the topic and with the open questions in the
field. Such engagement is facilitated by lecture notes like those by Lin~\cite{linlinNotes}, introductory reports~\cite{petersson2020quantum}, and through
case studies. This report serves as one such case study where we walk through the entire process,
(almost) from turning on the power switch to looking at solutions printed on the screen. 

At a high level a quantum computer is a device that operates on a quantum state (modeled by a normalized complex
valued vector) $\psib\in\mathbb{C}^N$, through the application of quantum gates. These gates are nothing but unitary transformations where each operation on the state is simply modeled by a multiplication from the left by a unitary matrix. For example, applying a $0\leftrightarrow2$ SWAP gate on a three-level quantum state is realized by the unitary matrix transformation
\begin{align}
    \psib_{out} = S_{02}\psib_{in},\quad
    S_{02}=\left(\begin{array}{cccc}
        0 & 0 & 1  \\
        0 & 1 & 0 \\
        1 & 0 & 0 
    \end{array}\right).
\end{align} 

So how is such a gate created in the quantum hardware, and what is the mathematics and data driven modeling that allows the realization of such a gate? There is not one answer to this question as the
details will depend on the approach and the hardware itself. The hardware we will use here is the quantum system consisting of a single 4-level transmon qudit that is part of the QuDIT test-bed at Lawrence Livermore National Laboratory (LLNL)\cite{LLNLtestbed}. Out of the four levels, three are essential (useful for quantum information processing) and one is a so called guard state. The approach we will take is to construct a tailored control pulse, unique to the $0\leftrightarrow2$ SWAP gate, that will be applied to the quantum device. Ideally, once the control pulse has been found it should be possible to use
indefinitely, without any deterioration in fidelity over time. Unfortunately, current quantum computing devices are noisy and their physical properties can drift over time due to various reasons, ranging from cosmic radiation to variations in the ambient temperature in the laboratory. As a result, the control pulses must be updated frequently to reflect the variation in physical properties of the quantum device, which enter as parameters in the differential equation model that is used for optimizing the controls.

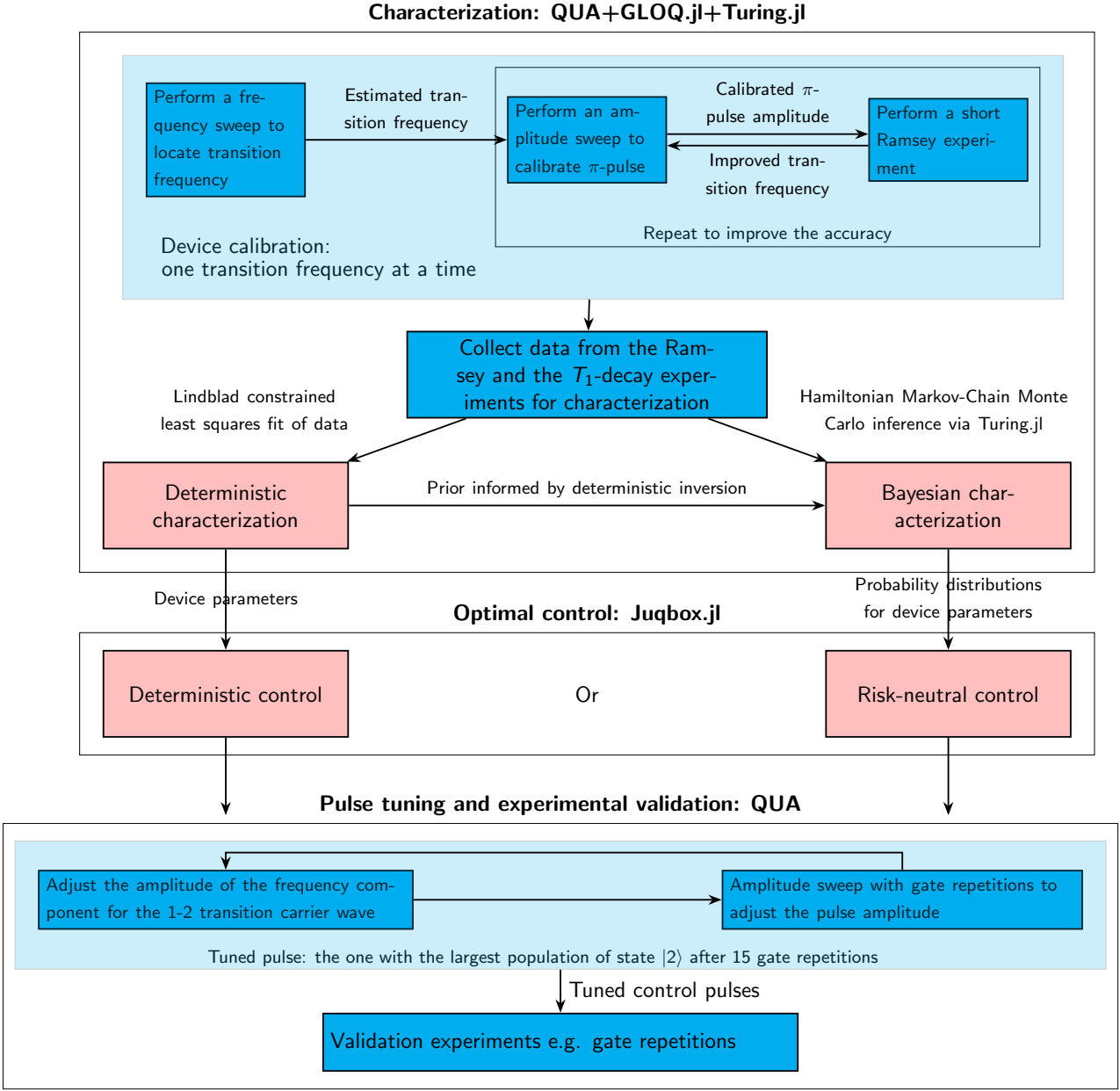
\begin{figure}[h!]
\begin{center}
\scalebox{0.87}{
\begin{tikzpicture}  
  \tikzset{every node}=[font=\sffamily\sansmath]
  \node[block_device_small,text width=2.5cm] (AmpSweep) {\footnotesize Perform an amplitude sweep to calibrate $\pi$-pulse};
  \node[block_device_small,left=of AmpSweep,xshift=-2.5cm,text width=2.5cm] (FreqSweep) {\footnotesize Perform a frequency sweep to locate transition frequency };
  \node[block_device_small,right=of AmpSweep,xshift=2.5cm,text width=2.5cm] (ShortRamsey) {\footnotesize Perform a short Ramsey experiment};
  \node[draw,inner xsep=2mm,inner ysep=8mm,yshift=-3mm,fit=(AmpSweep)(ShortRamsey)](RepeatCon){};
  \node[below=of RepeatCon,yshift=1.5cm](Repeat){\footnotesize{Repeat to improve the accuracy}};
  \node[below=of AmpSweep,yshift=+0.2cm](CalibrateDevice){\hspace{-12cm} Device calibration:};
  \node[below=of AmpSweep,yshift=-0.2cm](CalibrateDevice){\hspace{-9.45cm} one transition frequency at a time};
  \node[draw,fill=cyan,opacity=.2,text opacity=1,inner xsep=4mm,inner ysep=2mm,fit=(FreqSweep)(AmpSweep)(ShortRamsey)(CalibrateDevice)(RepeatCon)](Preparation){};
  \node[block_device,below=of CalibrateDevice,text width=6.0cm,yshift=0.25cm] (CollectData) {Collect data from the Ramsey and the $T_1$-decay experiments for characterization}; 
  \node[block_laptop,below left=of CollectData,yshift=0.25cm,xshift=-0.0cm] (DetChar) {Deterministic characterization};  
  \node[block_laptop,below right=of CollectData,yshift=0.25cm] (BayChar){Bayesian characterization};
  \node[draw,inner xsep=4mm,inner ysep=4mm,fit=(Preparation)(CollectData)(DetChar)(BayChar)(FreqSweep),label={:\textbf{Characterization: QUA+GLOQ.jl+Turing.jl}}](Char){};
  \node[block_laptop,below=of DetChar,yshift=-0.75cm] (DetCon){Deterministic control};
  \node[block_laptop,below=of BayChar,yshift=-0.75cm] (RNCon){Risk-neutral control}; 
  \node (ConChoice) at ($(DetCon)!0.5!(RNCon)$) {Or}; 
  \node[draw,inner xsep=4mm,inner ysep=3mm,fit=(DetCon)(RNCon)(ConChoice),label={:\textbf{Optimal control: Juqbox.jl}}](Con){};
  \node (CalibrationAmpSweep) [block_device_small,text width=6.0cm, below = of RNCon,yshift=-1.3cm,xshift=-0.8cm] {\footnotesize Amplitude sweep with gate repetitions to adjust the pulse amplitude };
  \node (CalibrationRescale) [block_device_small,text width=6.25cm, below = of DetCon,
  yshift=-1.3cm] {\footnotesize Adjust the amplitude of the frequency component for the $1$-$2$ transition carrier wave}; 
  \node[below=of CalibrationAmpSweep,yshift=0.8cm,xshift=-6.25cm](CalibrationResult){\footnotesize{Tuned pulse: the one with the largest population of state $|2\rgl$ after $15$ gate repetitions}};
  \node[draw,inner xsep=4mm,inner ysep=2mm,yshift=3mm,fit=(CalibrationAmpSweep)(CalibrationRescale)(CalibrationResult),fill=cyan,opacity=.2](Calibration){};
  \draw [arrow] (CalibrationAmpSweep.north) -- ++(0,0.3) -- ++(-11.77,0)--(CalibrationRescale.north);
 \draw [arrow](CalibrationRescale)--(CalibrationAmpSweep);
  \node(OtherExps)[block_device_small,below=of Calibration,yshift=0.25cm,text width=8.0cm]{Validation experiments e.g. gate repetitions};
  \draw [arrow](Calibration)--node[right]{Tuned control pulses}(OtherExps);
  \node[draw,inner xsep=2mm,inner ysep=3mm,fit=(Calibration)(OtherExps),label={:\textbf{Pulse tuning and experimental validation: QUA}}](Exps){};
  \draw [arrow](DetCon.south)-- ++(0,-1.35) (Exps);
  \draw [arrow] (RNCon.south) -- ++(0,-1.35) (Exps);
  \draw[arrow] (FreqSweep)-- node [text width=3.0cm,midway,above,text centered] {\footnotesize{Estimated transition frequency}}(AmpSweep);
  \draw[arrow] ([yshift=0.1cm]AmpSweep.east)-- node [text width=3.0cm,midway,above,text centered] {\footnotesize{Calibrated $\pi$-pulse amplitude}}([yshift=0.1cm]ShortRamsey.west);
  \draw[arrow] ([yshift=-0.1cm]ShortRamsey.west)-- node [text width=3.0cm,midway,below,text centered] {\footnotesize{Improved transition frequency}}([yshift=-0.1cm]AmpSweep.east);  
  \draw[arrow] (Preparation)--(CollectData);
  \draw[arrow] (CollectData)-- node [text width=5.0cm,midway,left,text centered,yshift=0.5cm,xshift=0.0cm] {\footnotesize{Lindblad constrained least squares fit of data}}(DetChar); 
  \draw[arrow] (CollectData)-- node [text width=7.0cm,midway,right,text centered,yshift=0.5cm,xshift=-0.75cm] {\footnotesize{Hamiltonian Markov-Chain Monte Carlo inference via Turing.jl}}(BayChar);
  \draw[arrow] (DetChar) -- node [text width=8.5cm,midway,above,text centered] {\footnotesize{Prior informed by deterministic inversion}}(BayChar);
  \draw[arrow] (DetChar)-- node [text width=3.5cm,midway,text centered] {\footnotesize{Device parameters}}(DetCon);
  \draw[arrow] (BayChar)-- node [text width=3.5cm,midway,text centered] {\footnotesize{Probability distributions for device parameters}}(RNCon);
\end{tikzpicture}
}
\end{center}
\caption{Flowcharts for characterization, optimal control, control pulse tuning and experimental
  validation. The steps performed on the quantum device are inside the blue boxes and the steps
  performed on a classical computer are inside the pink boxes. \label{fig:whole_flow}} 
\end{figure}

For the full ``turning on the power switch'' to ``getting numbers on the screen'' process there are additional steps to perform before and after the characterization and control optimization.  Figure
\ref{fig:whole_flow} illustrates the flow of the entire process described in this report. These include i) the calibration and characterization of the system parameters of the device, ii) the
optimization to find the control pulses for realizing a unitary gate, and iii) the tuning of control pulses and state tomography for verifying the unitary gate transformation.


The rest of the report is organized as follows. Section \ref{sec:prelim} describes the hardware and the mathematical models of it. Section \ref{sec:control} presents the key ideas and crucial details of deterministic and risk neutral optimal control, using control pulses based on carrier waves with envelopes parameterized by B-splines. Section \ref{sec:characterization}, the characterization of system parameters and their probability distributions is presented, which is needed for the risk neutral optimal control method. Finally, in Section \ref{sec:calibration}, we present how to apply, calibrate and validate the generated control pulse on the actual quantum device.

\section{Preliminaries}\label{sec:prelim}
In this section we give a brief description of the quantum device, outline the mathematical model for this device, state the rotating wave approximation, describe what a $\pi$-pulse is and what essential and guard states are. We also discuss the basic elements of the quantum measurement and state classification.

\subsection{The LLNL QuDIT quantum computing device}
The experiments in this study are performed on one of the quantum devices within the Quantum Device and Integration Testbed (QuDIT) at LLNL. This device is a tantalum-based superconducting transmon \cite{place2021new}.  
On a practical level, users operate the QuDIT testbed with codes written in Python. To operate the arbitrary waveform generator (an OPX instrument from Quantum Machines) we also use the {\tt QUA} \cite{QUA} programming language.


To drive the QuDIT device and read out the result, we use IQ mixers to generate microwave pulses. For generating control signals to manipulate the quantum device, the IQ mixer takes an intermediate-frequency (IF) envelope signal and mixes it with a local oscillator (LO) base signal. The IF signal consists of in-phase (I) and quadrature (Q) components, where the frequency content is on the order of a few hundred MHz. The frequency of the LO base signal is fixed and typically around a few GHz. 
For reading out the state of the quantum device, a measurement signal with a frequency on the order of a few GHz is down-converted to a few hundreds MHz via an IQ mixer. This demodulated signal is further analyzed with an OPX device to distinguish between different quantum states.

\subsection{Notation}\label{sec:notation}

This document uses a mix of matrix-vector and Dirac bra-ket notation. In matrix-vector notation,
column vectors are set in boldface font with lower case symbols, e.g. $\psib$ or $\vb$. Upper case letters denote matrices (operators),
but we follow the convention in quantum physics and denote the density matrix by $\rho$. The
Hermitian conjugate (conjugate transpose) of a matrix $A$ is denoted by $A^\dagger$ and the
Hermitian conjugate of a column vector $\vb$ is the row vector $\vb^\dagger$.

Dirac bra-ket notation~\cite{Nielsen-Chuang} is often used in the quantum physics literature. It is related to
matrix-vector notation through
\[
| v\rangle \equiv \vb,\quad \langle v | \equiv \vb^\dagger,\quad 
\langle u | A | v \rangle \equiv
\langle \ub, A \vb\rangle = \ub^\dagger A \vb,
\quad \langle u | v \rangle \equiv
\langle \ub,\vb\rangle = \ub^\dagger \vb.
\]
Here the standard $\ell_2$ scalar product and norm for vectors $\ub$ and $\vb$ in $\mathbb{C}^N$
are defined by
\begin{equation}\label{eq_sp+norm}
  \langle \ub, \vb \rangle = \sum_{j=0}^{N-1} {u}^*_j v_j,\quad \| \vb\| = \sqrt{\langle \vb, \vb \rangle},
\end{equation}
where $u_j^*$ denotes the complex conjugate of $u_j\in\mathbb{C}$.

The terms (linear) operator and matrix are used interchangeably, referring to a
linear mapping (morphism) between two vector spaces. We remark that there is no ambiguity in this notation because we only consider finite dimensional vector spaces. In that case, all linear operators can be represented by a matrix once a basis has been selected.

\subsection{Mathematical models\label{sec:model}}
Quantum systems that are isolated from their environment are called closed systems. The state of a closed system can be modeled by Schr\"{o}dinger's equation:
\begin{align}
    \dot{\psib} = -i H(t)\psib, \label{eq:schrodigner}
\end{align}
where $\psib\in\mathbb{C}^N, ||\psib||_2^2=1$ is the state vector of the quantum system and the Hamiltonian $H(t)\in\mathbb{C}^{N\times N}$ is a Hermitian matrix, i.e. $H^\dagger = H$. Although all quantum devices do interact with the environment, the Schr\"{o}dinger equation is often a very accurate model. In particular, solutions to the Schr\"{o}dinger equation are good approximations when the time duration of the control pulses needed to implement unitary gates are short. By short we mean times over which decoherence processes, if accounted for, would only cause minor perturbations to the Schr\"{o}dinger solution.

In practice, all quantum systems interact with their environment; such systems are called open. The state of an open quantum system is described by a density matrix $\rho \in \mathbb{C}^{N \times N}$. Lindblad's master equation~\cite{lindblad1976generators,Nielsen-Chuang} can be used to model the evolution of the density matrix under the assumption of Markovian interactions between the quantum system and its environment,
\begin{align}
    \dot{\rho} = -i\left(H\rho - \rho H\right) + \sum_{j=1}^{N^2 - 1} \left( {\cal L}_{j} \rho {\cal L}_{j}^\dagger -
\frac{1}{2}\left( {\cal L}_{j}^\dagger{\cal L}_{j}\rho + \rho{\cal L}_{j}^\dagger{\cal L}_{j} \right) \right).
\label{eq:lindblad}
\end{align}

For the superconducting transmon qudit under consideration $N = 3$ and the Hamiltonian is on the form
$H(t)=H_s+H_c(t)$, where $H_s$ and $H_c(t)$ denote the system and control Hamiltonian matrices, respectively. The system Hamiltonian $H_s$ is a diagonal matrix 
\begin{equation}
\label{eq:system_hamiltonian}
H_{s} =\left(
 \begin{array}{cccc}
0 & 0 & 0 & 0 \\
0 & \omega_{0,1} & 0 & 0 \\
0 & 0 & \omega_{0,1} + \omega_{1,2} & 0\\
0 & 0 &  0 & \omega_{0,1} + \omega_{1,2}+\omega_{2,3} \\
\end{array}
\right),
\end{equation}
where $\omega_{k,k+1}$, for $k=0,1,2$, denotes the angular transition frequency between quantum states 
$|k\rgl$ and $|k+1\rgl$. 

In the laboratory frame of reference, the control Hamiltonian $H_c(t)$ is on the form \cite{gerry2005introductory}:
\begin{equation}
    H_c(t)=(a+a^\dagger)f(t),
\end{equation}
where $a$ and $a^\dagger$ are the lowering and the raising operators. Further, $f(t)$ is the real-valued control function, given by
\begin{align}
f(t) &= 2\,\mbox{Re}\{e^{i\omega_d t} d(t)\} = 2p(t)\cos(\omega_d t)
- 2q(t) \sin(\omega_d t)\notag\\
& = 2I(t)\cos(\omega_d t)+2Q(t)\sin(\omega_dt).
    \label{eq:control-Hamiltonian-IQ}
\end{align}
Here, $\omega_d$ is the angular drive frequency, $d(t)$ is a slowly varying envelope function,
$p(t)=\textrm{Re}(d(t))$ and $q(t)=\textrm{Im}(d(t))$. The functions $I(t)=p(t)$ and $Q(t)=-q(t)$
are called the in-phase and the quadrature components of the control function.  These functions are
typically used as input signals to an IQ mixer, which generates the signal $f(t)$ that is sent to
the qudit device.


In the following, we will only consider two decoherence operators in Lindblad's equation: the decay operator
$\mathcal{L}_1$  and the dephasing operator  $\mathcal{L}_2$. They are defined by
\begin{equation}
{\cal L}_{1} = \begin{pmatrix}
0 & \sqrt{\gamma_{1,1}} & 0 & 0 \\
0 & 0 & \sqrt{\gamma_{1,2}} & 0  \\
0 & 0 & 0 & \sqrt{\gamma_{1,3}} \\
0 & 0 & 0 & 0
\end{pmatrix} \; \text{and} \; \;
{\cal L}_{2} = \begin{pmatrix}
0 & 0 & 0 & 0\\
0 &\sqrt{\gamma_{2,1}}& 0 & 0\\
0 & 0 & \sqrt{\gamma_{2,2}} & 0 \\
0 & 0 & 0 & \sqrt{\gamma_{2,3}}
\end{pmatrix}.
\label{eq:lindblad_terms}
\end{equation}
The parameter $\gamma_{1,k}$ is the decay rate for state $|k\rgl$. It is related to
the corresponding decay time by $\gamma_{1,k}=1/\sqrt{T_{1,k}}$. Similarily, the dephasing rate
$\gamma_{2,k}$ is determined by the pure dephasing time $T_{2}$ (see Appendix
\ref{sec:app:T2-gamma} for details).

\begin{figure}[t]
  \begin{center} 
  \includegraphics[width=0.6\textwidth]{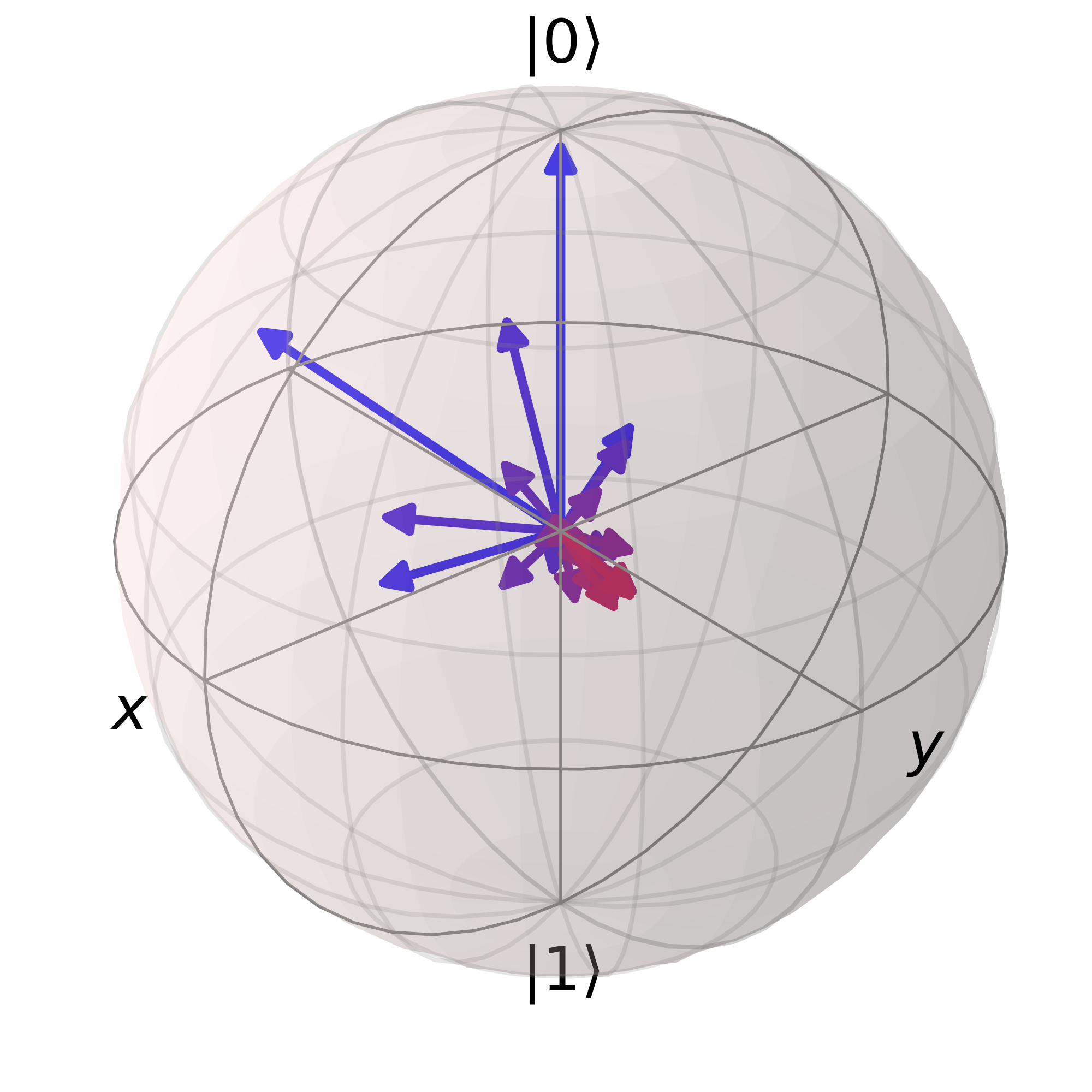}
  \caption{Bloch sphere representation of an evolution of the $|0\rgl$ state (from blue to red) due to decoherence effects.\label{fig:bloch}}
 \end{center}
\end{figure}

\pzc{The decay effect describes the energy dissipation due to the loss of energy from a quantum system. The pure dephasing effect damps out the off-diagonal elements of the density matrix that represent the phase relation between different states, without changing the population of the states. We refer to Chapter 8 of \cite{nielsen2002quantum} for more detailed descriptions of these effects. Both effects lead to decoherence of the quantum system. In Figure \ref{fig:bloch} we plot the state vector of a two level system on a Bloch sphere (c.f.~\cite{qiskit_url_bloch}) to demonstrate how the decoherence changes a quantum system. The decay time $T_1$ and the pure dephasing time $T_2$ represent the individual time scales of decay and pure dephasing, respectively. According to  \cite{tempel2011relaxation} the combined time scale of decoherence, $T_{2}^*$, is determined by $T_1$ and $T_2$:
\begin{equation}
\frac{1}{T_2^*} = \frac{1}{2T_1} + \frac{1}{T_2}.
\end{equation}
}

\subsubsection{Rotating wave approximation\label{sec:rotating_wave}}
The equations above are stated in the laboratory frame of reference and the search for an optimal control can, in principle, be performed there. However, as the transition frequencies and the drive frequency are high, typically in the \SI{}{\giga\hertz} range, this makes the time integration of the Sch\"{o}dinger equation computationally expensive. The expense being large due to the sampling requirements for the (highly) oscillatory controls and state vector. A more slowly varying description of the controls and the state vector can be obtained by applying the so called rotating wave approximation (RWA). 

The RWA consists of a change of variables followed by an approximation. First, we apply the rotating frame transformation 
\begin{align}
    R(t)=\exp(i\omega_d a^\dagger a t), \label{eq:rotating_frame_transformation}
\end{align}
and in the rotating frame, the Hamiltonian of the qudit device \eqref{eq:system_hamiltonian} becomes:
\begin{align}
    H_\textrm{rot}(t)&=R(t)(H_s+H_c(t))R(t)^\dagger-iR(t)\dot{R}(t)^\dagger\notag\\
     &=H_s^{rw}+p(a+a^\dagger)+iq(a-a^\dagger)
     +(p-iq)e^{-i2\omega_d t}a+(p+iq) e^{i2\omega_d t}a^\dagger,\label{eq:hamiltonian_rot}
\end{align}
where 
\begin{equation}
    H_s^{rw}=H_s- \omega_d \,a^\dagger a. \label{eq:rw_system_hamiltonian}
\end{equation}
The second step of the RWA is to drop the rapidly oscillating  terms $(p-iq)e^{-i2\omega_d t}a$ and $(p+iq) e^{i2\omega_d t}a^\dagger$ in the control Hamiltonian. This results in an approximation of the control Hamiltonian,  
\begin{align}
    \widetilde{H}_c(t)&=p(a+a^\dagger)+iq(a-a^\dagger)\notag\\
    &=I(t)(a+a^\dagger)-iQ(t)(a-a^\dagger). \label{eq:rwa_control}
\end{align}
All the computational parts of this report are performed in the rotating frame and we always make the rotating wave approximation.

\subsubsection{Essential and guard states}
Following \cite{palao2008protecting,petersson2021optimal}, we divide the states of the qudit into essential states and guard states. The essential states are the states used for quantum information processing. These states correspond to the lowest energy levels in the system. The guard states are states corresponding to higher energy levels. The purpose of these guard states is to act as a buffer between the essential states and even higher states, which are excluded from the computational model. During optimal control, population of the guard states is discouraged through a penalty term in the objective function.

\subsection{Control pulses for state preparation and characterization}
\subsubsection{The $\pi$ and $\pi/2$ pulses}
The $\pi$ pulse is defined as a pulse that transforms the state from one energy level to the next. For example, a $\pi_{k,k+1}$ pulse drives the quantum device from state $|k\rgl$ to $|k+1\rgl$ and vice versa. Similarly, $\pi/2_{k,k+1}$ brings the state from $|k\rgl$ to $(|k\rgl+|k+1\rgl)/\sqrt{2}$. The $\pi$ pulse on the QuDIT testbed is a fixed frequency and constant amplitude pulse with duration $152\,\mathrm{ns}$, both for the $0-1$ and the $1-2$ transitions. The amplitude of the pulse is calibrated by sweeping the amplitude and monitoring the measurement outcome. The pulse for transitioning from the $|0\rgl$ state to $|1\rgl$ state, is called the $\pi_{0,1}$ pulse. Similarly, either half of the $\pi$ pulse amplitude or half of the duration defines the $\pi/2$ pulse.
The constant amplitude of the $\pi$ pulse in the rotating frame makes an analytical model straightforward to drive (see Appendix~\ref{sec:rabi}).



\subsubsection{Ramsey, $T_1$ decay and Hahn-Echo experiments\label{sec:exps}}

The most commonly used experiments to determine the parameters in the Lindblad equation (\ref{eq:lindblad}) are the (a) Ramsey, (b) $T_1$ decay and (c) Hahn-Echo experiments. These experiments will be mentioned throughout the document. Here we only give an overview of typical results and state what they measure. We defer to Section \ref{sec:exp-protocol} for a more detailed description.  

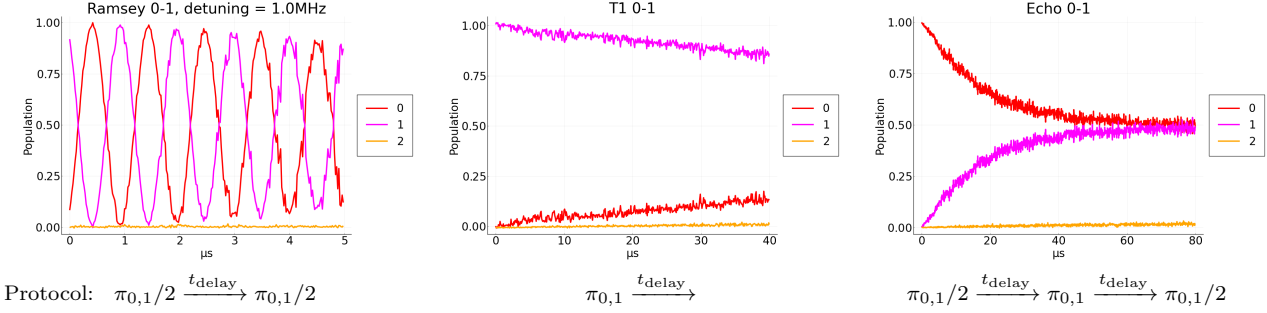
\begin{figure}[]
  \begin{center} 
  \begin{tikzpicture}
    \node (ImgRamsey){\includegraphics[width=0.31\textwidth]{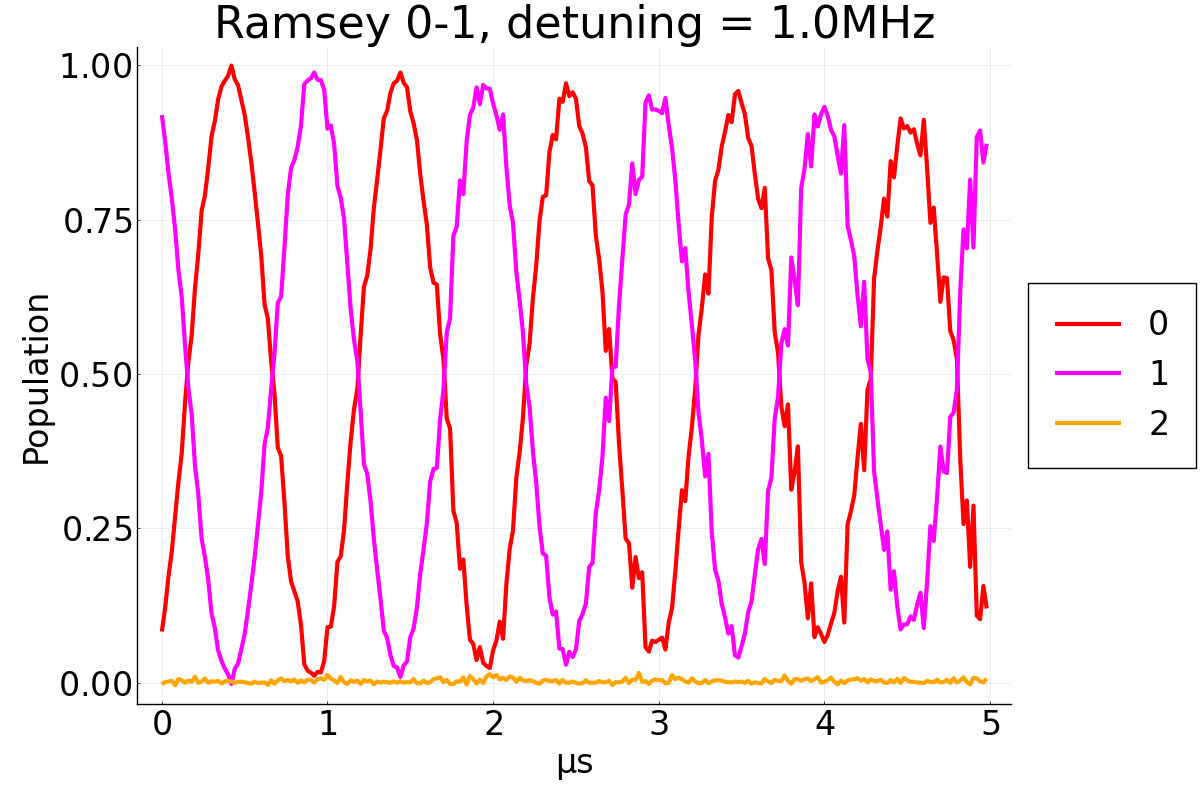}};
    \node(Ramsey)[below=of ImgRamsey,yshift=+1.2cm]{\scriptsize$\pi_{0,1}/2\xrightarrow{t_{\textrm{delay}}}\pi_{0,1}/2$};
    \node(Ramsey)[left=of Ramsey,yshift=-0.05cm,xshift=1.0cm]{\scriptsize Protocol:};
  \end{tikzpicture}
  \begin{tikzpicture}
  \node (ImgT1){\includegraphics[width=0.31\textwidth]{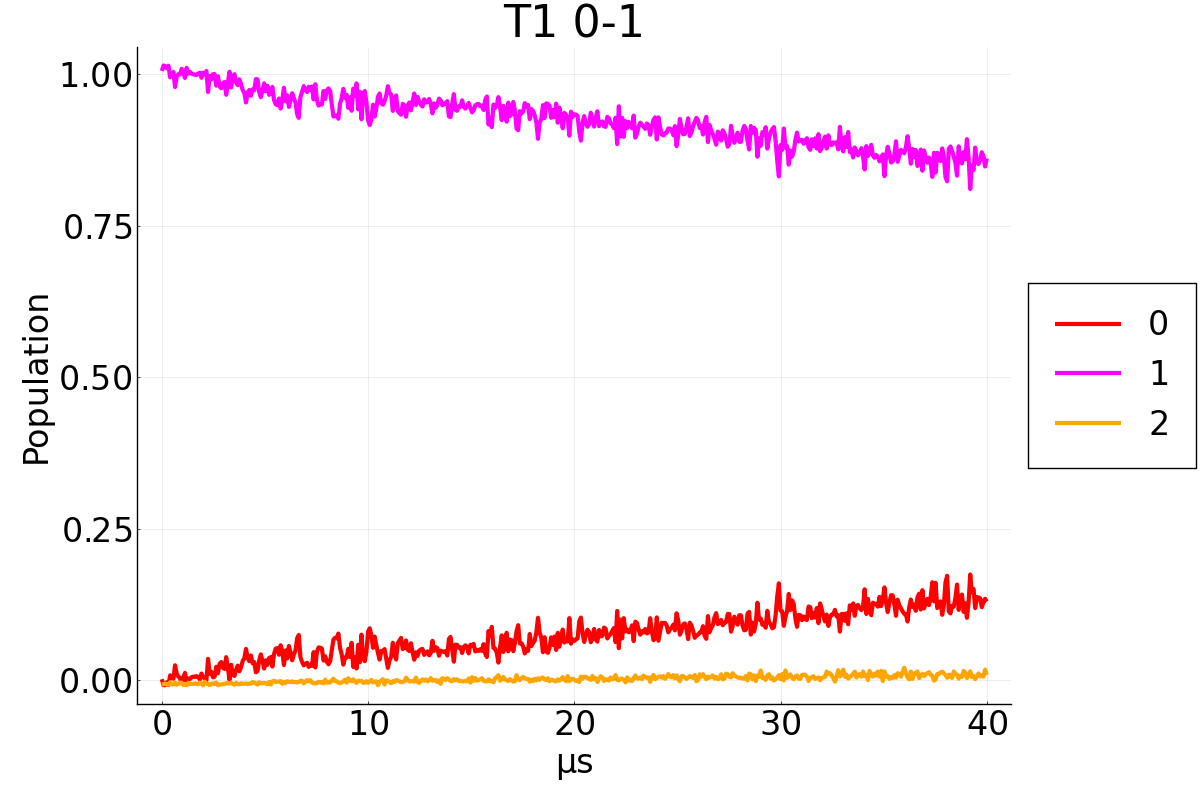}};
  \node(T1)[below=of ImgT1,yshift=+1.2cm]{\scriptsize$\pi_{0,1}\xrightarrow{t_{\textrm{delay}}}$};
  \end{tikzpicture}
  \begin{tikzpicture}
  \node(ImgEcho){\includegraphics[width=0.31\textwidth]{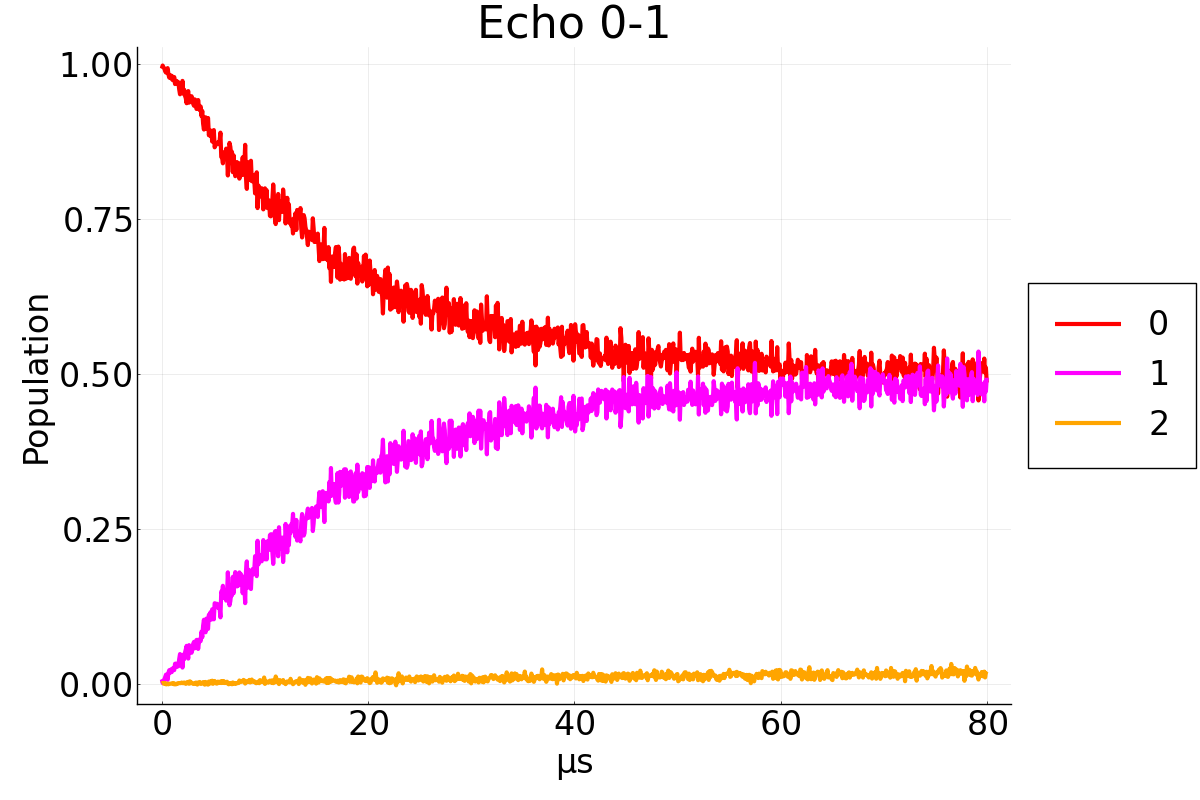}};
  \node(Echo)[below=of ImgEcho,yshift=+1.2cm]{\scriptsize$\pi_{0,1}/2\xrightarrow{t_{\textrm{delay}}}\pi_{0,1}\xrightarrow{t_{\textrm{delay}}}\pi_{0,1}/2$};
  \end{tikzpicture}
  \caption{Left: Ramsey 0-1 experiment with \SI{1}{\mega\hertz} nominal detuning. Middle: $T_1$ decay experiment for the $0$-$1$ transition. Right: Hahn-Echo experiment for the $0$-$1$ transition. The protocol of each experiment is displayed under experimental results, and $\pi_{0,1}$, $\pi_{0,1}/2$ stand for $\pi$ and $\pi/2$ pulses for the $0$-$1$ transition respectively.\label{fig:exps}}
 \end{center}
\end{figure}

The Ramsey experiment determines the decoherence time $T^{*}_{2,k}$ of the $|k\rgl$ state. 
A typical measurement of a Ramsey $0$-$1$ experiment is presented on the left in Figure \ref{fig:exps}. Note that the population of state $|0\rgl$ and $|1\rgl$ are both sinusoidal in the delay time, with decaying amplitudes. The frequency of the oscillation is equal to the detuning frequency, $\Delta_1=\omega_{0,1}-\omega_d$, and the envelope decays roughly as $\exp(-t/T_{2,1}^*)$. We can also estimate transition frequencies with the Ramsey experiment. Given the drive frequency $\omega_d$, the transition frequency $\omega_{0,1}$ follows by estimating $\Delta_1$ from data.

To find the decay time $T_1$, we measure the population change of a state as a function of delay time time. A typical result presented in the middle of Figure \ref{fig:exps}. The population of state $|1\rgl$ decays exponentially with time that follows the relation $\exp(-t/T_{1,1})$. Experimentally, the decay can be measured by changing the delay time between $\pi$ pulse and the readout. 


The Hahn-Echo experiment, very similar to Ramsey experiment, is designed to measure the pure dephasing time $T_2$. The state is intially prepared to a superposition state. The $\pi$ pulse applied between the two $\pi/2$ pulses cancels accumulated incoherent phase errors.
A typical result of this experiment is presented on the right of Figure \ref{fig:exps}. Here, the population of state $|1\rgl$ decays exponentially with respect to the delay time $t$, as $\exp(-t/T_{2,1})$, where $T_{2,1}$ is the decoherence time for the $0$-$1$ transition. 


\subsection{Accounting for parity events in the mathematical model\label{sec:parity}} 

The transmon qudit exhibits so called parity events \cite{riste2013millisecond}. These random events occur about every millisecond and perturb the transition frequencies of the device through flips of the charge parity. The perturbation of the transition frequency is called charge dispersion. It is larger for higher energy levels than it is for the lower ones. 

\begin{figure}[]
  \begin{center} 
  \includegraphics[width=0.49\textwidth]{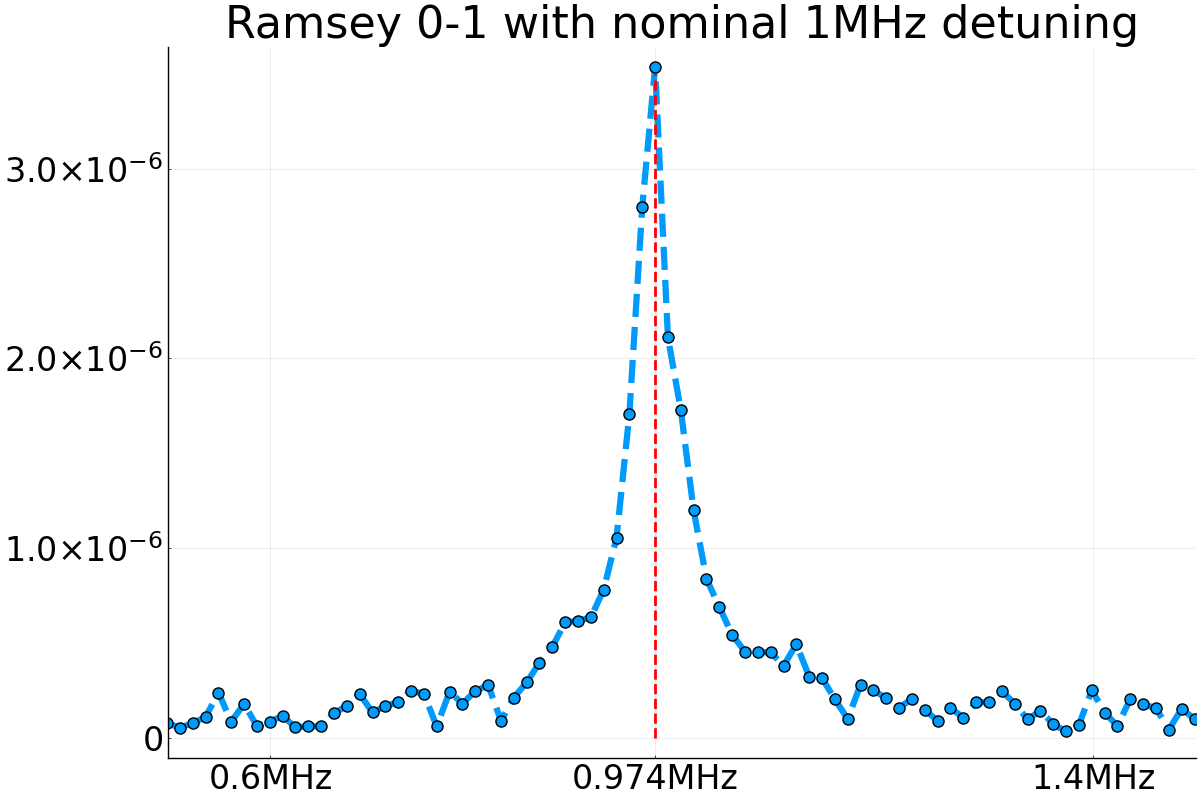}
  \includegraphics[width=0.49\textwidth]{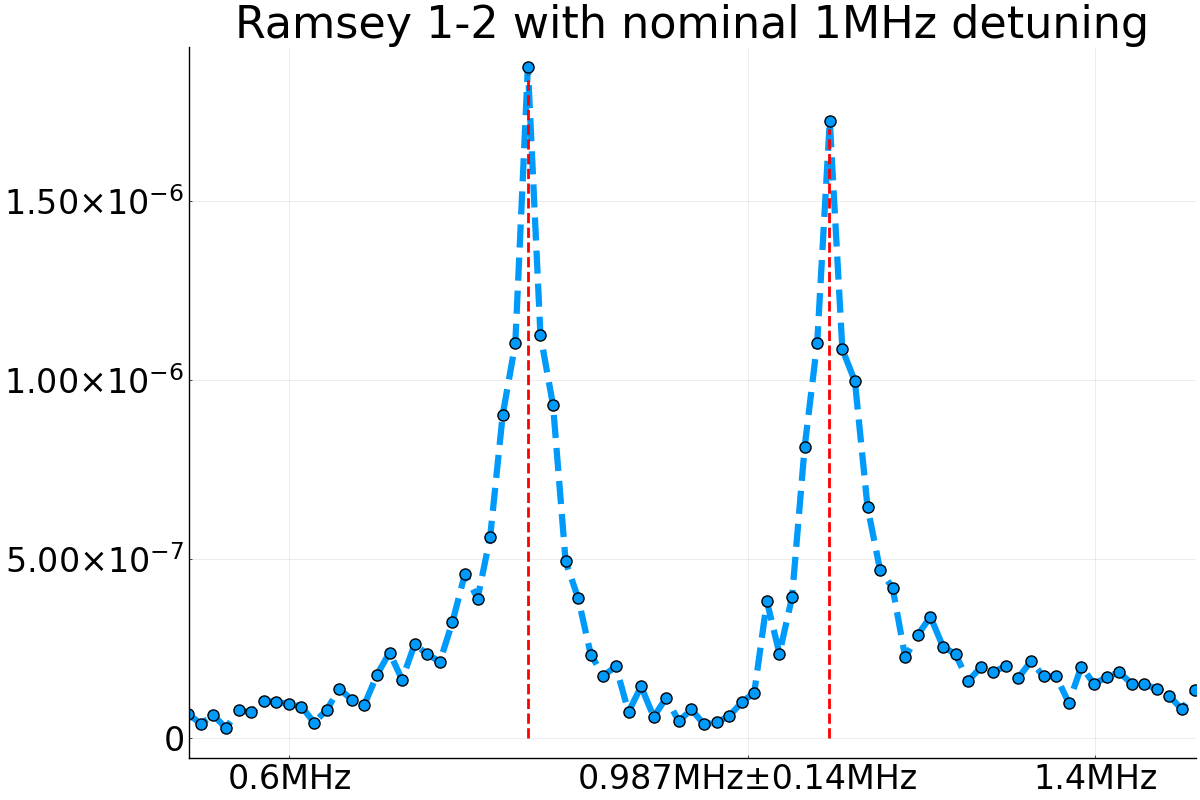}
  \caption{Ramsey curves with $80\mu s$ as the longest delay time in the frequency domain. Left: magnitude of FFT of the population for state $|1\rgl$. Right: magnitude of FFT of the population for state $|2\rgl$.\label{fig:ramsey_fft}}
 \end{center}
\end{figure}

The presence of parity events can be detected by measuring the transition frequency $\omega_{k,k+1}$ using a Ramsey experiment. Because Ramsey data oscillates with frequency $\Delta_k$, where $\Delta_k=\omega_{k,k+1}-\omega_d$ is the detuning relative to the transition frequency, it is expected that the spectral amplitude will exhibit a maximum at the detuning frequency. A typical spectrum of the Ramsey data for the $0$-$1$ transition is displayed on the left of Figure \ref{fig:ramsey_fft}. As can be seen, there is only a single peak. On the other hand, the spectrum on the right of the same Figure displays the same quantity for the $1$-$2$ transition. Here we note two distinctive peaks. Given these results, we surmise that the charge dispersion can be neglected for the $0$-$1$ transition, but must be included in the model for the $1$-$2$ transition.

Mathematically, the parity flip in the $1$-$2$ transition frequency can be modeled as follows:
\begin{align}
    \omega_{1,2} = \bar{\omega}_{1,2} + p\, \epsilon_{1,2}, \quad p\in\{-1,1\}, 
\end{align}
where $\bar{\omega}_{1,2}$ is the average 1-2 transition frequency, $\epsilon_{1,2}$ is the charge dispersion and $p\in\{-1,1\}$ is a discrete random variable called parity. The parity $p$ takes values $\pm 1$ with equal probability \cite{riste2013millisecond}. We define the frequency corresponding to the positive/negative parity flip as 
\[
\omega^\pm_{1,2}=\bar{\omega}_{1,2}\pm\epsilon_{1,2},
\]
and note that  $\omega_{1,2}^\pm$ are the frequencies corresponding to the two peaks in Figure \ref{fig:ramsey_fft}. Since $p$ has mean zero, the average 1-2 transition frequency and the charge dispersion are $\bar{\omega}_{1,2} = \frac{1}{2}(\omega_{1,2}^++\omega_{1,2}^-)$ and $\epsilon_{1,2}=\frac{1}{2}(\omega_{1,2}^+-\omega_{1,2}^-)$, respectively.

According to \cite{riste2013millisecond}, the approximate time between parity events is on the order of milliseconds. This time is much longer than the duration of a single shot of the experiments we perform here, which typically are over in a few microseconds. Although a parity event could take place during a single shot, the time scales are such that this would be rare. On the other hand, the population of different states is measured by averaging over 1000 repeated shots. In our experiments the waiting time between successive shots for the same delay time varies between \SI{0.05}{\milli\second} and \SI{0.1}{\milli\second}, depending on the particular experiment. As a result it is likely that different shots correspond to different values of the parity $p$. To account for both parities during the characterization we compute the density matrix as the average of the density matrices $\rho^{\pm}$. Here, $\rho^+$ and $\rho^-$ correspond to solving Lindblad's equation \eqref{eq:lindblad} with transition frequencies $\omega_{1,2}^+$ and $\omega_{1,2}^-$, respectively. 

\begin{figure}[htb]  
\begin{center}
\begin{tikzpicture}
   \node (img){\includegraphics[width=0.75\textwidth,trim={0.0cm 0.28cm 6.0cm 1.0cm},clip]{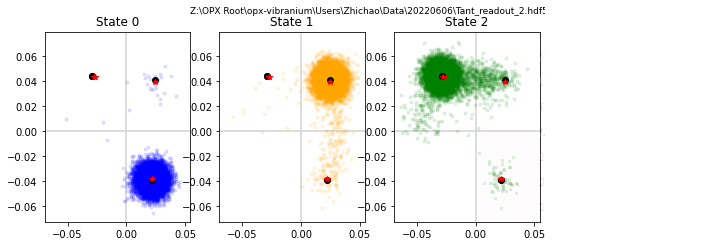}};
   \node[above =of img, node distance=0cm, minimum height=0.05cm,yshift=-1.25cm,xshift=0.5cm] {State $|0\rgl$\hspace{3.0cm}State $|1\rgl$\hspace{2.75cm}State $|2\rgl$};
   \node[below =of img, node distance=0cm, minimum height=0.05cm,yshift=1.0cm,xshift=0.5cm] {$I$\hspace{4.3cm}$I$\hspace{4.3cm}$I$};
   \node[left=of img, node distance=0cm, rotate=90, anchor=center,yshift=-1.2cm] {$Q$};
 \end{tikzpicture}
\end{center}
\caption{A typical measurement result preparing the device to states $|0\rgl$, $|1\rgl$ and $|2\rgl$ with $\pi$ pulses. $I$: in-phase component; $Q$: quadrature component. \label{fig:IQ-confusion}}
\end{figure}
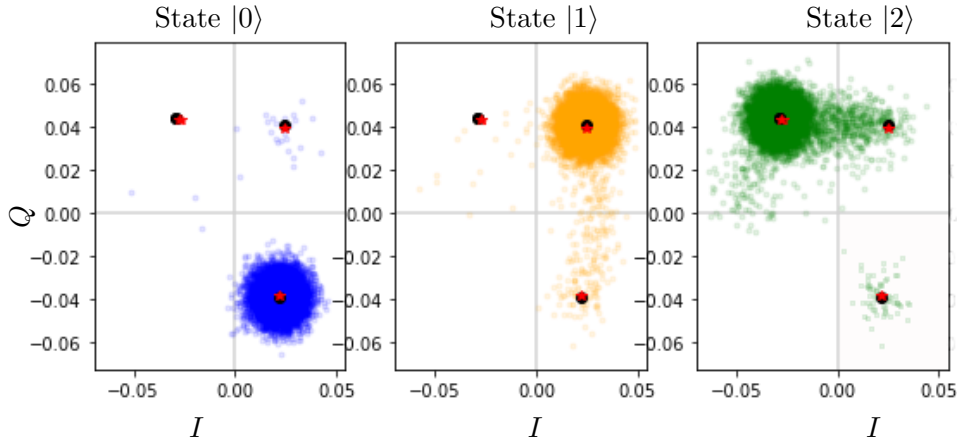

\subsection{Measurements and classification of quantum states\label{sec:measure-data}}
The elements of the state vector $\psib$ holds probability amplitudes and the density matrix $\rho$ holds population and coherence information of the state of the quantum
system. Unfortunately, neither of these representations can be directly measured. However, observable quantities such as position, momentum and energy can be measured. \pzc{The state of the quantum system can be inferred from the measurement of these observables. Here we start with the mathematical relation between measurements and population, then present how to classify quantum state based on measurement results in practice.}

\subsubsection{Classification of quantum state based on measurement results}
Repeated measurements are used to give statistics of the observables, and from this it is possible to infer the state of the quantum system, for example by statistical classifiers. 
Here, we present how a statistical classifier can be trained. 
Following \cite{blais2004cavity}, we drive the device with a series of $\pi$
pulses to prepare the QuDIT in different states, directly followed by measurement. 
In each shot, the readout pulse is demodulated into an in-phase component, $I$, and the quadrature component, $Q$, as shown in Figure \ref{fig:IQ-confusion}.
A typical training set, based on 80,000 shots, is presented in Figure
\ref{fig:IQ-confusion}. Each cluster in the $I$-$Q$ plane represents an energy level of the qudit. Using the readouts, we train a Gaussian mixture model (GMM) \cite{zhuang1996gaussian,xu2005survey} as a classifier that maps the given $(I,Q)$ point to a vector whose $k$-th element represents the probability of this point belonging to state $|k\rgl$. 

\pzc{Due to the errors in state preparation and the readout, the result of our measurement is not always accurate.}
The GMM provides a confusion matrix $C$ that can be used to account for \pzc{these errors}.
The confusion matrix can be seen as the transition probability matrix between measured and actual populations. 
The elements of this matrix, $c_{ij}$, represent the probability of measuring state $|i\rgl$ after the system is prepared in state $|j\rgl$. As an example, the confusion matrix provided by the GMM trained with the data in Figure \ref{fig:IQ-confusion} is
\begin{equation}
C = \left(
 \begin{array}{cccc}
            9.97125\mathrm{e}{-1}& 2.62500\mathrm{e-}{3} &2.50000\mathrm{e}{-4}\\
            1.67500\mathrm{e}{-2}& 9.81250\mathrm{e-}{1} &2.00000\mathrm{e}{-3}\\
            6.12500\mathrm{e}{-3}& 4.33750\mathrm{e-}{2} &9.50500\mathrm{e}{-1}
\end{array}
\right).
\end{equation}
The populations corresponding to the actual state vector $\psib$ can then be estimated from the measured populations, represented by the state vector $\phib$, by inverting the confusion matrix,
\begin{align}
    \begin{pmatrix}
    |\psi_0|^2\\
    |\psi_1|^2\\
    \vdots
    \end{pmatrix}
    = C^{-1}  \begin{pmatrix}
    |\phi_0|^2\\
    |\phi_1|^2\\
    \vdots
    \end{pmatrix}.
    \label{eq:error_mitigation}
\end{align}


 
\section{Constructing unitary gates by optimal control\label{sec:control}}
Our control pulses are generated by the Julia package {\tt Juqbox.jl} \cite{Juqbox-software, petersson2021optimal, petersson2020discrete} which utilizes $B$-splines acting as the envelope of carrier waves to parameterize the control function. The advantage of this approach over other popular methods such as the GRAPE algorithm \cite{khaneja2005optimal,johansson2012qutip,leung2017speedup}, which uses one control parameter per time step, is that the number of control parameters in {\tt Juqbox.jl} is independent and much smaller that the number of time steps; the resulting control functions also map directly onto the input of an IQ-mixer. 
Moreover, {\tt Juqbox.jl} provides a risk neutral optimization method that takes the uncertainty in the transition frequencies into account. This risk neutral approach provides an opportunity to design control pulses that are resilient to noise. 


\subsection{Optimal control with $B$-splines and carrier waves}
As mentioned in Section \ref{sec:rotating_wave}, we use the rotating wave approximation (RWA) throughout this document. 
Let $\balpha$ be vector that contains the control parameters. In {\tt Juqbox.jl} the optimal control pulse is determined by fixed carrier frequencies $\Omega_k$ and the control vector $\balpha$, which together define the envelope function 
\begin{align}
d(t;\balpha)=\sum_{k=1}^{N_c} (p_k(t;\balpha)+iq_k(t;\balpha))e^{it\Omega_{k}}, \ \ 
p_k(t;\balpha)=\sum_{b=1}^{N_b}\hat{S}_b(t)\alpha^{(p)}_{b,k}, \ \ 
q_k(t;\balpha)=\sum_{b=1}^{N_b}\hat{S}_b(t)\alpha^{(q)}_{b,k}.\label{eq:control_function}
\end{align}
Here $\Omega_k$ is the $k$-th carrier wave frequency, the basis functions $\hat{S}_b$ are piece-wise quadratic B-spline wavelets, $N_c$ is the number of carrier waves, and  $N_b$ is the number of B-spline coefficients. From \eqref{eq:control-Hamiltonian-IQ}, the envelope function corresponds to the laboratory frame control function
\begin{align}
    f(t) = 2\mbox{Re}\left\{
    \sum_{k=1}^{N_c} (p_k(t;\balpha)+iq_k(t;\balpha))e^{it(\omega_d+\Omega_{k})}
    \right\}.
\end{align}
In practice, the drive frequency $\omega_d$ is often fixed to equal the 0-1 transition frequency of the system. The carrier frequencies $\Omega_k$ therefore provide a natural way to drive the system at a different frequency, e.g., for generating detuned $\pi/2$ pulses during a Ramsey experiment, or for exciting higher energy levels in the system.

The objective function to be minimized by the optimal control
algorithm is defined as
\begin{align}
    \mathcal{G}(\balpha,H_s):=  {\cal J}_1(\balpha,H_s) + {\cal
    J}_2(\balpha,H_s),\label{eq:det_objective}
\end{align}
where $H_s$ is the system Hamiltonian, $\mathcal{J}_1$ is the gate infidelity and $\mathcal{J}_2$ measures leakage to guard states. Define the solution operator at time $t$ corresponding to the control Hamiltonian determined by $\balpha$ as $U_{\textrm{sol}}(t;\balpha)$, and let the duration of the control pulse be $T_g$. The gate infidelity 
\begin{align}
\mathcal{J}_1(\balpha,H_s) = \left(1-\left(\frac{1}{d_E}\left|\textrm{Tr}\big((V^\dagger U_\textrm{sol}(T_g;\balpha) \big)\right|\right)^2\right),
\end{align}
measures the difference between the target gate unitary $V$, 
and the unitary transformation determined by the control vector, $U_\textrm{sol}(T_g;\balpha)$. Here $d_E$ is the number of essential states. 
The leakage to guard states is defined as
\begin{align}
    \mathcal{J}_2({\balpha},H_s)=\frac{1}{T_g}\int_{0}^{T_g}\lgl U_\textrm{sol}(t;\balpha),WU_\textrm{sol}(t;\balpha)\rgl_F dt.
\end{align}
Here $W$ is a positive semi-definite weight matrix. In the case of a qudit, $W$ has zero elements in the upper left $d_E\times d_E$ corner and can have positive elements elsewhere.

In the following, we assume that the qudit device can be modeled by a system with three essential states and one guard state.  
We set the drive frequency of the rotating frame transformation in \eqref{eq:rotating_frame_transformation} as $\omega_d=\omega_{0,1}$.
 To design control pulses for the $0\leftrightarrow2$ SWAP gate, we use two carrier waves whose frequencies are $\Omega_1=0$ and $\Omega_2=\omega_{1,2}-\omega_{0,1}$ in the rotating frame. 

\subsection{Risk neutral optimization}\label{sec:risk-neutral}
In practice the coefficients of the system Hamiltonian are not precisely known, or may be subject
to time-dependent variations such as charge noise. For this reason, {\tt
  Juqbox.jl} also supports a risk neutral approach \cite{ge2021risk} for optimizing the control functions.
 
 Modeling the coefficients in the system Hamiltonian as random variables, collected in the vector $\zb$,  
 we now have that the system Hamiltonian, $H_s(\zb)$, is a function of $\zb$. In the risk neutral optimization mode, {\tt Juqbox.jl} uses the objective function
\begin{align}
\sum_{k=1}^{N_\textrm{quad}}w_k\mathcal{G}(\balpha,H_s(\zb_k)), \label{eq:riskneutral_obj}
\end{align}
where $\zb_k$ and $w_k$ are quadrature nodes and weights. This objective approximates 
\[
\mathbb{E}(\mathcal{G}(\balpha,H_s(\zb))),
\]
that is, the expectation of $\mathcal{G}(\balpha,H_s(\zb))$. 

\section{Characterization of the QuDIT device\label{sec:characterization}}

Before optimal control pulses can be designed with {\tt Juqbox.jl} (or any other optimal control code) it is necessary to first specify the Hamiltonian of the quantum device. In {\tt Juqbox.jl}, the system Hamiltonian can either be described by deterministic parameters (in our case the transition frequencies $\omega_{0,1}$, $\omega_{1,2}$ and $\omega_{2,3}$), or by probability distributions of the parameters in the Hamiltonian. The latter approach is needed to use the risk neutral optimization we described in Section~\ref{sec:risk-neutral}.

The characterization protocols described below take decoherence into account as quantified by the decay times ($T_{1,k}$) and dephasing times ($T^\ast_{2,k}$). These effects are included through a Lindblad master equation during the characterization. We remark that decoherence is not modeled in the {\tt Juqbox} code, but is available for optimal control in the code {\tt Quandary}.  

\begin{figure}[t]
  \begin{center} 
  \includegraphics[width=0.6\textwidth]{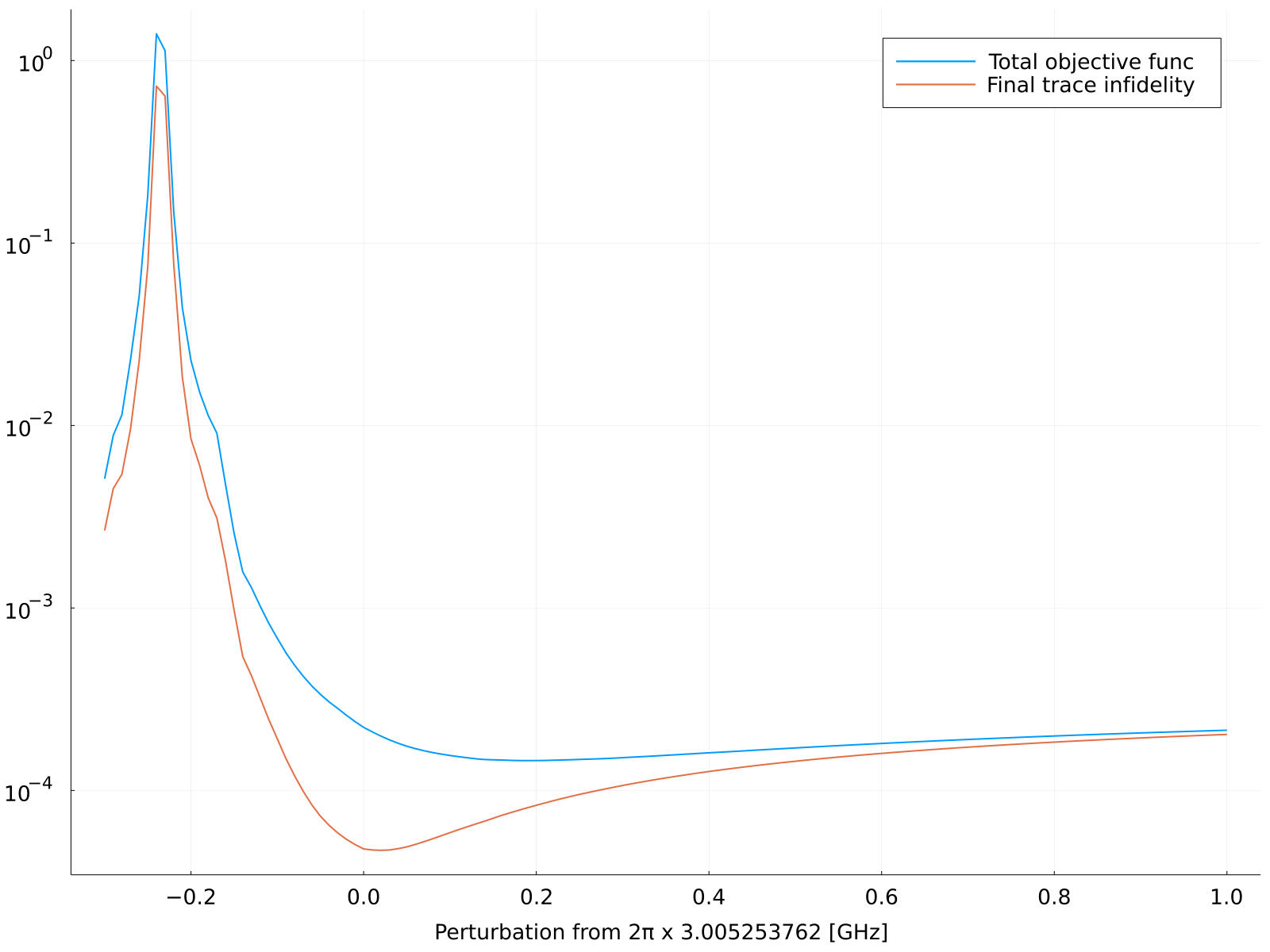}
  \caption{Total objective function and final trace infidelity as a function to the perturbation to the 2-3 transition frequency $\omega_{2,3}$.\label{fig:perturbation}}
 \end{center}
\end{figure}

In the QuDIT device we take levels $\{0, 1, 2\}$ to be essential and level 3 to be a guard level. A reason for only using levels $\{0, 1, 2\}$ is that it is difficult to accurately prepare and measure the QuDIT test bed in the state $|3\rgl$. As a consequence we cannot carry out the experiments required for characterizing the  2-3 transition frequency.  However, as the $|3\rangle$ state is not an essential state but a guard state, the value for $\omega_{2,3}$ does not need to be known with high accuracy. To support this claim we consider perturbations of the optimized control for a $0 \leftrightarrow 2$ SWAP gate using 3 essential and 1 guard level, with the transition frequencies $\omega_{0,1}$ and $\omega_{1,2}$ in Table \ref{tab:det}. For the guard state we use $\omega_{2,3}/2\pi = 3.005253762$~\SI{}{\giga\hertz} during the optimization of the control. We then perturb $\omega_{2,3}$ and perform forward simulations with the same control and evaluate the resulting objective function (\ref{eq:det_objective}). The results are displayed in Figure \ref{fig:perturbation}. We conclude that, unless the perturbation is such that $\omega_{2,3} \approx \omega_{1,2}$, the objective function is robust to perturbations of $\omega_{2,3}$.

\begin{figure}[h!]
\begin{center}
\begin{tikzpicture}
\tikzset{every node}=[font=\sffamily\sansmath]
\node (Calibrate) [block_device,text width=3cm] {Calibrate the device};
\node (Collect) [block_device,text width=3cm,right=of Calibrate] {Collect experimental data};
\node (Det) [block_laptop,right=of Collect,text width=3cm] {Deterministic characterization};
\node (MC) [block_laptop,right=of Det,text width=3cm] {Bayesian characterization with MCMC};
\draw[arrow] (Calibrate) -- (Collect);
\draw[arrow] (Collect) -- (Det);
\draw[arrow] (Det) --node [text width=1.5cm,midway,above,text centered,] {\footnotesize{Design priors}}(MC);
\end{tikzpicture}
\end{center}
\caption{Flowchart of the characterization procedure. MCMC stands for the Markov Chain Monte Carlo method. Blue: performed on the quantum device. Pink: performed on the classical device.\label{fig:ch_flow}}
\end{figure}
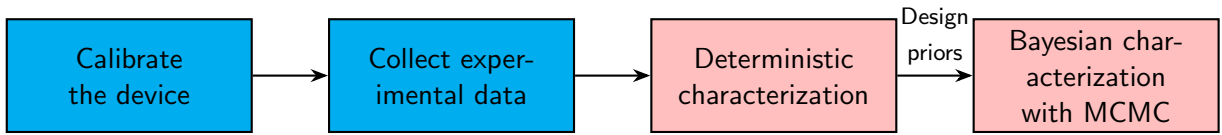

Because the value of $\omega_{2,3}$ is less critical and as the qudit device cannot be accurately prepared and/or measured in state  $|3\rgl$ we use a Lindblad equation with 3 essential levels and no guard level for the characterization
The characterization is carried out in multiple steps as outlined in Figure \ref{fig:ch_flow}. The first step is a  calibration of already implemented $\pi$-pulses on the device, described in the next subsection. After this calibration, we can  collect experimental data that can then be used to perform a deterministic or a Bayesian characterization as described below.  

\subsection{Calibrating $\pi$-pulses}
The initial step in the device characterization is to calibrate $\pi$-pulses. This calibration is carried out before the characterization to account for drift in device parameters that occurs on a timescale of hours to days. It follows a well-known procedure, described for example in \cite{qiskit_url_calibration}. The upshot of the calibration will be approximate transition frequencies and amplitudes of the $\pi$ pulses. 
 

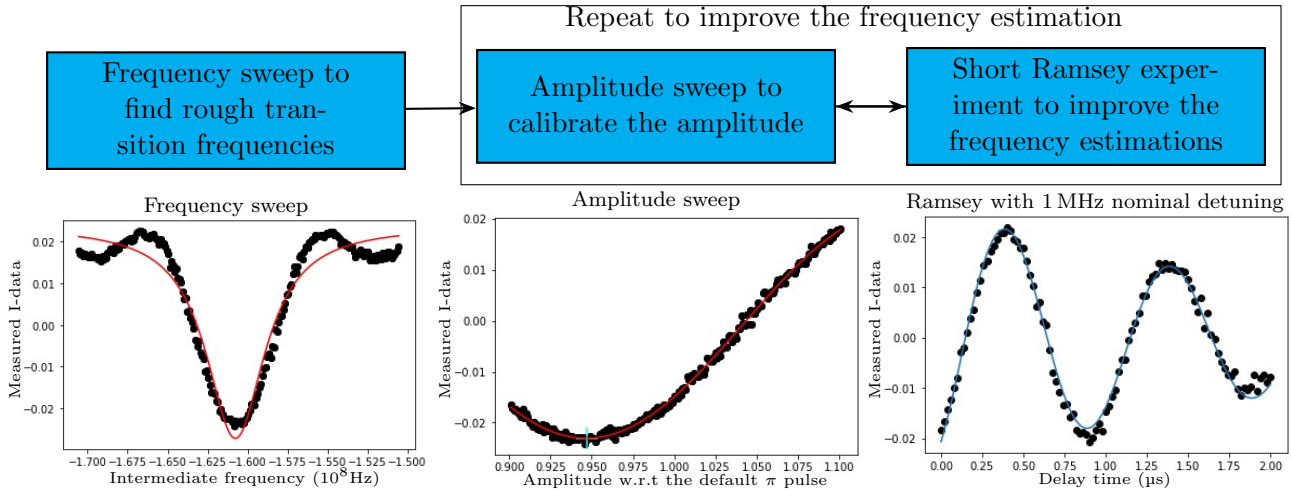
\begin{figure}[t]
\centering
\begin{tikzpicture}[node distance=2cm]
\node (FreqSweepPic) [] {\includegraphics[width=.32\textwidth,trim={0.65cm, 0.62cm, 0.1cm, 0.12cm},clip]{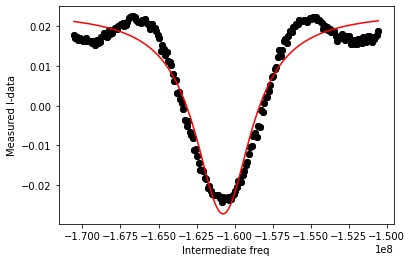}};
\node(FreqSweepPicYLabel)[left=of FreqSweepPic,rotate=90,yshift=-2.05cm,xshift=1.25cm]{\tiny Measured I-data};
\node(FreqSweepPicXLabel)[below=of FreqSweepPic,yshift=2.3cm,xshift=0.25cm]{\tiny Intermediate frequency ($10^8$\SI{}{\hertz})};
\node(FreqSweepName)[above=of FreqSweepPic,yshift=-2.25cm]{\scriptsize Frequency sweep};
\node (AmpSweepPic) [right=of FreqSweepPic,,xshift=-2cm] {\includegraphics[width=.32\textwidth,trim={0.65cm, 0.62cm, 0.1cm, 0.12cm},clip]{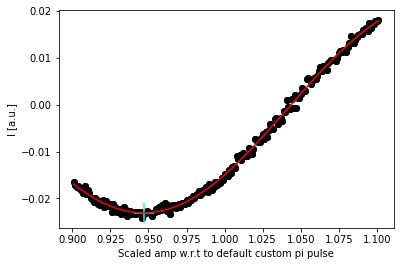}};
\node(AmpSweepPicYLabel)[left=of AmpSweepPic,rotate=90,yshift=-2.05cm,xshift=1.25cm]{\tiny Measured I-data};
\node(AmpSweepPicXLabel)[below=of AmpSweepPic,yshift=2.3cm,xshift=0.25cm]{\tiny Amplitude w.r.t the default $\pi$ pulse};
\node(AmpSweepName)[above=of AmpSweepPic,yshift=-2.25cm]{\scriptsize Amplitude sweep};
\node (RamseyPic) [right=of AmpSweepPic,xshift=-2cm] {\includegraphics[width=.32\textwidth,trim={0.65cm, 0.62cm, 0.1cm, 0.12cm},clip]{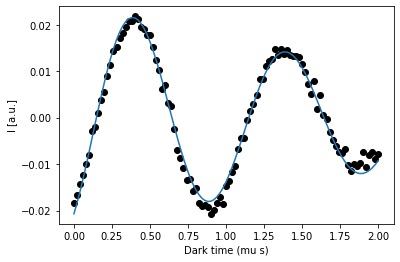}};
\node(RamseyPicYLabel)[left=of RamseyPic,rotate=90,yshift=-2.05cm,xshift=1.25cm]{\tiny Measured I-data};
\node(RamseyPicXLabel)[below=of RamseyPic,yshift=2.3cm,xshift=0.25cm]{\tiny Delay time ($\SI{}{\micro\second}$)};
\node(RamseyPicName)[above=of RamseyPic,yshift=-2.25cm,xshift=0.12cm]{\scriptsize Ramsey with \SI{1}{\mega\hertz} nominal detuning};
\node (FreqSweep) [block_device,text width=4.5cm,above=of FreqSweepPic,yshift=-1.5cm] {Frequency sweep to find rough transition frequencies};
\node (AmpSweep) [block_device,above=of AmpSweepPic,text width=4.5cm,yshift=-1.5cm] {Amplitude sweep to calibrate the amplitude};
\node (Ramsey) [block_device,text width=4.5cm,above=of RamseyPic,yshift=-1.5cm] {Short Ramsey experiment to improve the frequency estimations};
\node(UntilCon)[above=of AmpSweep,yshift=-1.9cm,xshift=2.5cm]{Repeat to improve the frequency estimation};
 \node[draw,inner xsep=2mm,inner ysep=4mm,yshift=1.5mm,fit=(AmpSweep)(Ramsey)](RepeatCon){};
\draw[arrow](FreqSweep)--node [text width=2cm,midway,above,text centered] {}(AmpSweep);
\draw[arrow](AmpSweep)--node [text width=2cm,midway,text centered] {}(Ramsey);
\draw[arrow](Ramsey)--(AmpSweep);
\end{tikzpicture}
\caption{The flowchart for device calibration. Top: basic steps of the calibration. Bottom: experimental data and fitted $I$-data curves for the $1$-$2$ transitions. The functional forms of the curves are defined in Appendix \ref{sec:app-curve-fitting}.\label{fig:calibration_device_flow}}
\end{figure}

\pzc{The calibration consists of a frequency sweep and an iteration between an amplitude sweep and a short Ramsey experiment. Here we take the  the $1$-$2$ transition as an example to illustrate the procedure in detail.}
The calibration starts by sweeping over different values of the drive frequency to approximate the $1$-$2$ transition frequency as the frequency at which the $I$-coordinate in the readout is minimized, see the leftmost graph in Figure \ref{fig:calibration_device_flow}. The calibration then proceeds by iterating between an amplitude sweep and a short Ramsey experiment. In the amplitude sweep the drive frequency and the pulse duration are held fixed while the amplitude is varied. The amplitude for the $\pi$ pulse is then chosen as the one that minimizes the $I$-coordinate in the readout. A $\pi/2$ pulse then follows by halving the duration of the $\pi$ pulse. During the short Ramsey experiment this $\pi/2$ pulse is used together with a detuned drive frequency. The recorded population data is then fitted (in the least squares sense) to a trigonometric function with decaying amplitude, from which an improved value of the $\omega_{1,2}$ frequency can determined (see also Appendix \ref{sec:app-curve-fitting}). The iteration is terminated when the change in the transition frequency falls below a user specified tolerance. 


For completeness, the detailed experimental protocol of the device calibration is described in Protocol \ref{prot:pi}, and summarized in Figure \ref{fig:calibration_device_flow}. 

The calibration for the $0$-$1$ transition can be done in a similar manner. However, in this case the $Q$-coordinate in the readout is used to distinguish between the states $|0\rgl$ and $|1\rgl$, see  Figure \ref{fig:IQ-confusion}. 

We follow a state-by-state strategy when calibrating the device, where we first calibrate the $\pi_{01}$ pulse and estimate the $0$-$1$ transition frequency, $\omega_{0,1}$. Then, based on the calibrated $\pi_{01}$ pulse, we can calibrate the $\pi_{12}$ pulse and estimate $\omega_{1,2}$. The experiments presented here resulted in the estimates: $\omega_{0,1}/2\pi \approx 3.4486698$\SI{}{\giga\hertz} and $\omega_{1,2}/2\pi \approx 3.2402576$\SI{}{\giga\hertz}.

\clearpage 

\subsection{Collection of experimental characterization data \label{sec:exp-protocol}}

Assuming a set of projective measurement operators (observables) $\{\Pi_k\}$, where $\Pi_k = |k\rangle\langle k|$, the probability of observing the state $|j\rangle$ is
\begin{align}
    \hat{P}_j = \mbox{Tr}(\Pi_j \rho) = \langle j | \rho | j \rangle = |\langle j | \psi \rangle|^2 = |\psi_j|^2.
\end{align}
The first two equalities apply in the general case when the system is described by the density matrix $\rho$. The latter two equalities hold for pure states, corresponding to $\rho = |\psi\rangle\langle \psi|$.

After the device calibration we collect experimental data for the characterization. All the experimental data are populations of the different states at different times  (measured and classified following Section \ref{sec:measure-data}). The {\bf simulated} population of state $|k\rgl$ corresponding to the delay time $t = j\Delta t$ is denoted
\[ 
\hat{P}_k(j\Delta t; \mathbf{y}) = \langle k | \rho(j\Delta t) | k \rangle,\quad k=0,1,2.
\] 
Here, the argument $\mathbf{y}$ denotes dependence on the parameters in the physical model. 
The {\bf experimentally determined} population is obtained following the  procedure in Section \ref{sec:measure-data}. More specifically, the measurement of shot number $s$ is recorded as a coordinate in the $I$-$Q$ plane, $(I_s,Q_s)$. The trained GMM is then used to map this coordinate to a probability population vector, which is then multiplied by the inverse of the confusion matrix (see equation \eqref{eq:error_mitigation}) to mitigate measurement errors. Let the resulting probability vector be $(P_0^{(s)}(j\Delta t),\dots,P_2^{(s)}(j\Delta t))^T$. Finally, the {\bf experimentally determined} population of state $|k\rgl$ is averaged over all shots for that delay time,
\[
P_k(j\Delta t) = \frac{1}{S}\sum_{s=1}^S(P_k^{(s)}(j\Delta t)),\quad k=0,1,2,
\]
where $S$ is the total number of shots. Note that the ordering of the individual populations is irrelevant for the averaged population.

In the following, the experimental data is obtained from $T_1$-decay experiments and Ramsey experiments. Although we do not perform Hahn echo experiments here, we note that such experiments could also be included.

\begin{figure}[ht]
\begin{center}
\begin{tikzpicture}
\tikzset{every node}=[font=\sffamily\sansmath]
\node (StatePreparationT1) [block_device_small,text width=3.0cm] {Prepare  state $|k\rgl$};
\node (FreePropagationT1) [block_device_small,text width=3.35cm,right=of StatePreparationT1,xshift=-0.5cm] {Free propagation for the delay time $t_d$};
\node (MeasureT1) [block_device_small,text width=2.0cm,right=of FreePropagationT1,xshift=-0.5cm] {Measure};
\node (T1Name) [text width=8.0cm,above=of FreePropagationT1,yshift=-1.0cm] {$T_1$-decay experiment, $t_d$: the delayed time};
\draw[arrow](StatePreparationT1)--(FreePropagationT1);
\draw[arrow](FreePropagationT1)--(MeasureT1);
\node[draw,inner xsep=2mm,inner ysep=4mm,yshift=1.5mm,fit=(StatePreparationT1)(FreePropagationT1)(MeasureT1)](T1){};
\node (StatePreparationRamsey) [block_device_small,text width=3.0cm,below=of StatePreparationT1,yshift=-0.5cm,xshift=-1.0cm] {Prepare  state $|k\rgl$};
\node (HalfPiOneRamsey) [block_device_small,text width=4.0cm,right=of StatePreparationRamsey,xshift=-0.5cm] {Apply $\frac{\pi_{k,k+1}}{2}$ with the  frequency $\omega_d$};
\node (FreePropagationRamsey) [block_device_small,text width=3.35cm,right=of HalfPiOneRamsey,xshift=-0.5cm] {Free propagation for the delay time $t_d$};
\node (HalfPiTwoRamsey) [block_device_small,text width=4.0cm,below=of FreePropagationRamsey,yshift=0.5cm] {Apply $\frac{\pi_{k,k+1}}{2}$ with the  frequency $\omega_d$};
\node (MeasureRamsey) [block_device_small,text width=3.0cm,left=of HalfPiTwoRamsey,xshift=0.5cm] {Measure};
\node (RamseyName) [text width=8.0cm,above=of HalfPiOneRamsey,yshift=-1.1cm] {Ramsey experiment, $t_d$: the delayed time, $\omega_d$: a detuned drive frequency.};
\node[draw,inner xsep=2mm,inner ysep=6.5mm,yshift=4mm,fit=(StatePreparationRamsey)(HalfPiTwoRamsey)](Ramsey){};
\draw[arrow](StatePreparationRamsey)--(HalfPiOneRamsey);
\draw[arrow](HalfPiOneRamsey)--(FreePropagationRamsey);
\draw[arrow](FreePropagationRamsey)--(HalfPiTwoRamsey);
\draw[arrow](HalfPiTwoRamsey)--(MeasureRamsey);
\node (HalfPiOneEcho) [block_device_small,text width=3.0cm,below=of MeasureRamsey] {Apply $\frac{\pi_{k,k+1}}{2}$};
\node (StatePreparationEcho) [block_device_small,text width=3.0cm,left=of HalfPiOneEcho,xshift=0.5cm] {Prepare state $|k\rgl$};
\node (FreePropagationOneEcho) [block_device_small,text width=3.35cm,right=of HalfPiOneEcho,xshift=-0.5cm] {Free propagation for the delay time $t_d$};
\node (PiEcho) [block_device_small,text width=3.0cm,yshift=0.5cm,below=of FreePropagationOneEcho] {Apply $\pi_{k,k+1}$};
\node (FreePropagationTwoEcho) [block_device_small,text width=3.35cm,left=of PiEcho,xshift=0.5cm] {Free propagation for the delay time $t_d$};
\node (HalfPiTwoEcho) [block_device_small,text width=3.0cm,left=of FreePropagationTwoEcho,xshift=0.5cm] {Apply $\frac{\pi_{k,k+1}}{2}$};
\node (MeasureEcho) [block_device_small,text width=3.0cm,below=of HalfPiTwoEcho,yshift=0.5cm] {Measure};
\node (EchoName) [text width=9.0cm,above=of HalfPiOneEcho,yshift=-1.0cm] {Hahn-Echoes experiment, $t_d$: the delayed time};
\draw[arrow](StatePreparationEcho)--(HalfPiOneEcho);
\draw[arrow](HalfPiOneEcho)--(FreePropagationOneEcho);
\draw[arrow](FreePropagationOneEcho)--(PiEcho);
\draw[arrow](PiEcho)--(FreePropagationTwoEcho);
\draw[arrow](FreePropagationTwoEcho)--(HalfPiTwoEcho);
\draw[arrow](HalfPiTwoEcho)--(MeasureEcho);
\node[draw,inner xsep=3mm,inner ysep=4mm,yshift=1.5mm,xshift=1mm,fit=(StatePreparationEcho)(MeasureEcho)(PiEcho)](Echo){};
\end{tikzpicture}
\end{center}
\caption{The flowchart of $T_1$-decay, Ramsey and Hahn-Echo experiments for $|k\rangle$ to $|k+1\rangle$ transition. \label{fig:characterization_exp_flow}}
\end{figure}
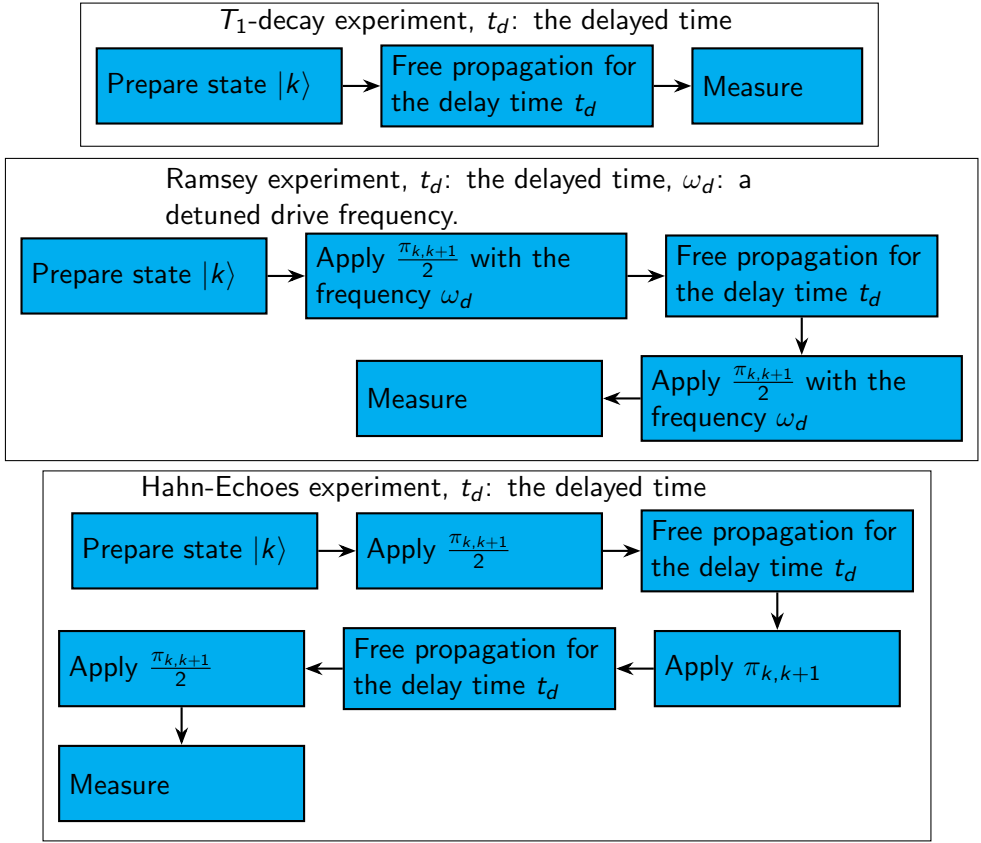

For reference, the experimental protocols are presented in Section \ref{sec:exps}. The protocol of the $T_1$-decay experiment in Protocol \ref{prot:T1}, the Ramsey experiment in Protocol \ref{prot:Ramsey} and the Hahn echo experiment in Protocol \ref{prot:Hahn-Echo}. These protocols are summarized as flowcharts in Figure \ref{fig:characterization_exp_flow} and described in even greater detail in \cite{qiskit_url_calibration}. 

The drive frequencies for the Ramsey experiments are chosen as the estimated transition frequencies from the calibration, reduced by a \SI{1}{\mega\hertz} nominal detuning, resulting in $\omega_d/2\pi = 3.4476698$ \SI{}{\giga\hertz} and $\omega_d/2\pi = 3.2392576$ \SI{}{\giga\hertz}, for the 0-1 and 1-2 Ramsey experiments, respectively. The step size in the delay time was \SI{20}{\nano\second} for both Ramsey experiments and the longest delay time was \SI{5}{\micro\second}.

The drive frequencies for the 0-1 and 1-2 $T_1$-experiments are chosen to be the same as the estimated transition frequencies from the calibration, see Table~\ref{tab:det}. Here, the step size in the delay time was \SI{80}{\nano\second}. The longest delay time was \SI{40}{\micro\second} for the $0$-$1$ experiment and \SI{20}{\micro\second} for the $1$-$2$ experiment. 
Each data point is acquired by taking the average over $1000$ measurements (shots).  



\subsection{System characterization}
Both the deterministic and Bayesian characterization rely on a forward model, which in this context is the Lindblad equation (\ref{eq:lindblad}). Solutions to the forward model are obtained by numerical simulation using the characterization software ${\tt GLOQ.jl}$ \cite{GLOQ}.  In the forward simulation, we use the RWA and vectorize the Lindblad equation 
\begin{align}
\frac{d}{dt}\textrm{vec}(\rho) = - i (I\otimes H - H^T \otimes I) \textrm{vec}(\rho)+\sum_{j=1}^2 \hat{\mathcal{L}}_j\textrm{vec}(\rho), \label{eq:GLvec}
\end{align}
where 
\begin{equation}
\hat{\mathcal{L}}_j = \mathcal{L}_j\otimes\mathcal{L}_j-\frac{1}{2}\left(I\otimes (\mathcal{L}_j^T\mathcal{L}_j)
+(\mathcal{L}_j^T\mathcal{L}_j)\otimes I
\right).
\end{equation}
As the  $\pi$-pulses used in the Ramsey and $T_1$-experiments have constant amplitude (in the rotating frame), we can use the matrix exponentiation approach to efficiently integrate the system in time. We take parity events into account through the model described in Section \ref{sec:parity}. In each forward Lindblad simulation we thus solve (\ref{eq:GLvec}) twice, once for each parity, followed by averaging the results.   

\subsubsection{Deterministic characterization}
In the deterministic characterization, we determine the device parameters 
\[
\by=({\omega}_{0,1},{\omega}_{1,2}^+,{\omega}_{1,2}^-,{\gamma}_{1,1},{\gamma}_{1,2},{\gamma}_{2,1},{\gamma}_{2,2})^T,
\]
by minimizing the mismatch between the results of forward simulations and the experimental data. Here, the mismatch between the simulated and experimentally determined populations is defined by
\begin{align}
    J(\by) = \sum_{\textrm{experiments}}\sum_{n=0}^2\sum_{j=0}^{N_{\rm T}} \left(\hat{P}_n(j\Delta t;\by)-P_n(j\Delta t)\right)^2\Delta t_e,
\end{align}
where $\Delta t_e$ is the step size in the delay time for experiment $e$. The sum over experiments refers to the Ramsey and $T_1$-decay experiments, for the $0$-$1$ and the $1$-$2$ transitions (in total four experiments). 

When numerically solving this minimization problem, we  add constraints for the lower and upper bounds of the parameters in $\by$. The constrained problem is solved by the multilevel optimizer {\tt{fminbox()}} in the {\tt{Optim.jl}} package \cite{mogensen2018optim}. The outer loop of {\tt{ fminbox()}} adds a barrier penalty to $J(\by)$ to impose the constraints and adaptively updates the size of this penalty. In the inner loop, the penalized problem is solved by the L-BFGS method.

\begin{table}[h]
\begin{center}
\begin{tabular}{|c|c|c|c|c|c|c|c||c|c|}
\hline
\footnotesize{${\omega}_{01}/(2\pi)$} &  \footnotesize{${\omega}^-_{12}/(2\pi)$}  &  \footnotesize{${\omega}^+_{12}/(2\pi)$} & $T_{1,1}$ &  $T_{1,2}$& 
$T_{2,1}$&  $T_{2,2}$  \\
\hline
\hline
\SI{3.448646}{\giga\hertz} & \SI{3.240105}{\giga\hertz} & \SI{3.240403}{\giga\hertz} &  \SI{258.39}{\mu\second} & \SI{100.79}{\mu\second} & 
\SI{38.44}{\mu\second} & \SI{29.94}{\mu\second}\\
\hline
\end{tabular}
\caption{System parameters determined  by deterministic inversion.  \label{tab:det}}
\end{center}
\end{table}

\begin{figure}[tb]
\centering
  \includegraphics[width=0.49\textwidth]{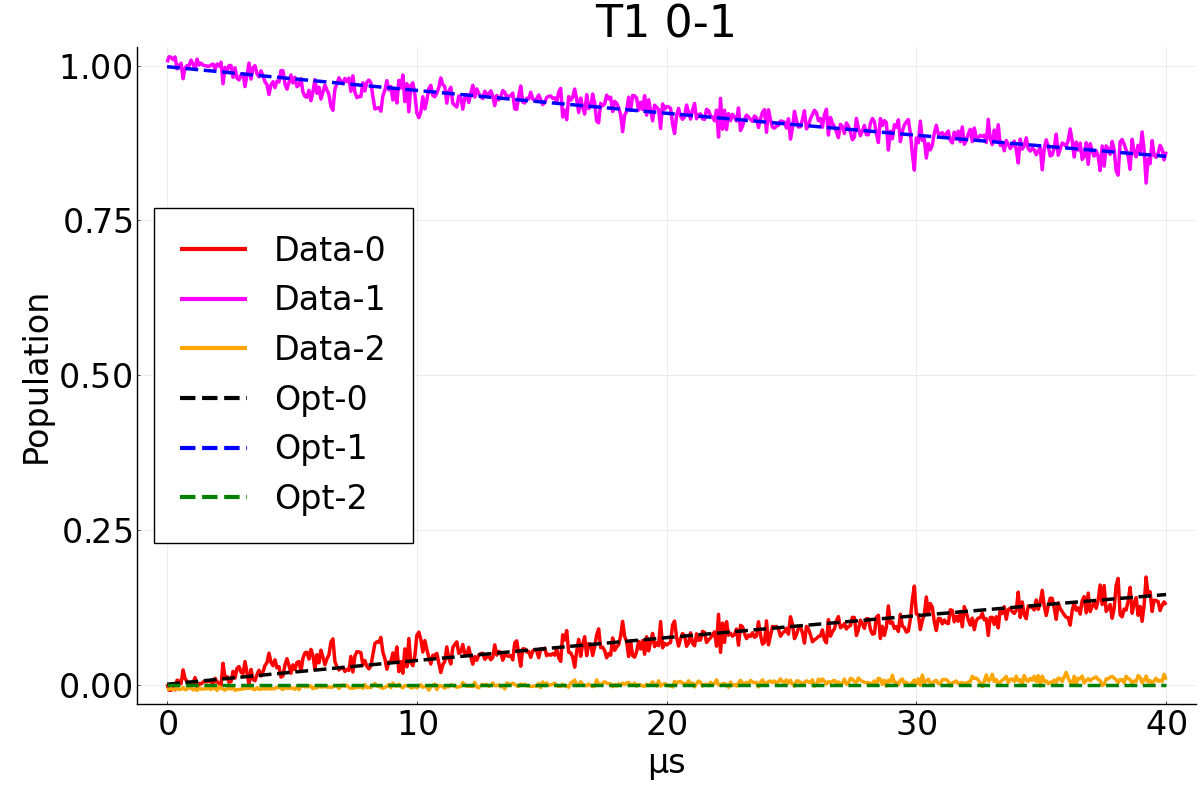}
    \includegraphics[width=0.49\textwidth]{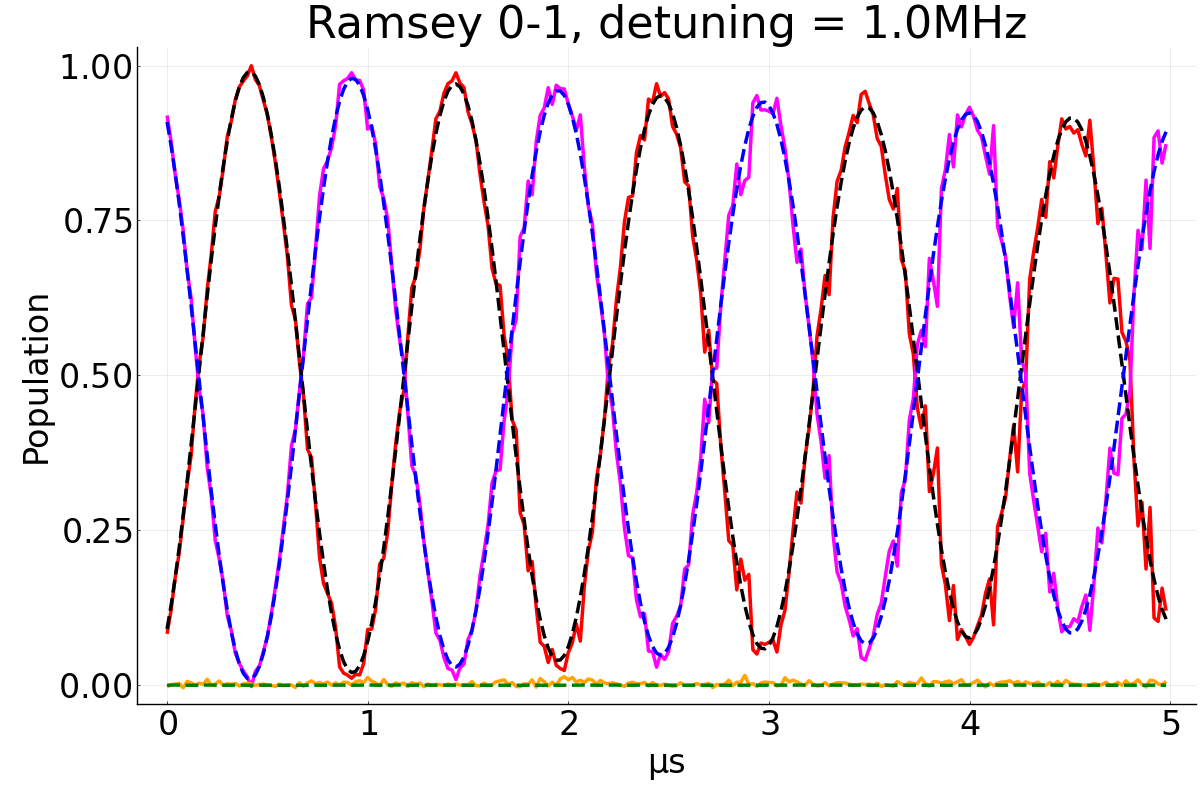}
  \includegraphics[width=0.49\textwidth]{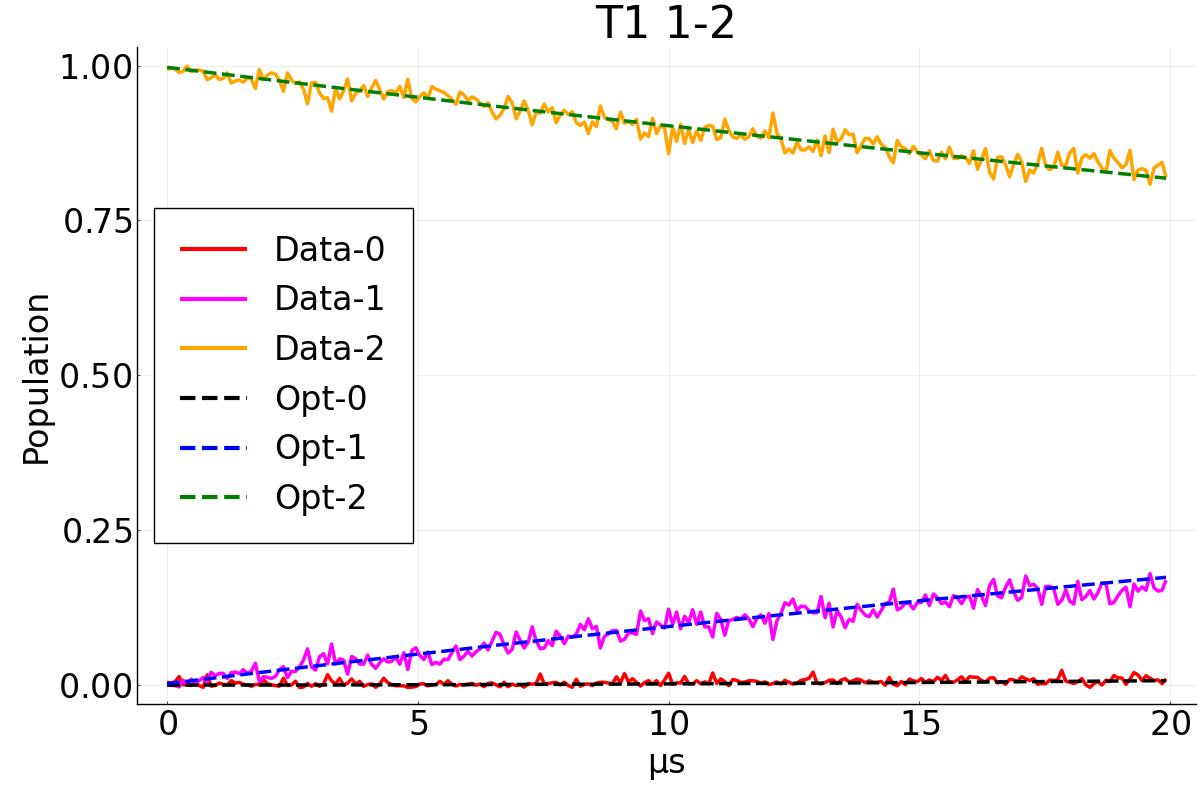}
  \includegraphics[width=0.49\textwidth]{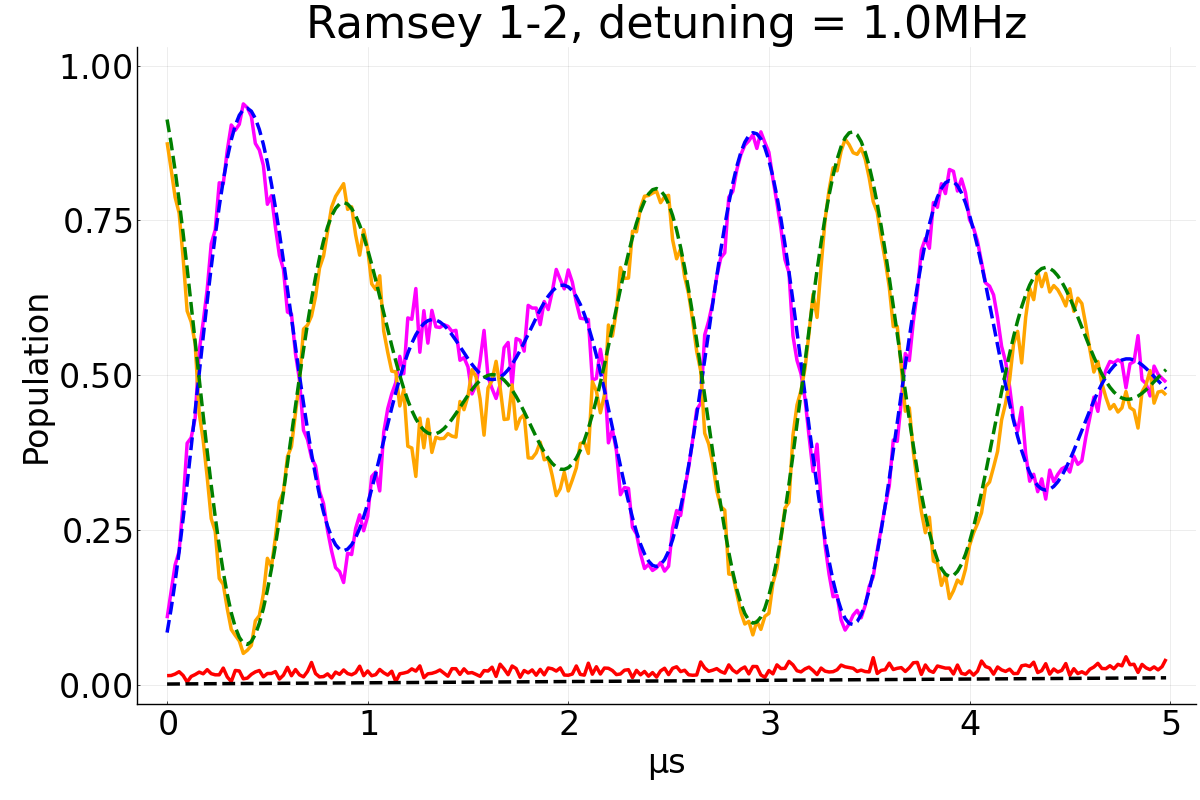}
\caption{Comparison between experimental data and forward Lindblad simulation using  parameters from Table \ref{tab:det} (deterministic characterization results).\label{fig:det_vs_exp}}
\end{figure}

The result of the deterministic characterization is summarized in Table \ref{tab:det} (see Section \ref{sec:model} for notation). Note that $T_{1,k}$ and $T_{2,k}$ are obtained from $\gamma_{1,k}$ and $\gamma_{2,k}$ as described in Section \ref{sec:model} and Appendix \ref{sec:app:T2-gamma}.  The experimental data and the results from Lindblad simulations, using the parameters in Table \ref{tab:det}, are compared in Figure \ref{fig:det_vs_exp}. We note very good qualitative agreement, including the beating in the Ramsey 1-2 experiment (between times 1.0 and 2.5 $\mu$s). We remark that the beating can not be captured if a single value of $\omega_{1,2}$ is used in the Lindblad model.

\subsubsection{Bayesian characterization\label{sec:bayesian}}
To use the risk neutral optimization in {\tt Juqbox.jl} we need probability distributions of the transition frequencies. Because {\tt Juqbox.jl} does not model decoherence, we can make the simplifying assumption that $T_{1,k}$ and $T_{2,k}$ are fixed and take the values determined by the deterministic characterization, see Table \ref{tab:det}. 

In Bayesian characterization, $\omega_{0,1}$ and $\omega_{1,2}^\pm$ are modeled as random variables with unknown probability distributions. Let $\zb=(\omega_{0,1},\omega_{1,2}^+,\omega_{1,2}^-)^T$, and let $D$ be the {\bf Ramsey} experimental data obtained by the procedure described above (the $T_1$ data is not used here). The goal of the Bayesian characterization is to (computationally) approximate the posterior distribution ${\rm Pr}(\zb | D)$ by sampling.  

The sampling is based on Bayes' rule,
\begin{align}
    {\rm Pr}(\zb| D ) = \frac{{\rm Pr}(\zb) {\rm Pr}( D |\zb)}{{\rm Pr}(D)},\label{eq:Bayes}
\end{align}
which relates the posterior distribution to the product of the prior, ${\rm Pr}(\zb)$, which encodes the best available estimate of the distribution of $\zb$, and the likelihood of observing the data, ${\rm Pr}(D|\zb)$, given the model parameters $\zb$. The formula also contains the normalization factor ${\rm Pr}(D)$, which is the marginalized distribution of data. However, the latter is not of practical importance because ${\rm Pr}(\zb| D)$ can always be normalized to integrate to one. 

To simplify the description of our modeling choices for the likelihood and priors, suppose for a moment that we only have a single experimental population (observation), say $P_0(5\Delta t)$. We then assume that this data is well described by the Lindblad model and that the Lindblad model produces a simulated population $\hat{P}_0(5\Delta t; \zb)$, which includes a modeling error (noise) so that 
$$
P_0(5\Delta t) = \hat{P}_0(5\Delta t; \zb) + E.
$$
Here the random variable $E$ models the error. We now make the assumption that the errors at each delay time $j\Delta t$, and for each state, are independent. Further, we assume that each of the errors is normally distributed with zero mean. We also temporarily assume that we know the variance $\sigma^2$ (we will relax this assumption below). We thus have
$$
E \sim  \mathcal{N}(0,\sigma^2) = {\rm Pr}_{\rm noise}(\varepsilon) = \frac{1}{\sigma \sqrt{2 \pi}} e^{-\frac{\varepsilon^2}{2 \sigma^2}}.
$$
Making the additional assumption that the noise does not depend on $\zb$ we have that for a fixed realization $\zb^\ast$  
$$
{\rm Pr}(\varepsilon | \zb^\ast) = {\rm Pr}_{\rm noise}(\varepsilon).
$$
The data, $P_0(5\Delta t)$, still depends on the noise but its dependence is simple, it is normally distributed around $\hat{P}_0(5\Delta t; \zb^\ast)$ 
$$
{\rm Pr}(P_0(5\Delta t) | \zb^\ast) = {\rm Pr}_{\rm noise}(P_0(5\Delta t)-\hat{P}_0(5\Delta t; \zb^\ast)).
$$
This is our model for the likelihood that an experimental data $P_0(5\Delta t)$ will be observed given fixed values of $\omega_{0,1},\omega_{1,2}^+,\omega_{1,2}^-$.

 Of course, we will observe data at many delay times $t_1=\Delta t,t_2=2\Delta t,\ldots, t_{N_{\rm T}}=N_{\rm T}\Delta t$ but as the errors are assumed to be independent this simply means that the (joint) likelihood takes the form 
\begin{multline*}
{\rm Pr}(P_0(1\Delta t), P_0(2\Delta t), \ldots | \zb^\ast) = \prod_{i=1}^{N_{\rm T}} {\rm Pr}_{\rm noise}(P_0(i\Delta t)-\hat{P}_0(i\Delta t; \zb^\ast)) \\
= \frac{1}{(\sqrt{2\pi}\, \sigma)^{N_T}}\exp\left(-\frac{1}{2\sigma^2} \sum_{i=1}^{N_{T}} (P_0(i\Delta t)-\hat{P}_0(i\Delta t; \zb^\ast))^2\right).
\end{multline*}
Incorporating the data for multiple states and multiple experiments into the likelihood can be done in the same way.  

We only need the likelihood up to a multiplicative constant to use Bayes rule and thus note that if the variance of the noise is considered known have 
$$
{\rm Pr}(P_0(5\Delta t) | \zb^\ast) \propto e^{-\frac{(P_0(5\Delta t)-\hat{P}_0(5\Delta t; \zb^\ast))^2}{2\sigma^2}}.
$$  
However, in reality we don't know the variance, $\sigma^2$, and instead model it as another parameter (called hyperparameter due to the fact that it is not part of the original physical model). We then have that the likelihood is
$$
{\rm Pr}(P_0(5\Delta t) | \zb^\ast,\sigma^2) \propto \frac{1}{\sigma} e^{-\frac{(P_0(5\Delta t)-\hat{P}_0(5\Delta t; \zb^\ast))^2}{2\sigma^2}}.
$$  

We now turn to the likelihood for the real model. Taking both the different delay times and states into account, the likelihood of the population measurements of state $|n\rgl$ for delay time $j\Delta t$ in the Ramsey $|k\rangle$ to $|k+1\rangle$ experiment is taken to be
\begin{align}
    P_{n}(j\Delta t) \sim \mathcal{N}(\hat{P}_n(j\Delta t;\zb),\sigma^{\textrm{Ramsey}}_{k,k+1}),\quad n=0,1,2,\quad k=0,1. \label{eq:proposal}
\end{align}
Note here that we allow the hyper parameters $(\sigma^{\textrm{Ramsey}}_{k,k+1})^2$ to be different for different $k$. A simplifying assumption would have been to take them to be the same for all $k$. 

The total likelihood is then the product of all of the individual likelihoods,
\[
{\rm Pr}(D|\zb)=C \prod_{k=0}^1\prod_{n=0}^2\prod_{j}\exp\left(-\frac{1}{2}\left(P_n(j\Delta t)-\hat{P}_n(j\Delta t) \right)^2/\left(\sigma_{k,k+1}^{\textrm{Ramsey}}\right)^2\right),
\]
where $C$ is a normalizing constant.

The priors for $\omega_{0,1}$ and $\omega_{1,2}^{\pm}$ are selected based on the deterministic characterization. In particular, the priors are taken to be truncated Gaussian distributions $\mathcal{TN}(\mu,\sigma,l,u)$ with mean $\mu$, standard deviation $\sigma$ and truncated outside the interval $[l,u]$. We thus have 
\begin{align}
    \omega_{0,1}\sim \mathcal{TN}(\bar{\omega}_{0,1},\sigma_{01},l_{01},u_{01}),\;\omega_{1,2}^\pm\sim \mathcal{TN}(\bar{\omega}^\pm_{12},\sigma_{12}^\pm,l_{12}^\pm,u_{12}^\pm), \label{eq:prior}
\end{align}
where we set the average transition frequencies in the prior, $\bar{\omega}_{0,1}$ and $\bar{\omega}_{1,2}^\pm$ to be the values from the deterministic characterization, see Table \ref{tab:det}. The standard deviation and the limits of the truncated intervals in the prior are taken to be
\[
 \sigma_{01}=\sigma_{12}^\pm=2 \pi \times \SI{50}{\kilo\hertz},
\]
\[
 l_{01}=\bar{\omega}_{01}- 2\pi \times \SI{125}{\kilo\hertz},\ \  u_{01}=\bar{\omega}_{01}+2\pi \times \SI{125}{\kilo\hertz},
\]
\[
  l_{12}^\pm =\bar{\omega}^\pm_{12}- 2\pi \times \SI{125}{\kilo\hertz},\ \  u_{12}^\pm=\bar{\omega}^\pm_{12}+2\pi \times \SI{125}{\kilo\hertz}.
\]
These hyper-parameters are tuned based on numerical experiments following the criteria below. The posterior determined by the chain should be narrow enough compared with the prior (see the top row of Figure \ref{fig:distributions_12}) so that a large enough sample space is explored. When the posterior is too wide, the sample space is enlarged by increasing the standard deviation and the size of the truncation interval. Secondly, the correlation between successive samples in the sampled Markov chain should be small enough (see Figure \ref{fig:markov-chains}). Strong correlation means that the chain explores the sample space slowly. Such slow exploration implies that the standard deviation and the size of the truncation interval of the prior may be too small. Thirdly, a small acceptance ratio indicates that the sample space may be too large and the prior may be too wide. In that case, the standard deviation and the size of the truncation interval should be reduced.

The priors for the hyper-parameters $(\sigma^{\textrm{Ramsey}}_{k,k+1})^2$ (for $k = 0,1$) in the likelihood function are modeled as inverse Gamma distributions. The choice of this prior for $(\sigma^{\textrm{Ramsey}}_{k,k+1})^2$ can be motivated as follows. Going back to the problem with a single data $P_0(5\Delta t)$, suppose $\zb$ was a deterministic parameter vector. Then the mean of the likelihood would be some constant and 
$$
{\rm Pr}(P_0(5\Delta t)|\sigma^2) \propto \frac{1}{\sigma} e^{-\frac{(P_0(5\Delta t)-{\rm CONST})^2}{2\sigma^2}}. 
$$  
The functional form of the above expression (when thought of as a function of $\sigma^2$) is the functional form of the inverse Gamma distribution
$$
{\rm Pr}(\sigma^2) \propto \left(\frac{1}{\sigma^2}\right)^{\alpha+1} e^{-\frac{\beta}{ \sigma^2}}. 
$$ 
The inverse Gamma function is said to be a conjugate prior to the likelihood when it has the same functional form as the likelihood.

A Markov Chain Monte Carlo (MCMC) algorithm \cite{kaipio2006statistical} is used to approximate the posterior distribution given by (\ref{eq:Bayes}). The basic idea of MCMC algorithms is to construct a Markov chain whose stationary distribution is the desired posterior, i.e., ${\rm Pr}(\zb | D)$. In {\tt GLOQ.jl} we use the {\tt Turing.jl}~\cite{ge2018turing} implementation of a particular MCMC algorithm called the No-U-Turn-Sampler (NUTS) \cite{hoffman2014no}. This sampler is a Hamiltonian Monte Carlo (HMC) method that results in smaller correlations between successive samples, compared with standard MCMC methods. It also converges faster towards the stationary distribution, resulting in a shorter Markov chain. Compared with standard HMC, NUTS provides a dynamical termination criteria when finding new samples and adapts the step size on the fly. As a  result, the NUTS technique requires less tuning than standard MCMC approaches.

\subsubsection*{Bayesian characterization results}
\begin{figure}[tb]
\centering
  \includegraphics[width=0.325\textwidth]{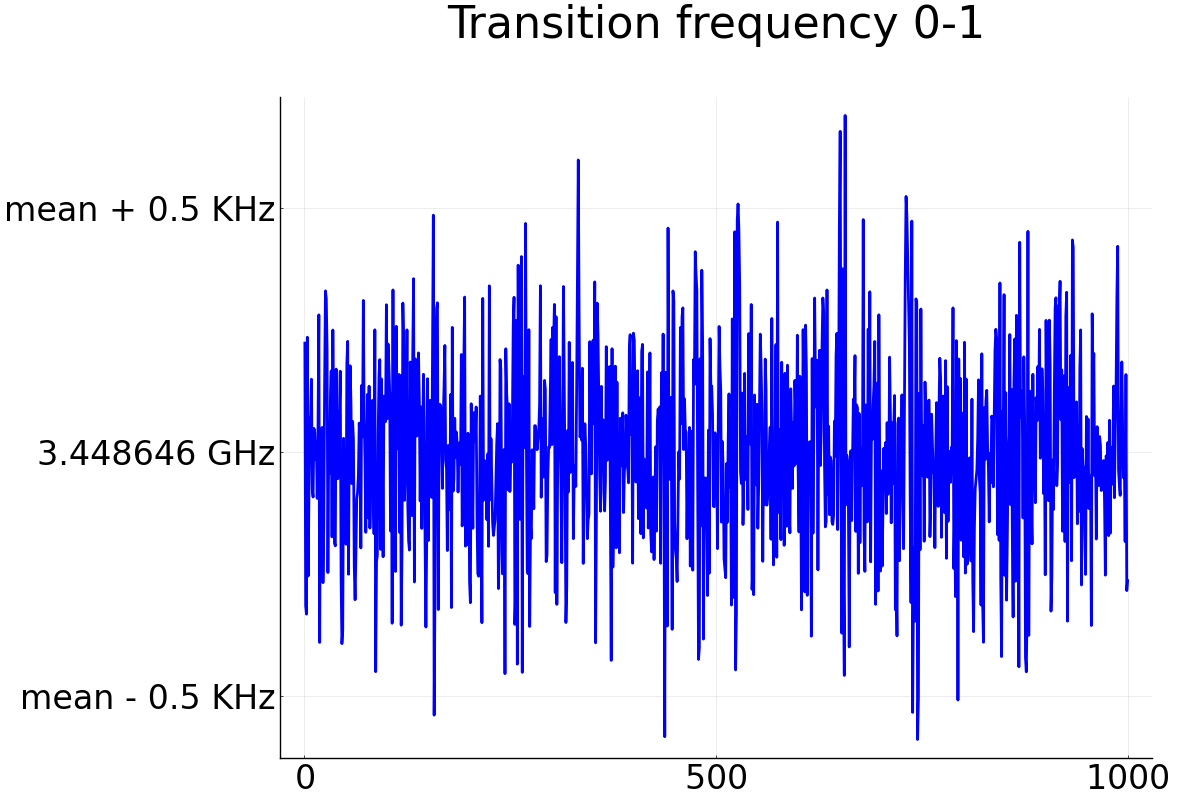}
  \includegraphics[width=0.325\textwidth]{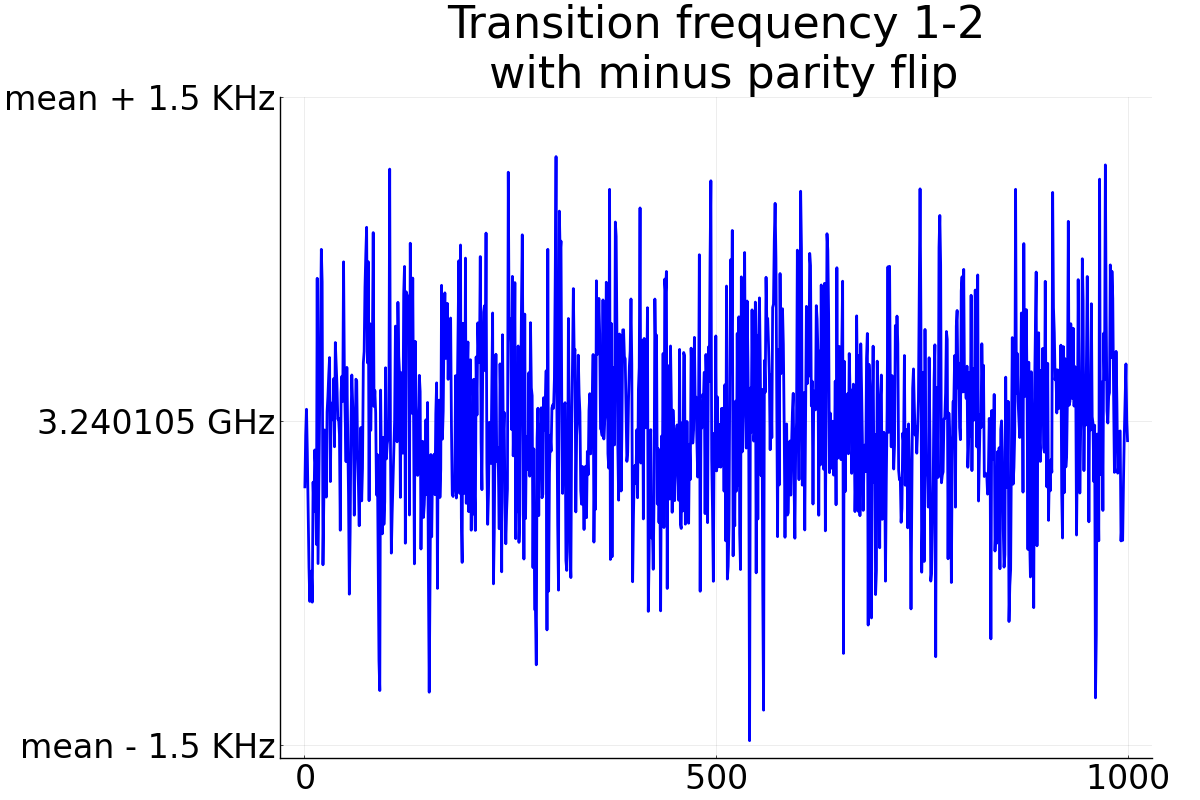}
  \includegraphics[width=0.325\textwidth]{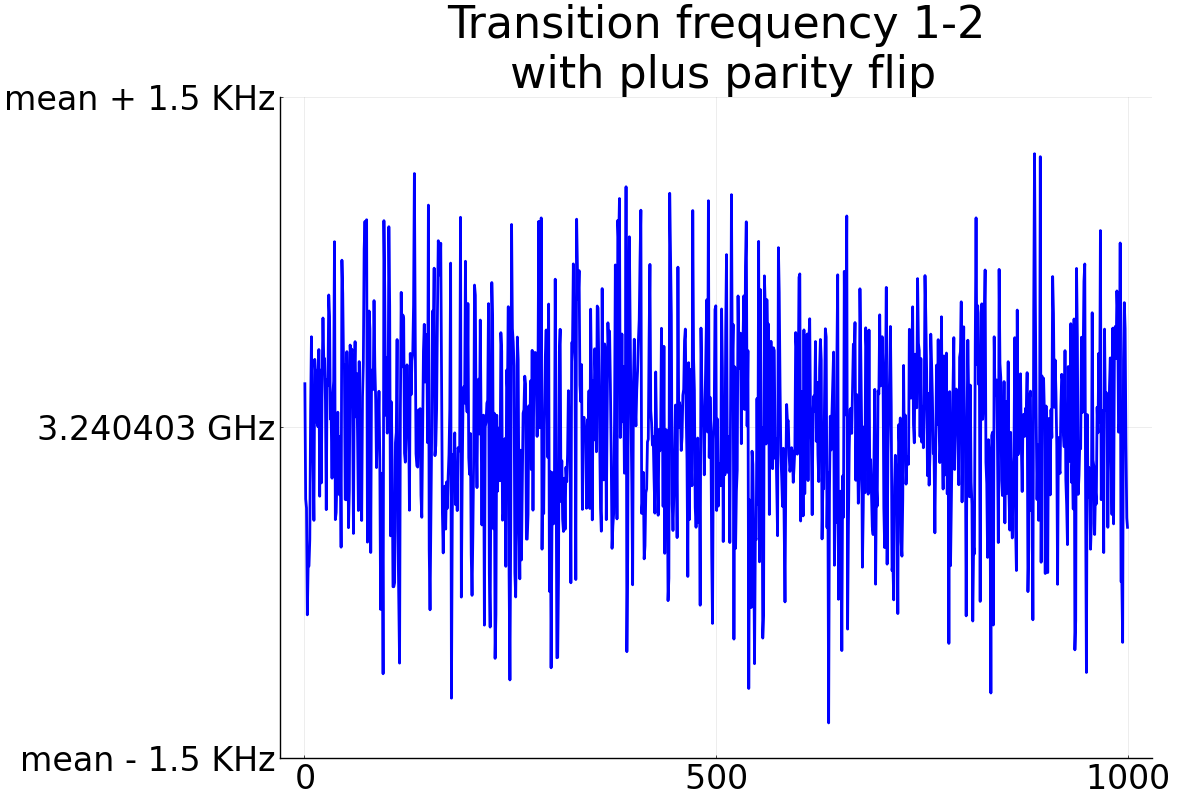}
\caption{Markov chains for the transition frequencies drawn by the NUTS sampler.\label{fig:markov-chains}}
\end{figure}

We now present the results from a Bayesian characterization using the NUTS sampler in {\tt turing.jl}. In total, we draw $1000$ samples with the target acceptance rate $0.65$. Figure \ref{fig:markov-chains} presents the sampled Markov chains. 
The initial part of our chains looks similar to the latter part of the chain, indicating that no burn-in is needed. We also observe that the correlation between successive samples is weak (less than $5\times 10^{-14}$) and these chains explore the sample space many times. These observations indicate that we have a sufficient number of effective samples in the chain. The $\hat{r}$ value is used as a diagnostic of the convergence of the Markov chain (see \cite{gelman1992inference} and {\tt Turing.jl} documentation \cite{ge2018turing}). The closer $\hat{r}$ is to $1$, the better the convergence is. The $\hat{r}$ values for $\omega_{0,1}, \omega_{1,2}^-$, and $\omega_{1,2}^+$, are 0.9990, 1.0004 and 1.0005, and for the hyper parameters $\sigma_{01}$ and $\sigma_{12}^-=\sigma_{12}^+$, $\hat{r}$ is 0.9994 and 1.0016, respectively.

\begin{figure}[tb]
\begin{center}
  \includegraphics[width=0.49\textwidth]{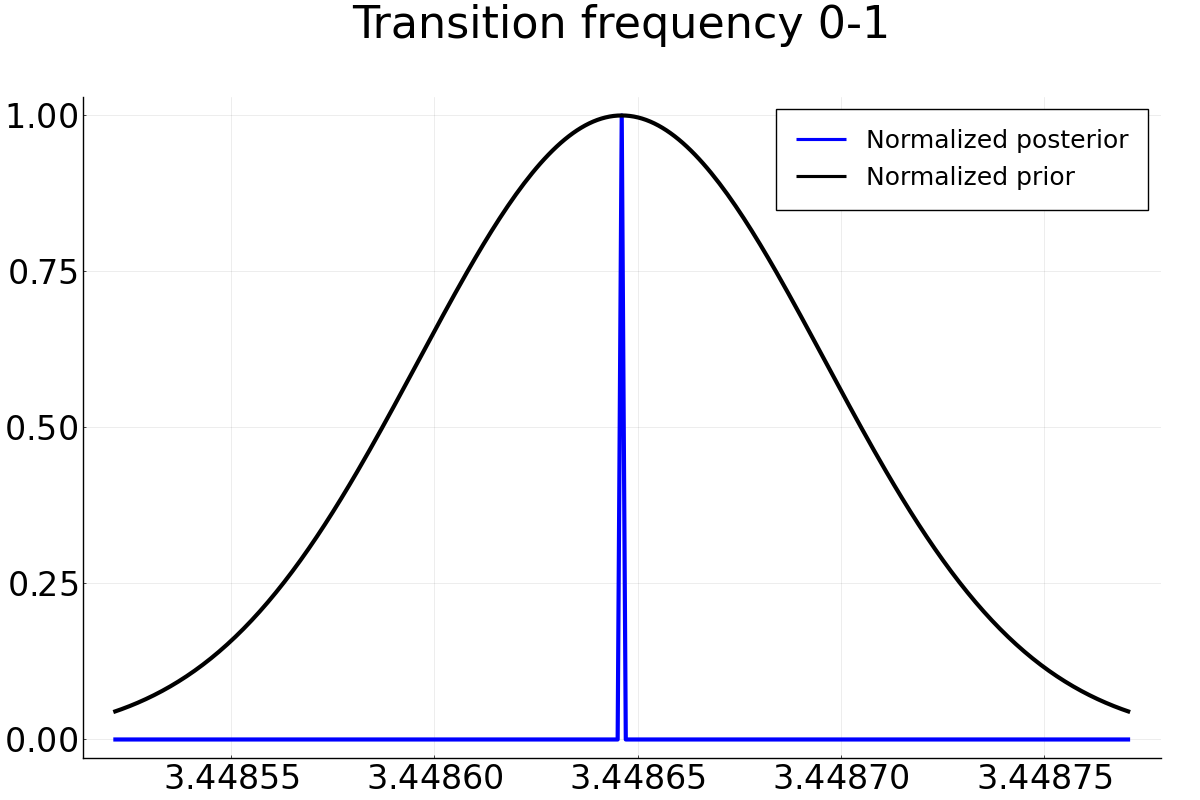}
\includegraphics[width=0.49\textwidth]{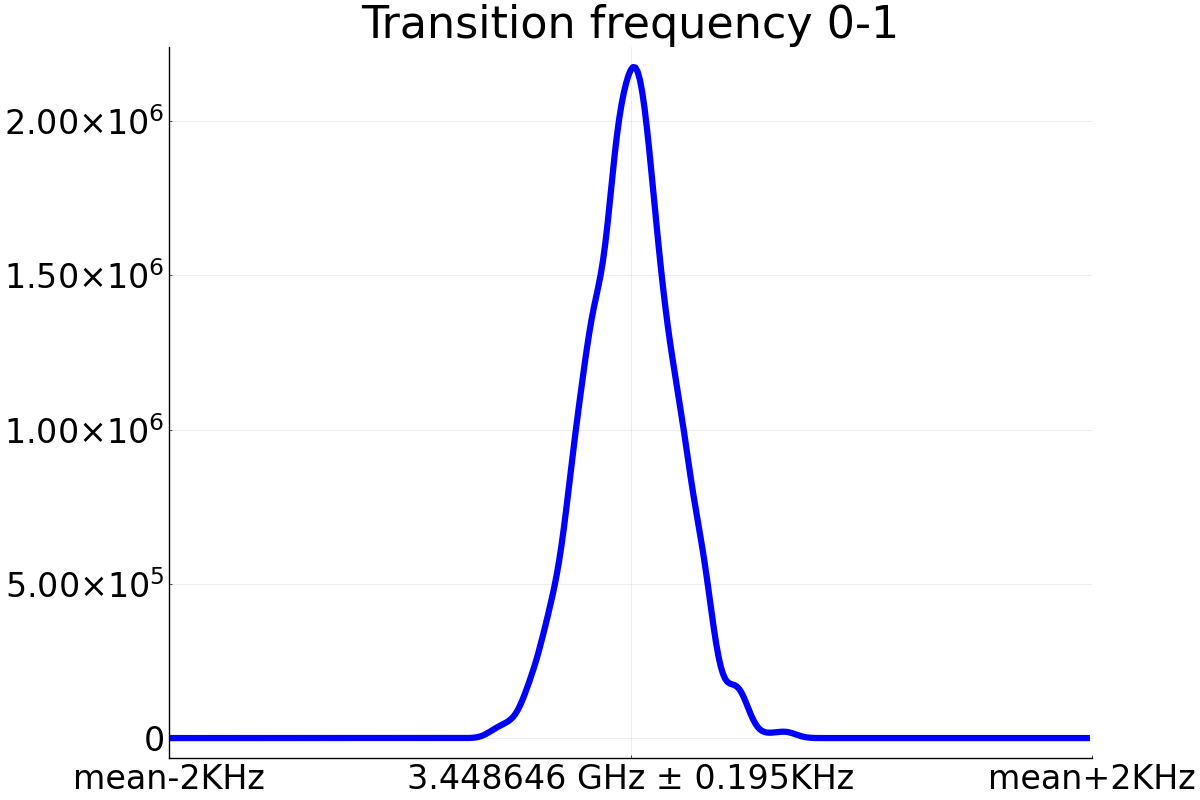}
\caption{Probability density distribution for $\omega_{01}/(2\pi)$. One the left, we compare the normalized posterior and the normalized prior. On the right, we present the $\textrm{mean}\pm \textrm{std}$ under the $x$-axis.\label{fig:distributions_01}}
\end{center}
\end{figure}

\begin{figure}[htb]
\centering
  \includegraphics[width=0.49\textwidth]{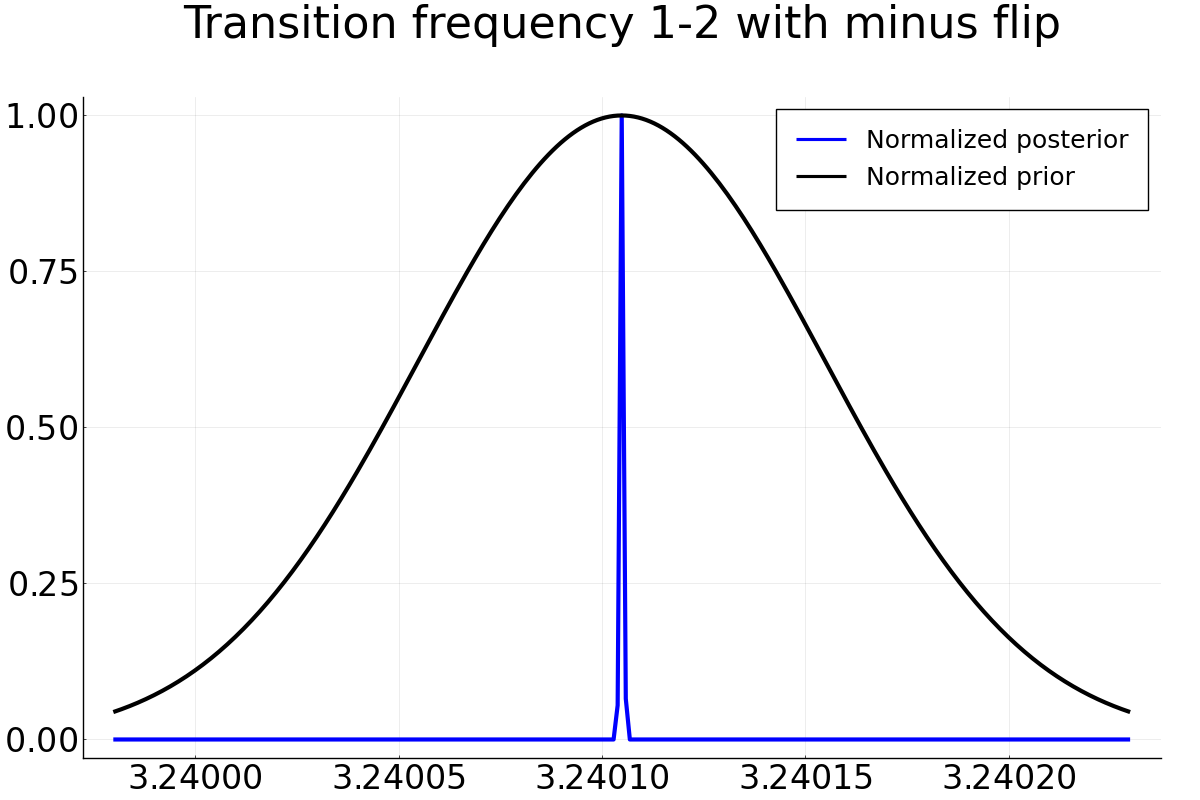}
  \includegraphics[width=0.49\textwidth]{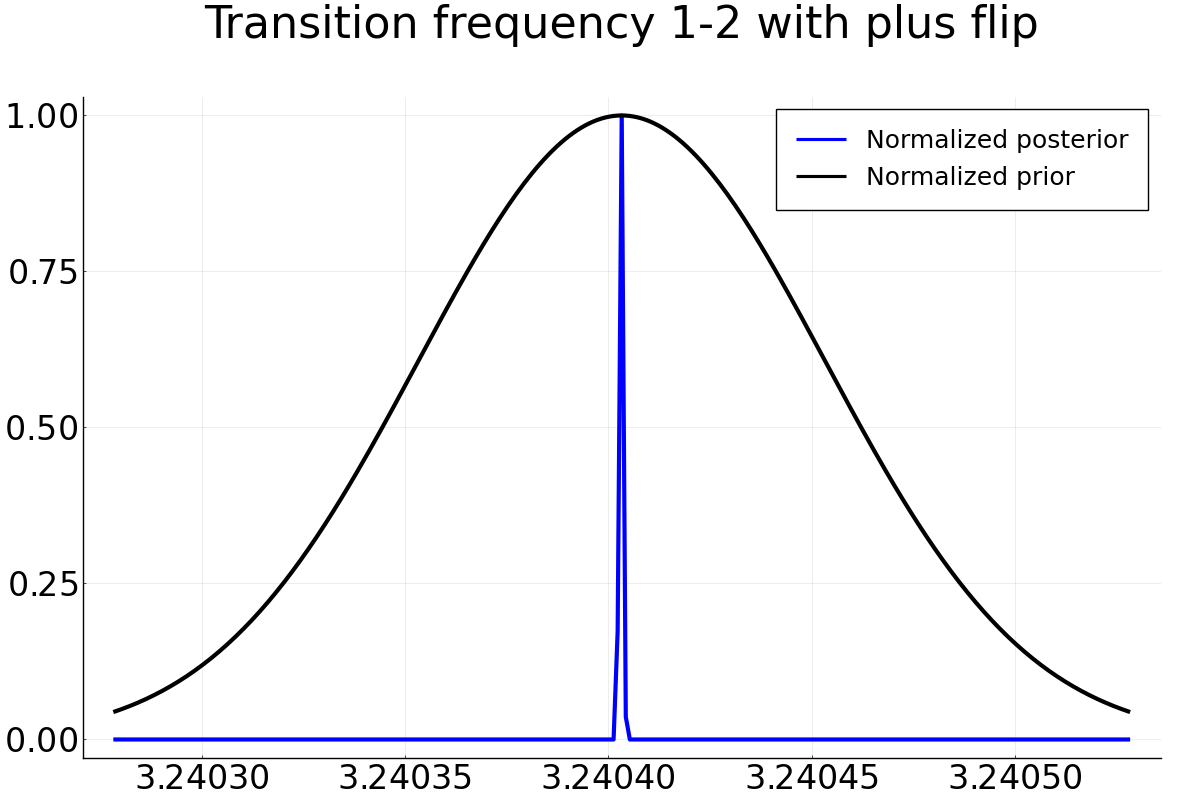}
  
  \vspace{0.4cm}
  
  \includegraphics[width=0.49\textwidth]{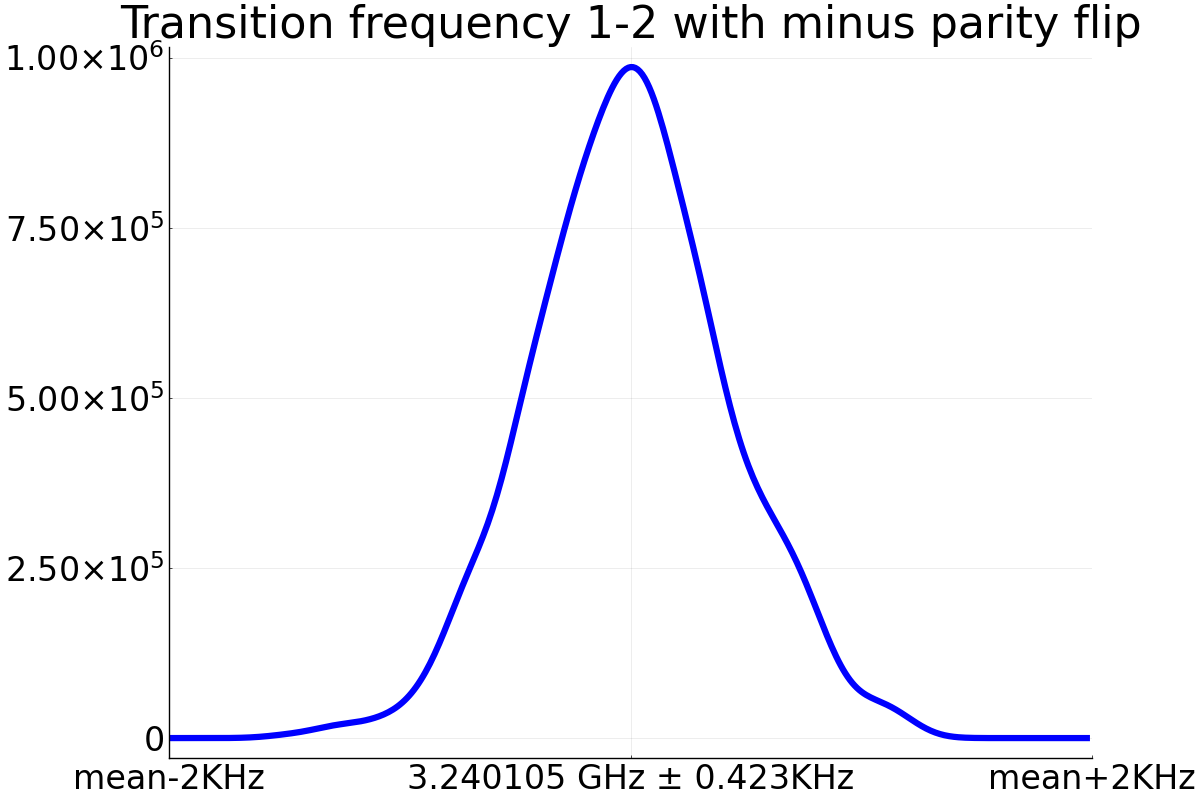}
  \includegraphics[width=0.49\textwidth]{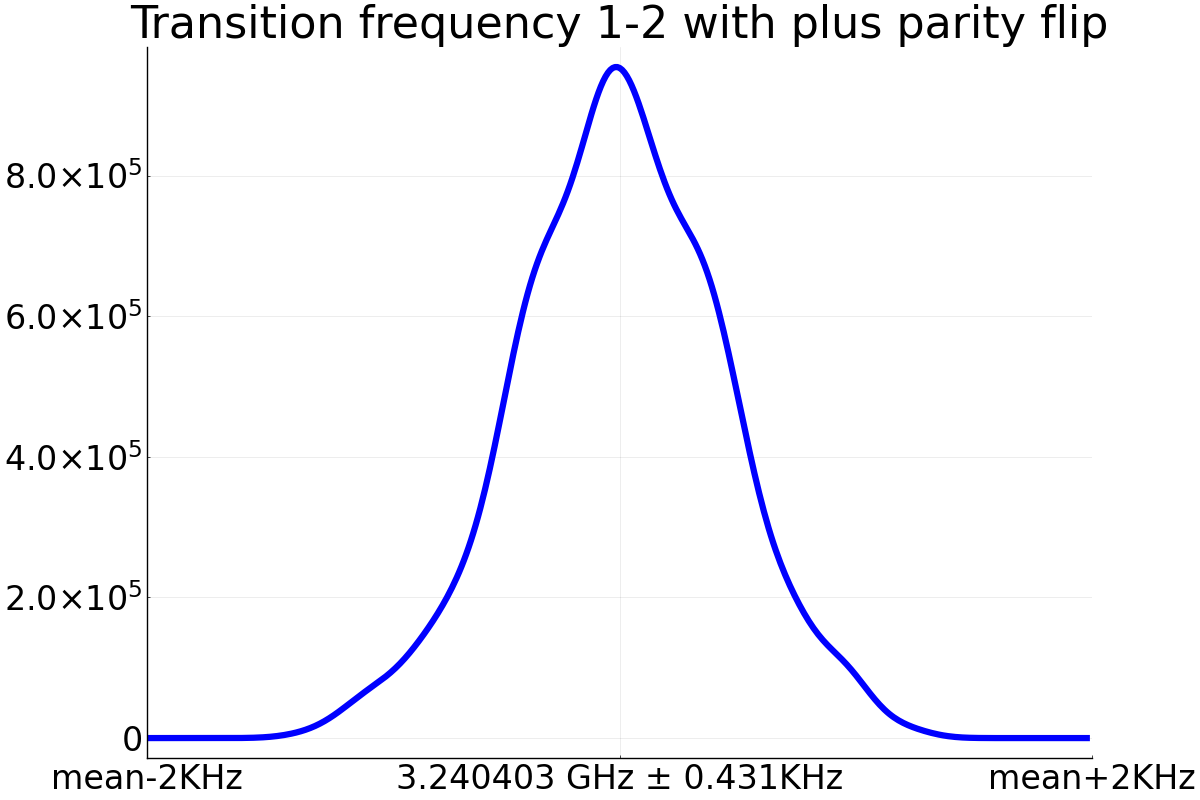}
  
\caption{Top row:  the  normalized posterior and the normalized prior for $\omega_{12}^\pm/(2\pi)$. Bottom row: probability density distribution $\omega^\pm_{12}/(2\pi)$, and we present the $\textrm{mean}\pm \textrm{std}$ under the $x$-axis. \label{fig:distributions_12}}
\end{figure}

The posterior of the transition frequencies are presented in Figures \ref{fig:distributions_01} and \ref{fig:distributions_12}. The posteriors are very concentrated compared to the corresponding priors, which indicates that the sample space of the transition frequencies have been well explored by the Markov chain.

The expectations and standard deviations for $\omega_{0,1}$ and $\omega_{1,2}^\pm$ are presented in Table \ref{tab:bayesian}. It can be noted that the expected values for each of the three parameters are identical to those obtained in the deterministic inversion (displayed in Table \ref{tab:det}). It can further be noted that the standard deviation in each of the three cases are quite small (on the order of \SI{}{\kilo\hertz}) compared to the expectation of the respective frequency (on the order of \SI{}{\giga\hertz}).  

\begin{table}[h]
\begin{center}
\begin{tabular}{|c|c|c|c|c|c|c|c||c|c|}
\hline
\footnotesize{$(\bar{\omega}_{01}\pm\sigma({\omega_{01}}))/(2\pi)$} & \footnotesize{$(\bar{\omega}^-_{12}\pm\sigma({\omega^-_{12}}))/(2\pi)$}  & \footnotesize{$(\bar{\omega}^+_{12}\pm\sigma({\omega^+_{12}}))/(2\pi)$}  \\
\hline
\hline
\SI{3.448646}{\giga\hertz}$\pm$\SI{0.195}{\kilo\hertz} & \SI{3.240105}{\giga\hertz}$\pm$\SI{0.423}{\kilo\hertz}&\SI{3.240403}{\giga\hertz}$\pm$\SI{0.431}{\kilo\hertz} \\
\hline
\end{tabular}
\caption{Expectations and standard deviations of transition frequencies determined by the Bayesian characterization.  \label{tab:bayesian}}
\end{center}
\end{table}

\begin{figure}[htb]
\centering
  \includegraphics[width=0.75\textwidth]{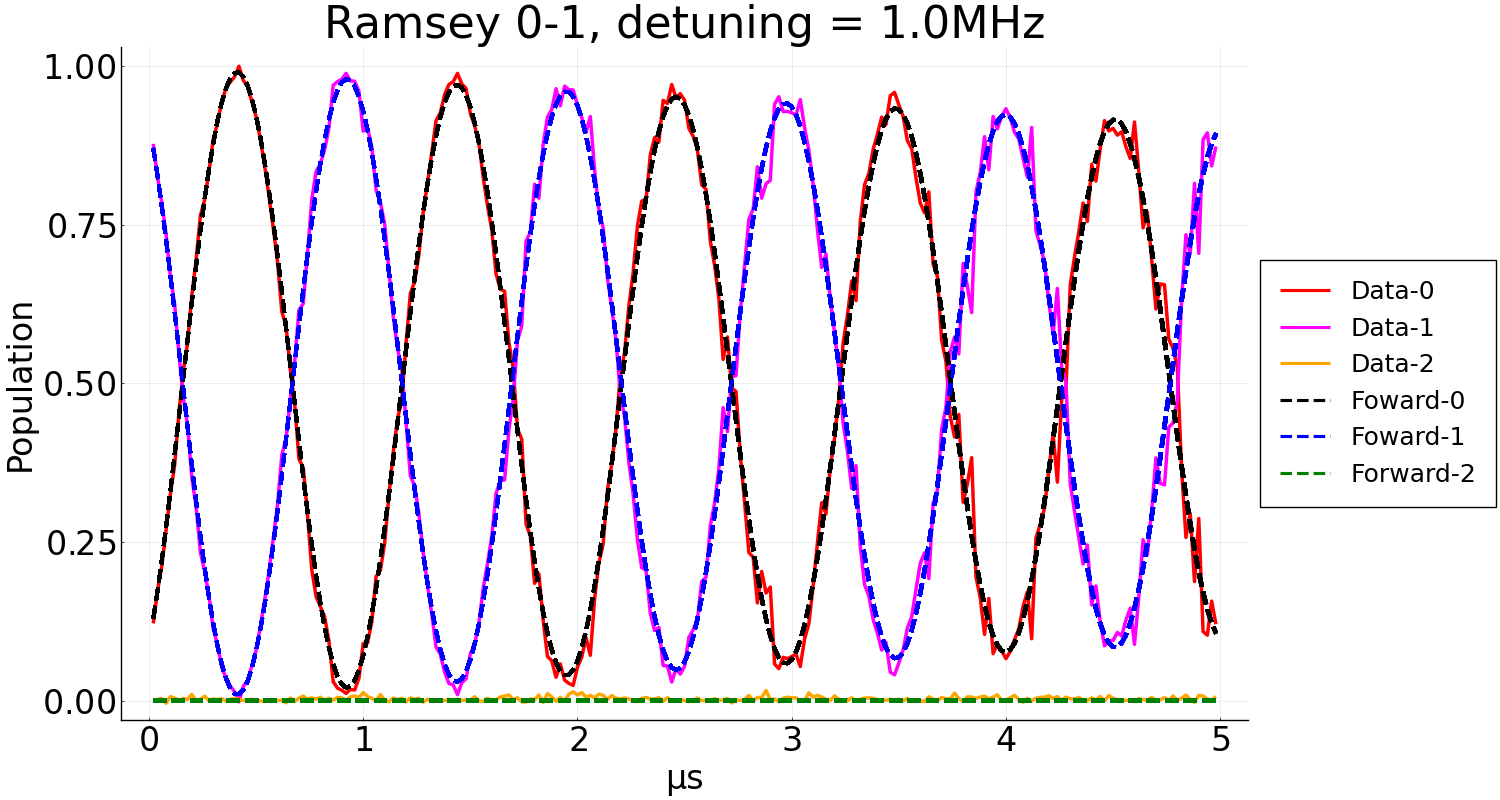}\\ 
  \includegraphics[width=0.75\textwidth]{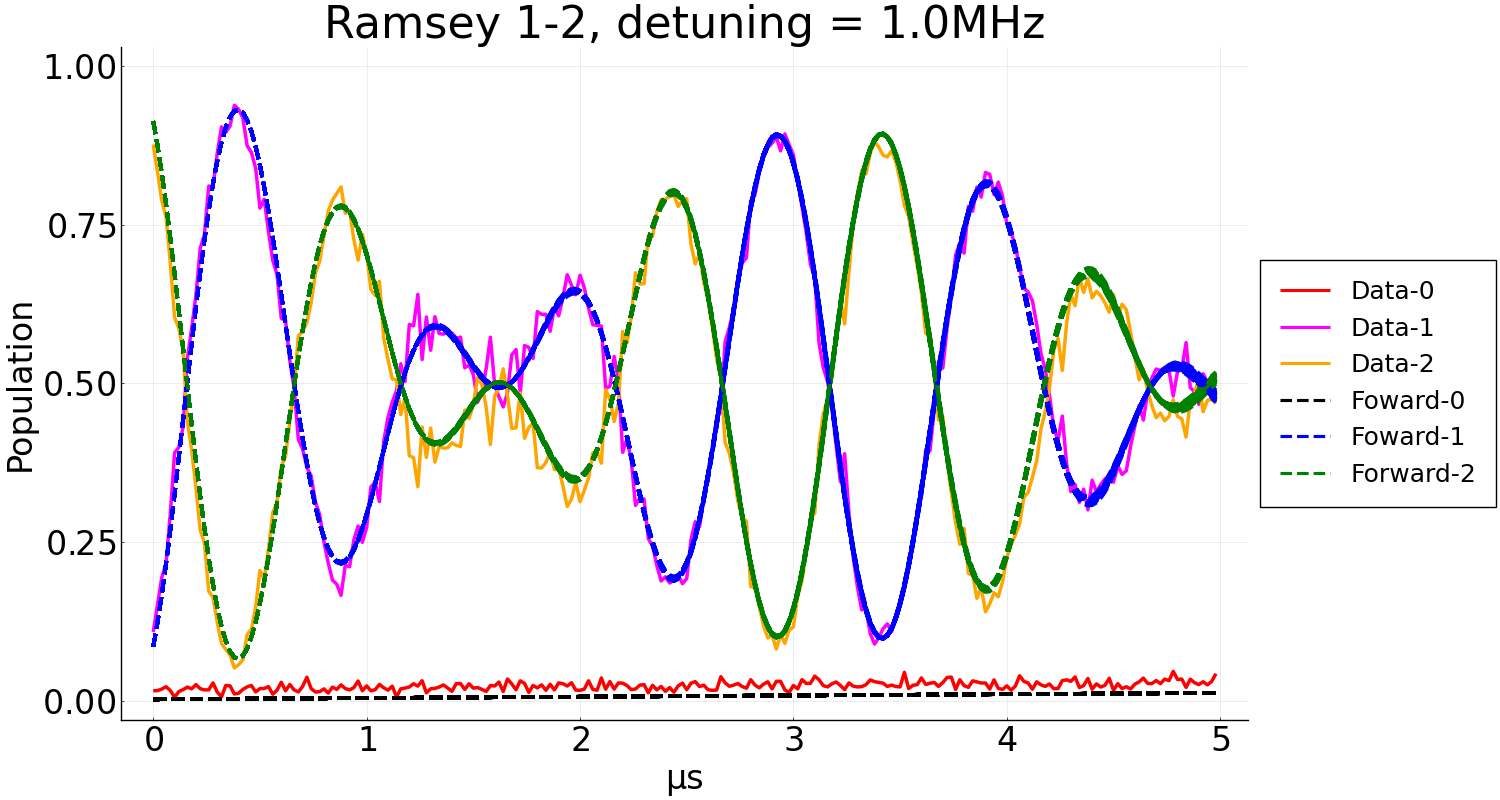}
\caption{Comparison between experimental data and forward simulation results corresponding to 200 samples of transition frequencies sampled from the Markov chain. The thickness of the lines (noticeable in the Ramsey 1-2 case) is a result of the variation in transition frequencies.}\label{fig:bayesian-vs-exp}
\end{figure}

The MCMC simulations show good qualitative agreement with experimental data. In Figure \ref{fig:bayesian-vs-exp} we overlay 200 forward simulations on top of the experimental data. Here, each forward simulation is based on fixed transition frequencies, drawn from individual samples of the Markov chains.

\clearpage 
\section{Tuning and experimental validation of the control pulse on  QuDIT\label{sec:calibration}}
Having characterized the parameters (transition frequencies) in the Hamiltonian we are now ready to use {\tt Juqbox.jl} to generate control pulses for the $0\leftrightarrow2$ SWAP gate. As we have both deterministic values and probability densities for the transition frequencies we can use both the deterministic and risk neutral optimization methods.

Up to this point, we have assumed that the control pulses can be directly applied to the quantum device. However, in most cases, the control pulses are generated at room temperature outside the refrigerator that contains the quantum device. In practice, the control pulse is transmitted into the fridge through wires. This transmission distorts the control signal, basically by a frequency dependent attenuation. As a result, the optimized control pulses must be corrected to compensate for this distortion before they are applied the qudit device. This tuning procedure is discussed below. In the following we also present one deterministic and two risk neutral control pulses for realizing a $0\leftrightarrow2$ SWAP gate, and discuss how to convert their {\tt Juqbox.jl} control vectors to an input format used by the IQ-mixer. Finally, we present experimental validation of the implementation of the $0\leftrightarrow2$ SWAP gate in a few different scenarios.  

\subsection{Deterministic and risk neutral control pulses}
Using  {\tt Juqbox.jl} we design one deterministic and two risk neutral pulses for the $0\leftrightarrow2$ SWAP gate. For all of these we use two carrier waves where the frequencies are ${\omega}_{0,1}$ and $\bar{\omega}_{1,2}=\frac{1}{2}(\omega_{1,2}^++\omega_{1,2}^-)$, as reported in Table \ref{tab:det}. 

As mentioned in Section \ref{sec:characterization}, the performance of the control pulses is insensitive to the choice of $\omega_{2,3}$, so we (somewhat arbitrary) set $\omega_{2,3}/(2\pi) =\SI{3.005}{\giga\hertz}$. As we will compare the experimental fidelity of the three pulses we keep their duration to be the same: \SI{256}{\nano\second}. 

\begin{figure}[h]
  \begin{center} 
  \includegraphics[width=0.49\textwidth]{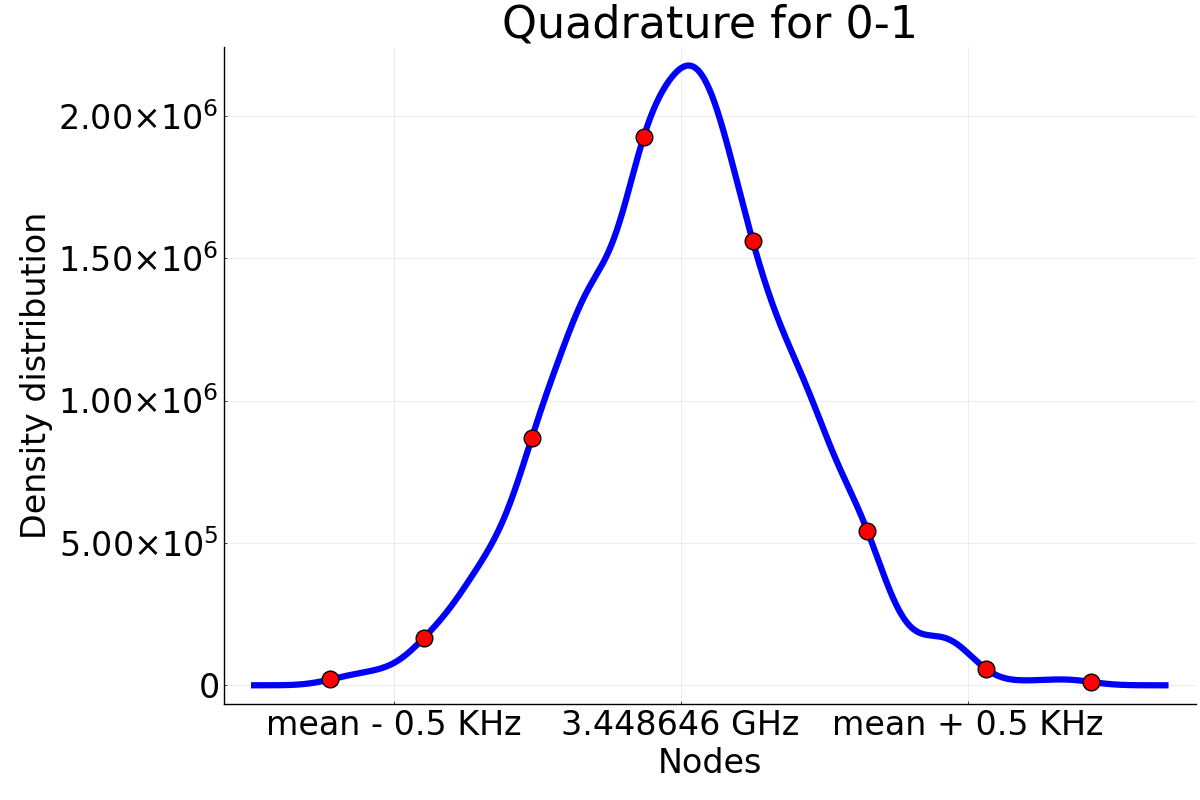}
  \includegraphics[width=0.49\textwidth]{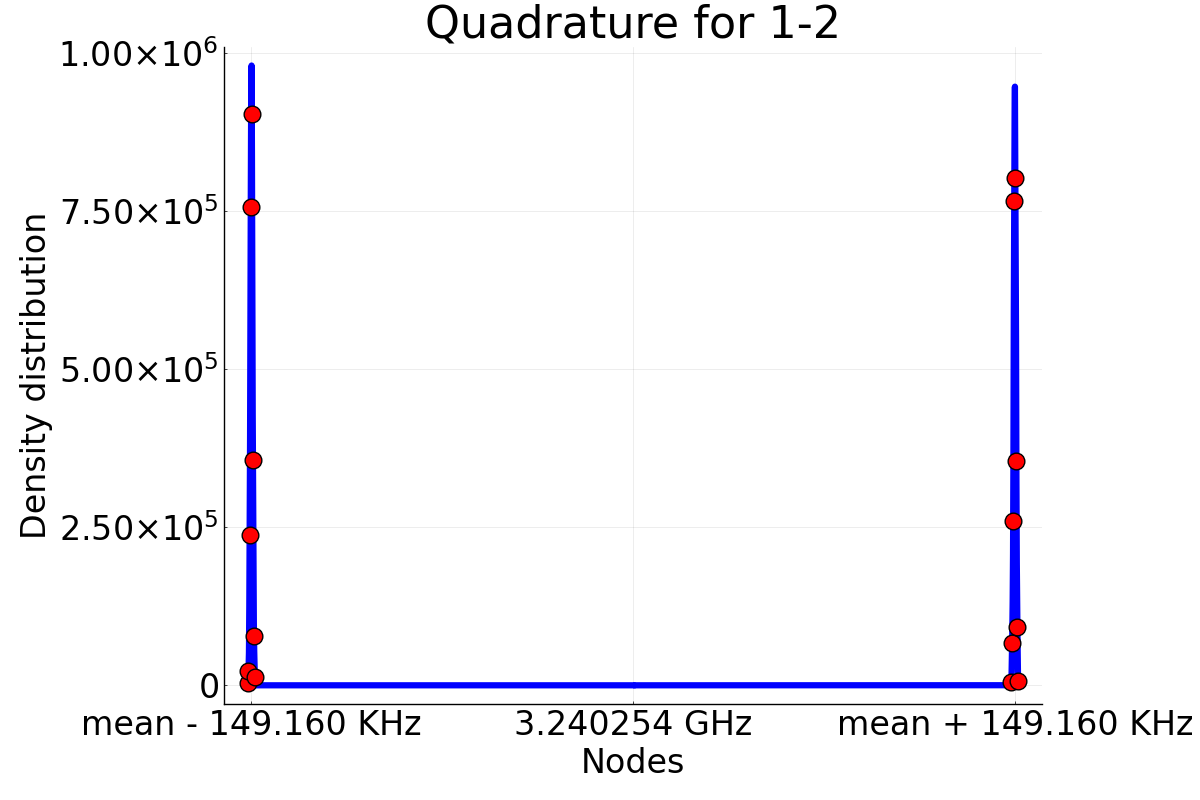}
  \caption{Quadrature nodes for transition frequencies and the underlying probability distribution. Dots: quadrature nodes. Solid line: the underlying probability distribution.  Left: $\omega_{01}/(2\pi)$. Right: $\omega_{12}/(2\pi)$.\label{fig:quad_points}}
  \end{center}
\end{figure}

\paragraph{Deterministic pulse.} The deterministic control pulse does not account for the presence of parity events and simply uses $\bar{\omega}_{1,2}$ (calculated from Table \ref{tab:det}) as the $1$-$2$ transition frequency. In this case, the deterministic objective \eqref{eq:det_objective} function is minimized and {\tt Juqbox.jl} is able to find a solution that gives $99.99\%$ fidelity.

\paragraph{2-point risk-neutral pulse.} In the risk neutral optimization method, the objective function (\ref{eq:riskneutral_obj}) is the expected value (average) of the deterministic objective function, with respect to the distribution of the parameters in the Hamiltonian. The expected value is approximated by quadrature. In this case we only consider the distribution of the 1-2 transition frequency and use the deterministic value of $\omega_{0,1}$ from Table \ref{tab:det}. To approximate the expected value of $\omega_{1,2}$ we use the minimal two point quadrature rule with quadrature nodes ${\omega}_{1,2}^\pm$ from Table \ref{tab:det} and equal quadrature weights. This minimal quadrature allows us to take the parity event into account. Using this setup, {\tt Juqbox.jl} finds a control pulse with $99.99\%$ average fidelity. 

\paragraph{128-point risk neutral pulse.} In our final example we consider a more advanced application of the risk neutral technique. Here we use the posterior for both $\omega_{0,1}$ and $\omega_{1,2}$ in the risk neutral objective function. The expectation is approximated by a tensor product Gauss quadrature rule with $8 \times 16 = 128$  points. The two one-dimensional quadrature rules are constructed following \cite{johnson2019accurate}. The quadrature nodes for $\omega_{0,1}$ and $\omega_{1,2}$ are presented in Figure \ref{fig:quad_points}. 

Using this setup, {\tt Juqbox.jl} finds an optimal control with $99.87\%$ average fidelity. We note that it is natural that the fidelity is slightly lower here than in the previous cases, because the control pulse needs to give high fidelity for many different values of the parameters ($\omega_{0,1}$ and $\omega_{1,2}$), rather than being very high for one particular case.   

\begin{figure}[h]
  \begin{center} 
  \includegraphics[width=0.32\textwidth]{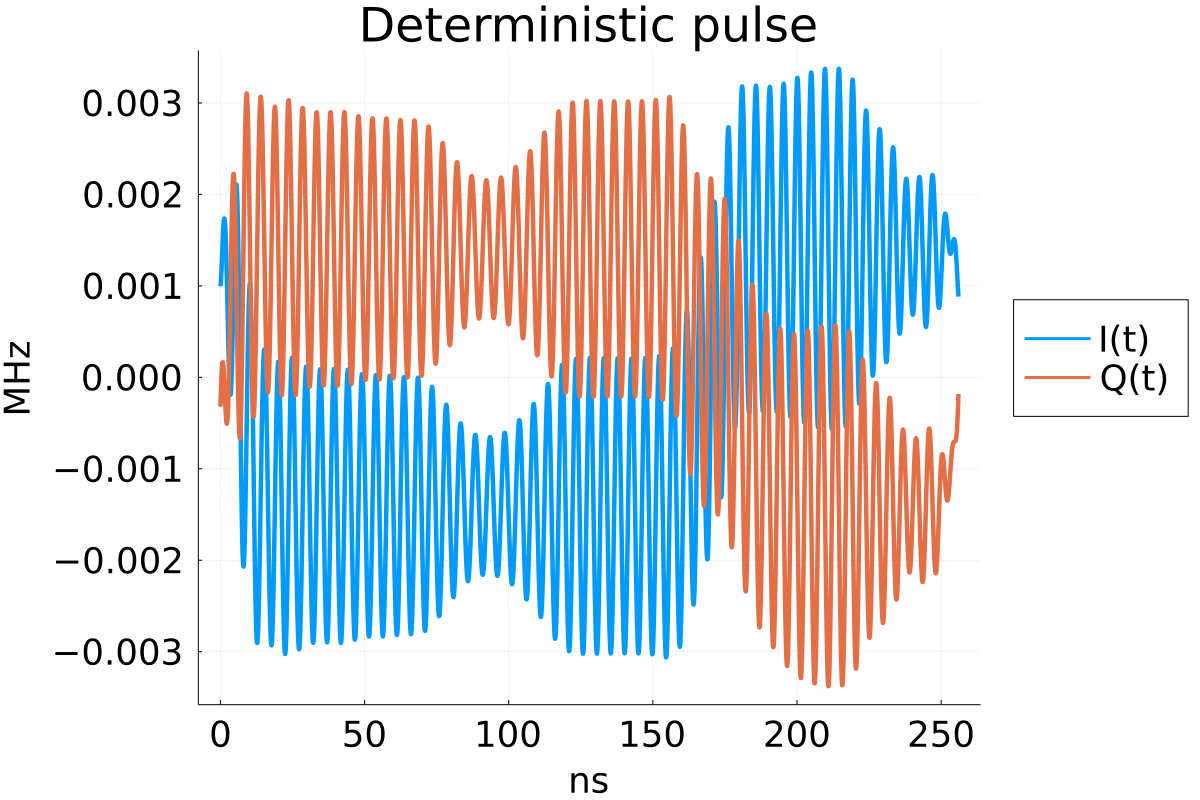}
  \includegraphics[width=0.32\textwidth]{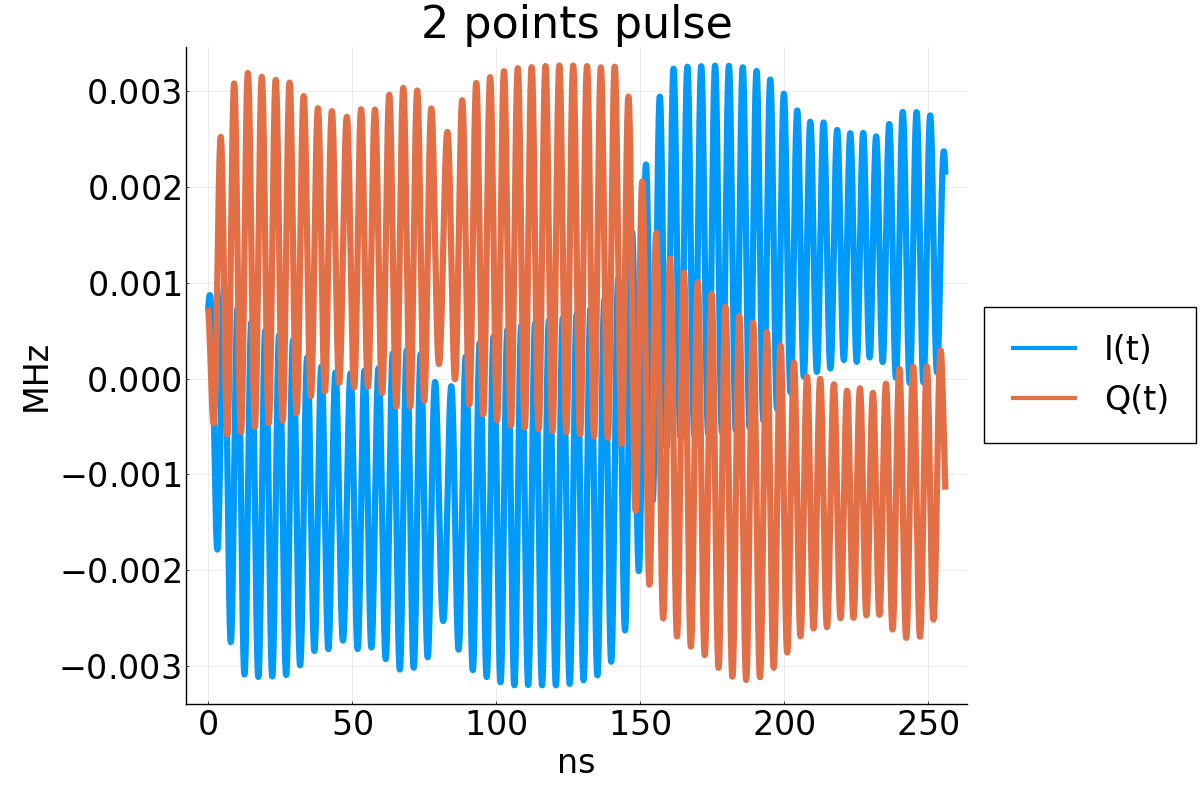}
  \includegraphics[width=0.32\textwidth]{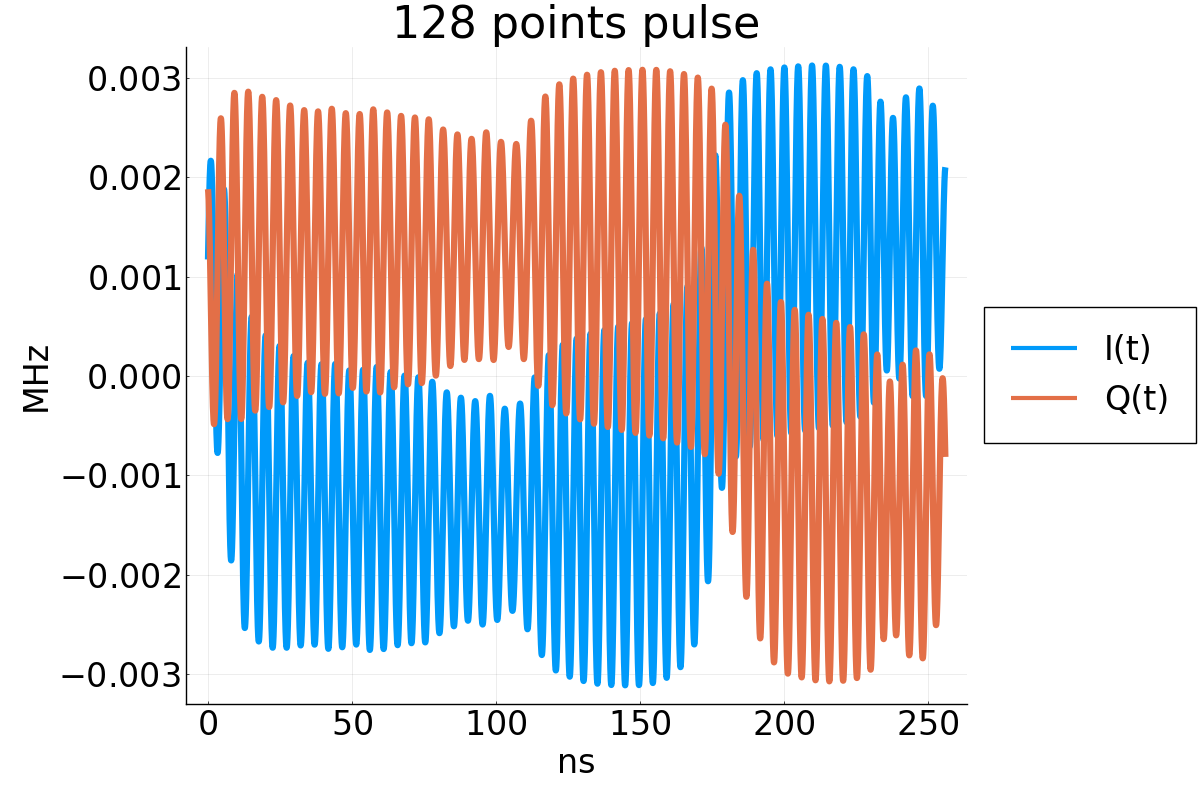}
  
  \vspace{0.3cm}
  
  \includegraphics[width=0.32\textwidth]{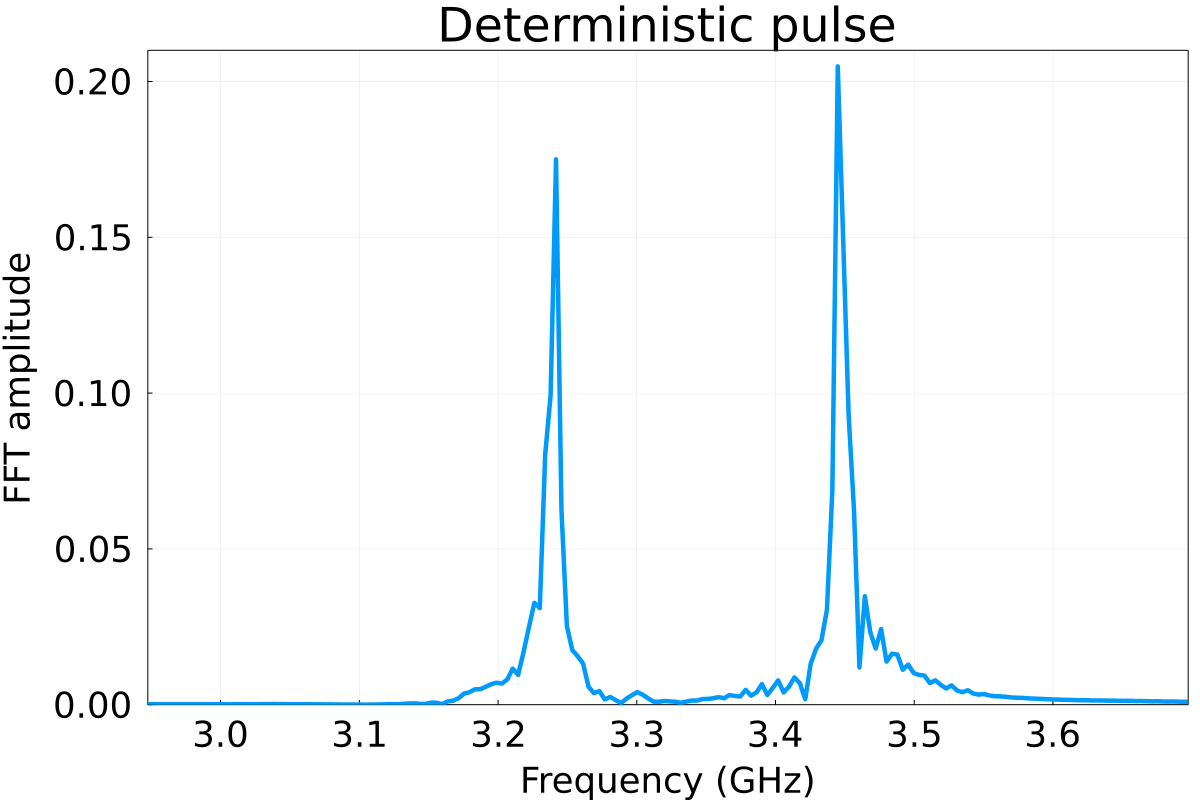}
  \includegraphics[width=0.32\textwidth]{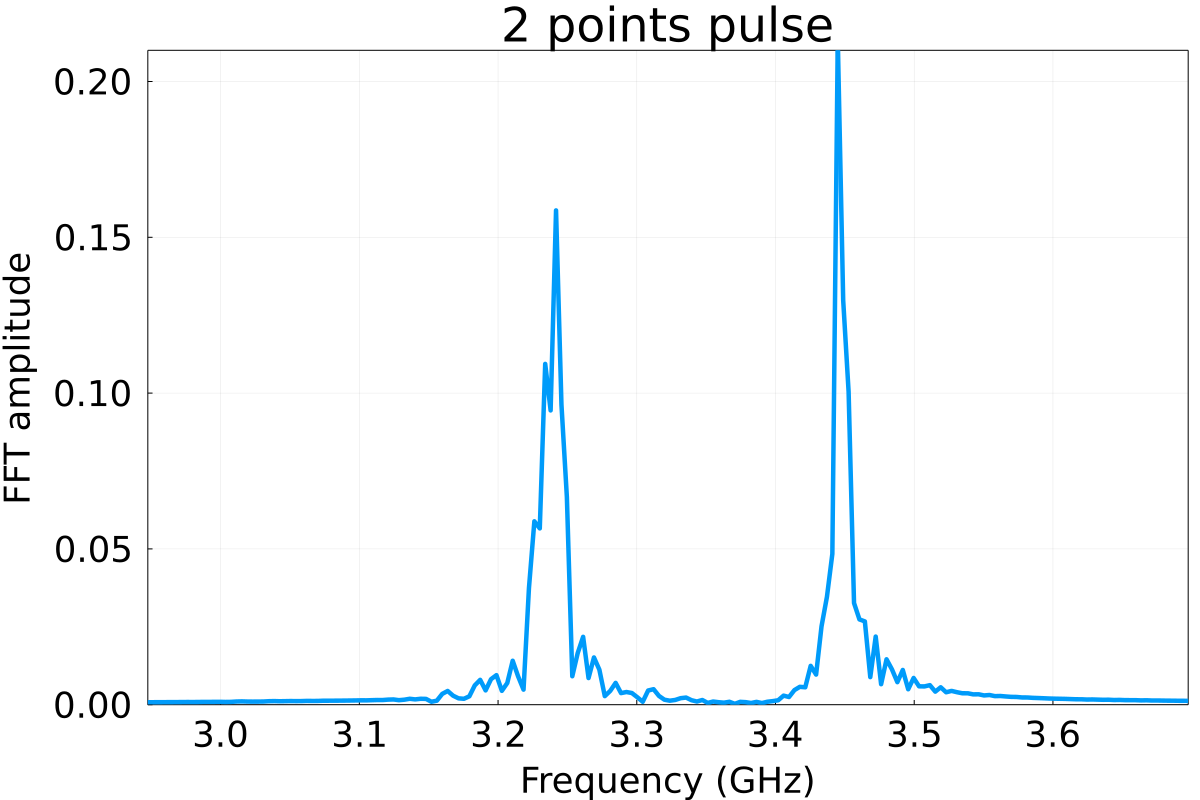}
  \includegraphics[width=0.32\textwidth]{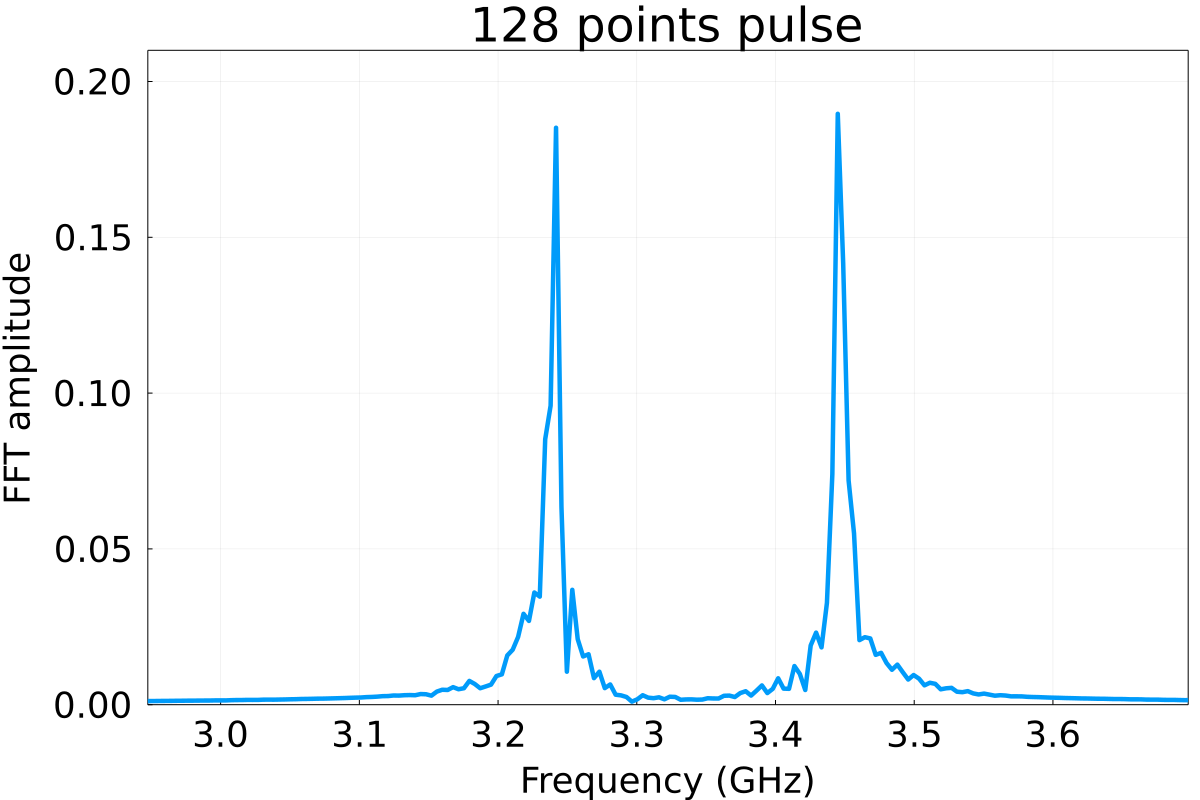}
  \caption{Top: the wave forms of $I(t)$, $Q(t)$ for the different pulses for realizing the $0$-$2$ SWAP gate. Bottom: Fourier amplitudes of the lab frame control pulses (defined by equation (\ref{eq:control-Hamiltonian-IQ}). \label{fig:control-pulses}}
  \end{center}
\end{figure}

In Figure \ref{fig:control-pulses}, we present the $I(t)$ and $Q(t)$ components of the three optimized control pulses and the Fourier transform of the corresponding pulses in the lab frame (defined by equation \eqref{eq:control-Hamiltonian-IQ}). The energy in the control pulses is concentrated around the carrier wave frequencies for all cases, but the three pulses have sightly different Fourier amplitudes. 

\subsection{Converting {\tt Juqbox.jl} controls to signals on the QuDIT device\label{sec:iq-mixer}}
The output of {\tt Juqbox.jl} is the control vector $\balpha$ consisting of the the coefficients of the B-splines with carrier waves (see equation (\ref{eq:control_function}). These coefficients must be converted to $I(t)$, $Q(t)$ signals to be played on the qudit.

Either an IQ-mixer or an arbitrary waveform generator (AWG) can be used to generate the control pulses for a quantum device. An IQ-mixer takes $I(t)$ and $Q(t)$ in 
equation \eqref{eq:control-Hamiltonian-IQ} as the input, and up-converts them to the lab frame control pulse for the quantum device. The AWG devices directly uses the lab frame control pulse as its input. 

The qudit device used in this study is controlled by an IQ-mixer; we proceed by describing how we use it together with the output from {\tt Juqbox.jl}. First, compute $I(t)$ and $Q(t)$ based on equation \eqref{eq:control_function} and equation \eqref{eq:control-Hamiltonian-IQ}. Then, sample from $I(t)$ and $Q(t)$ with a sample rate of $32$ samples per nanosecond, and store the sampled values separately in two arrays. The IQ-mixer reads the stored samples of $I(t)$ and $Q(t)$ and generates control pulses. \pzc{After the IQ-mixer reads the $I$-$Q$ samples, in practice, we also need to tune the control pulse to compensate for imperfect transmission of the signal from the warm side of the laboratory (the IQ-mixer) to the cold side, where the quantum device resides. This tuning step will be discussed in more details in the following subsection.}

\subsection{Final tuning of the control pulses for the $0\leftrightarrow2$ SWAP gate}
Before carrying out the experimental validation we tune the control pulse to account for distortion of the control pulses during transmission from the warm to the cold side. 

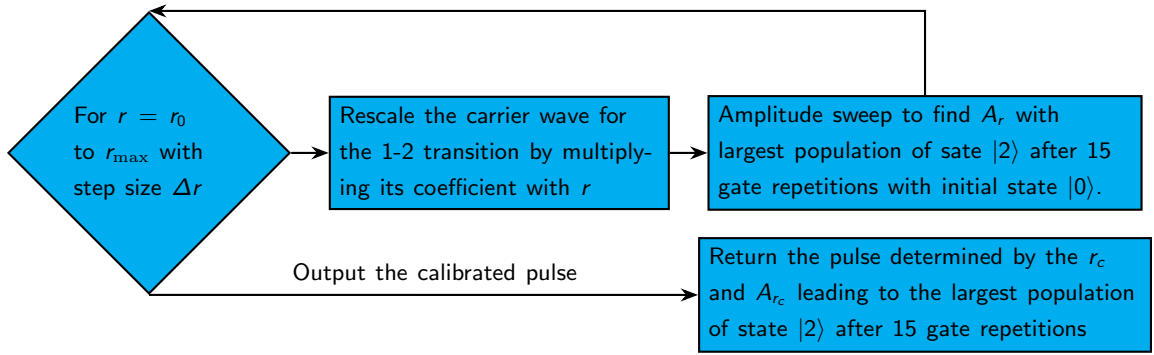
\begin{figure}[h]
\begin{center}
\begin{tikzpicture}
\tikzset{every node}=[font=\sffamily\sansmath]
\node (For) [diamond,text width=2.0cm,draw, thick,fill=cyan] {\footnotesize For $r=r_0$ to $r_\textrm{max}$ with step size $\Delta r$};
\node (Rescale) [block_device_small,text width=4.25cm, right = of For,xshift=-0.5cm ] {\footnotesize Rescale the carrier wave for the $1$-$2$ transition by multiplying its coefficient with $r$};
\node (AmpSweep) [block_device_small,text width=5.5cm, right = of Rescale,xshift=-0.5cm ] {\footnotesize Amplitude sweep to find $A_r$ with largest population of  sate $|2\rgl$ after $15$ gate repetitions with initial state $|0\rgl$.};
\node (Output) [block_device_small,text width=5.75cm, below = of AmpSweep,yshift=0.65cm ] {\footnotesize Return the pulse determined by the $r_c$ and $A_{r_c}$ leading to  the largest population of state $|2\rgl$ after $15$ gate repetitions};
\draw [arrow] (AmpSweep.north) -- ++(0,1.0)|-(For.north);
\draw [arrow](For)--(Rescale);
\draw [arrow](Rescale)--(AmpSweep);
\draw [arrow](For.south)--node [text width=4.5cm,midway,above,text centered,xshift=0.15cm] {\footnotesize{Output the calibrated pulse}}(Output.west);
\end{tikzpicture}
\end{center}
\caption{The flowchart for the calibration of $0\leftrightarrow2$ SWAP gate.\label{fig:pulse_calibration_flow}}
\end{figure}

In our $0\leftrightarrow2$ SWAP gate, we use two carrier waves whose frequencies correspond to the $0$-$1$ and the $1$-$2$ transition frequencies, respectively. These two carrier waves are  parameterized by B-splines. To fine tune the pulse and account for non-linearity we multiply the B-spline coefficients of the carrier wave corresponding to the $1$-$2$ transition with a constant $r$, and then rescale the control pulse by multiplying all the coefficients of the B-splines with the pulse amplitude $A$. This procedure changes the envelope function $d(t)$ in \eqref{eq:control_function} to $\hat{d}(t;r,A)$: 
\begin{equation}
    \hat{d}(t;r,A)=A\left[ (p_0(t;\balpha)+iq_0(t;\balpha))e^{i\Delta_0 t}
                            +r(p_1(t;\balpha)+iq_1(t;\balpha))e^{i\Delta_1 t} \right].
    \label{eq:calibration_rescale}
\end{equation}

Given the control vector $\balpha$, we tune the control pulse by finding proper values of $r$ and $A$. The experimental protocol to determine $r$ and $A$ is given in Protocol \ref{prot:freq_filter} and is also summarized as a flowchart in Figure \ref{fig:pulse_calibration_flow}. 

Using Protocol \ref{prot:freq_filter} we calibrate the three control pulses, yielding three pairs of $r_c$ and $A_{r_c}$ reported in Table \ref{tab:calibration}.
\begin{table}[h]
\begin{center}
\begin{tabular}{|c|c|c|c|c|c|c|c||c|c|}
\hline
       &Deterministic & $2$ points & $128$ points \\ \hline
       $r_c$ & $0.626$       & $0.633$    & $0.637$\\ \hline
       $A_{r_c}$ & $0.997$   & $0.996$    & $1.003$      \\ \hline
\end{tabular}
\caption{Values of $r_c$ and $A_{r_c}$ resulting from the tuning of different pulses. \label{tab:calibration}}
\end{center}
\end{table}

\subsection{Experimental validation - single gate}
We perform single gate measurements to investigate the performance of the three control pulses. In these tests and the tests in next subsection, all the pulses are tuned with ``their own'' values of $(r_c,A_{r_c})$ (from Table \ref{tab:calibration}).

\begin{figure}[h!]
  \begin{center} 
  \includegraphics[width=0.38\textwidth]{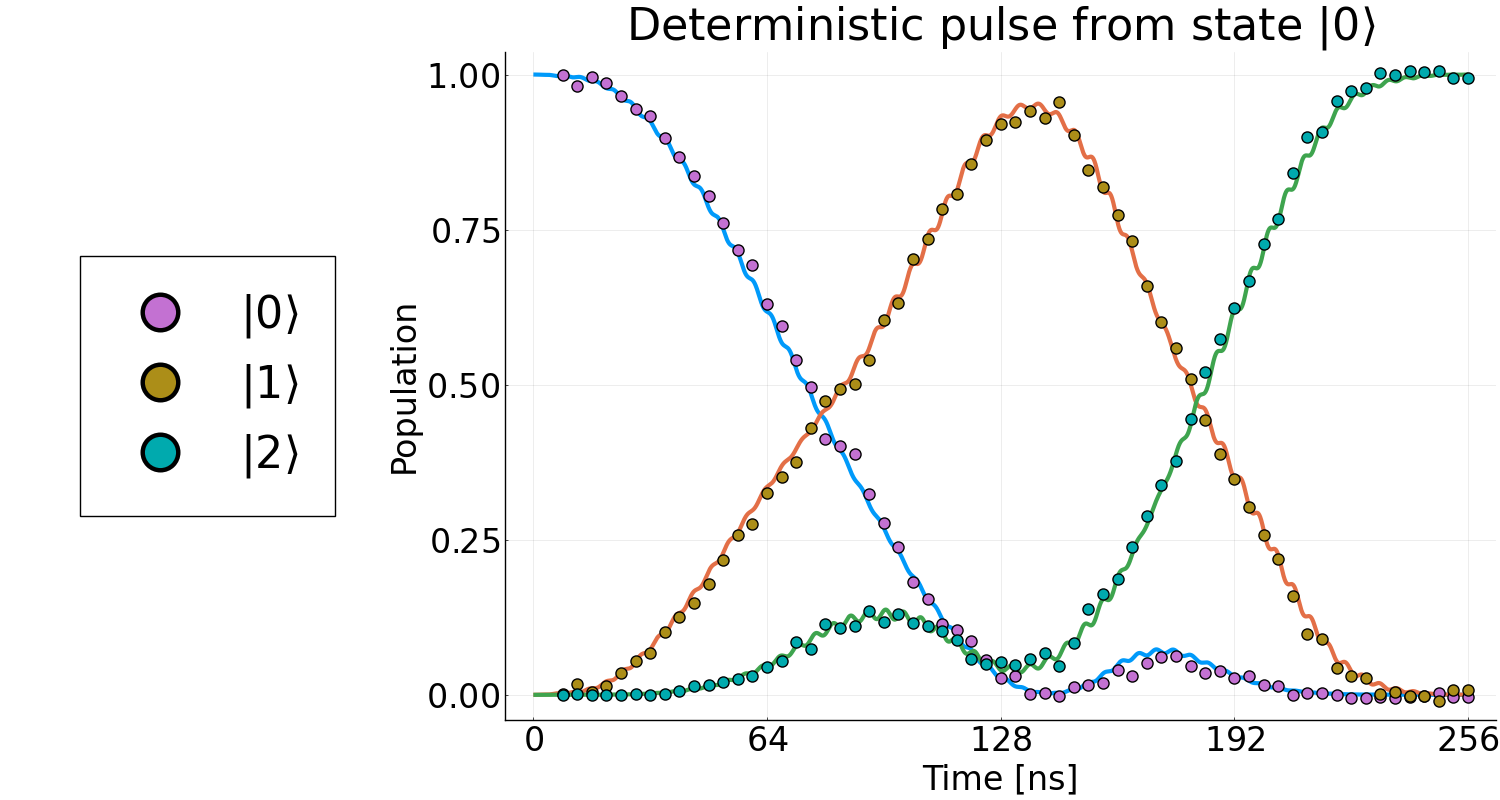}
  \includegraphics[width=0.30\textwidth]{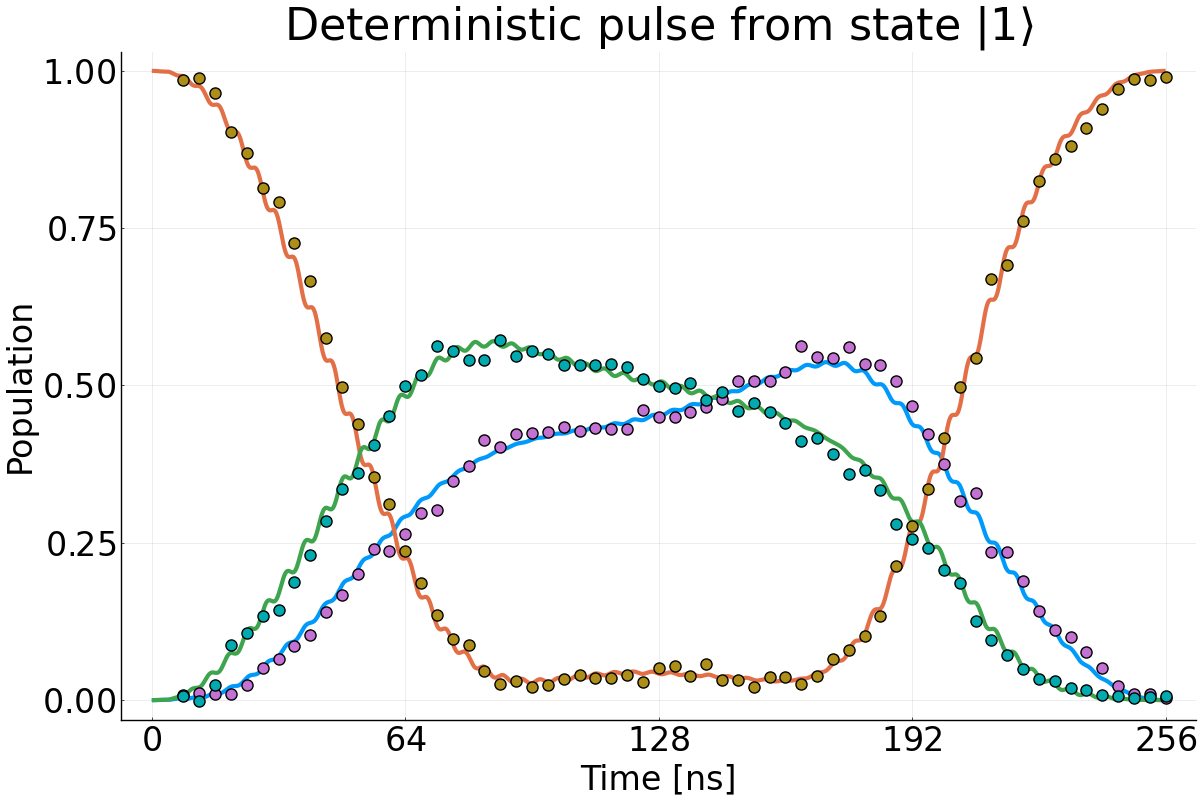}
  \includegraphics[width=0.30\textwidth]{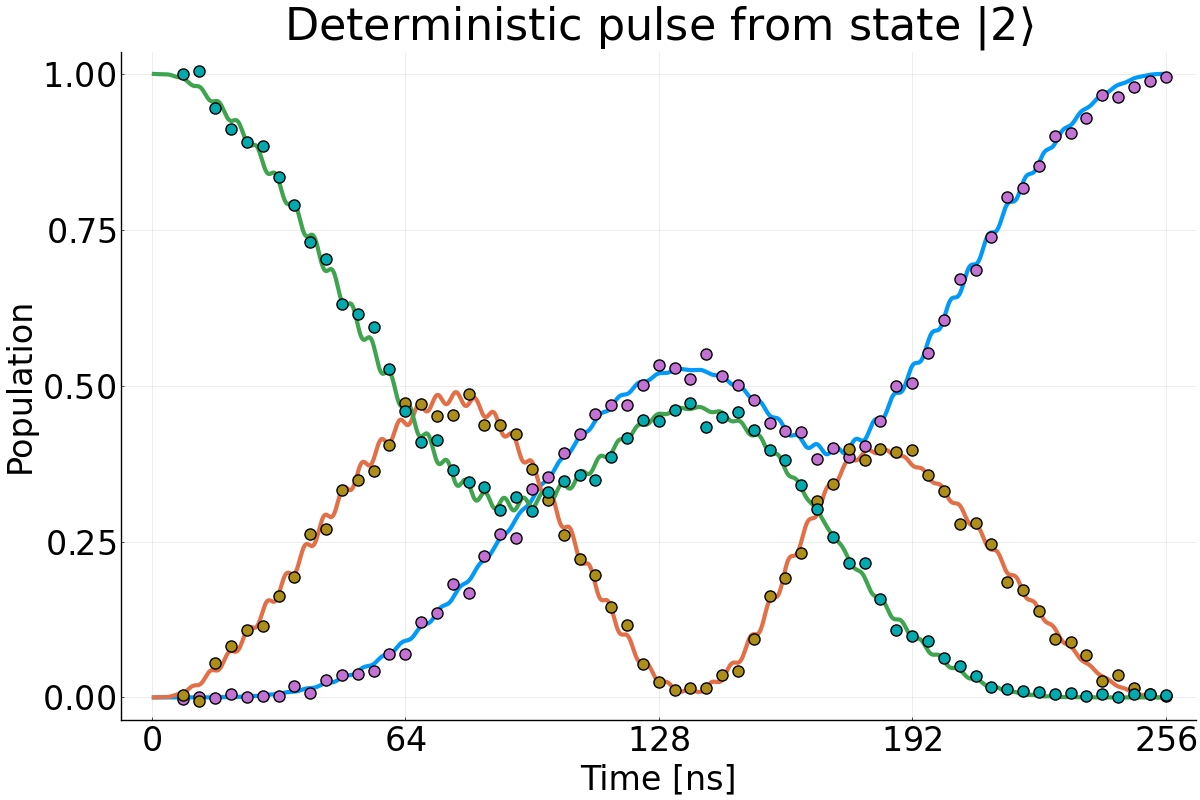}
  \includegraphics[width=0.38\textwidth]{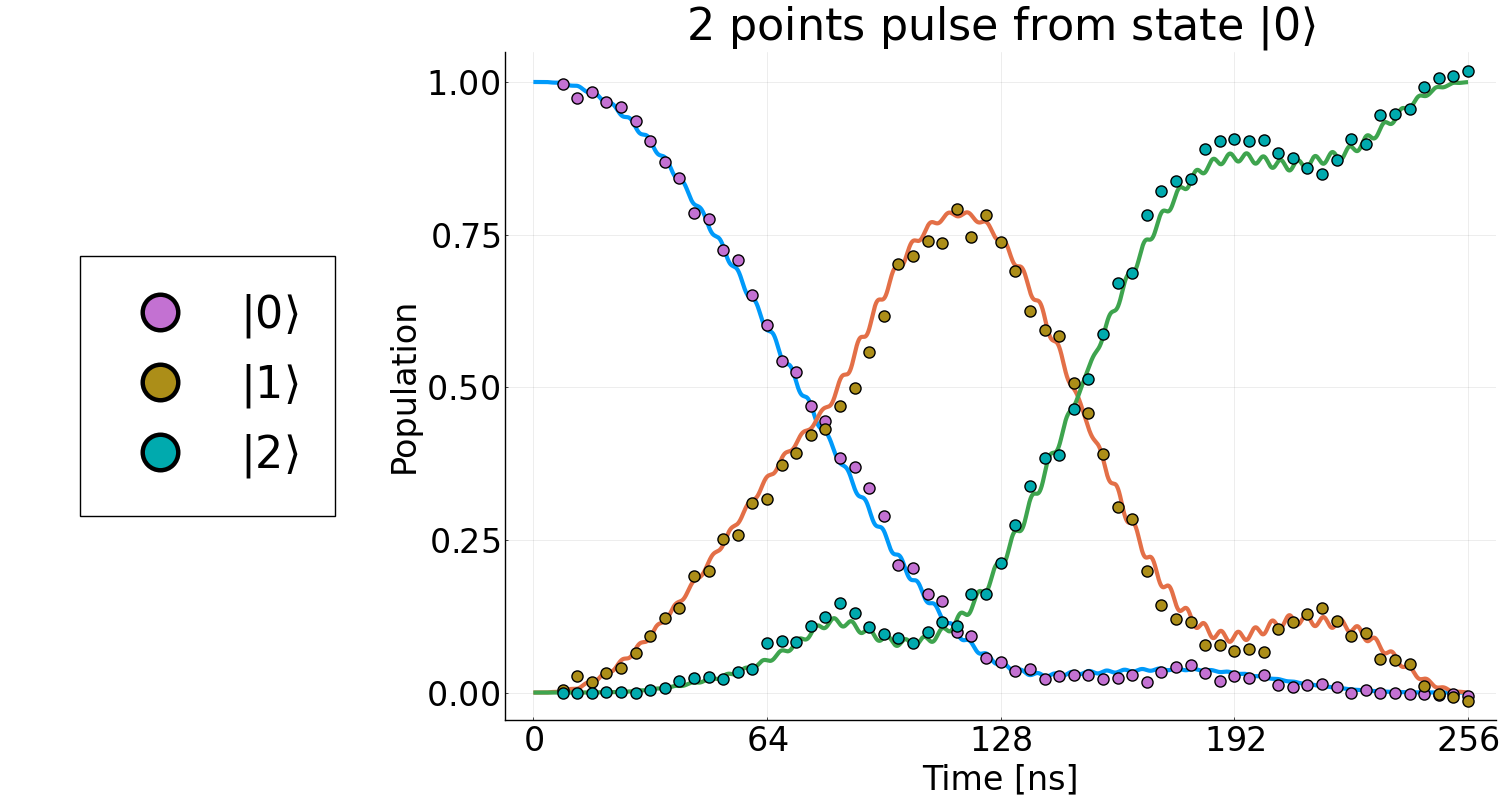}
  \includegraphics[width=0.30\textwidth]{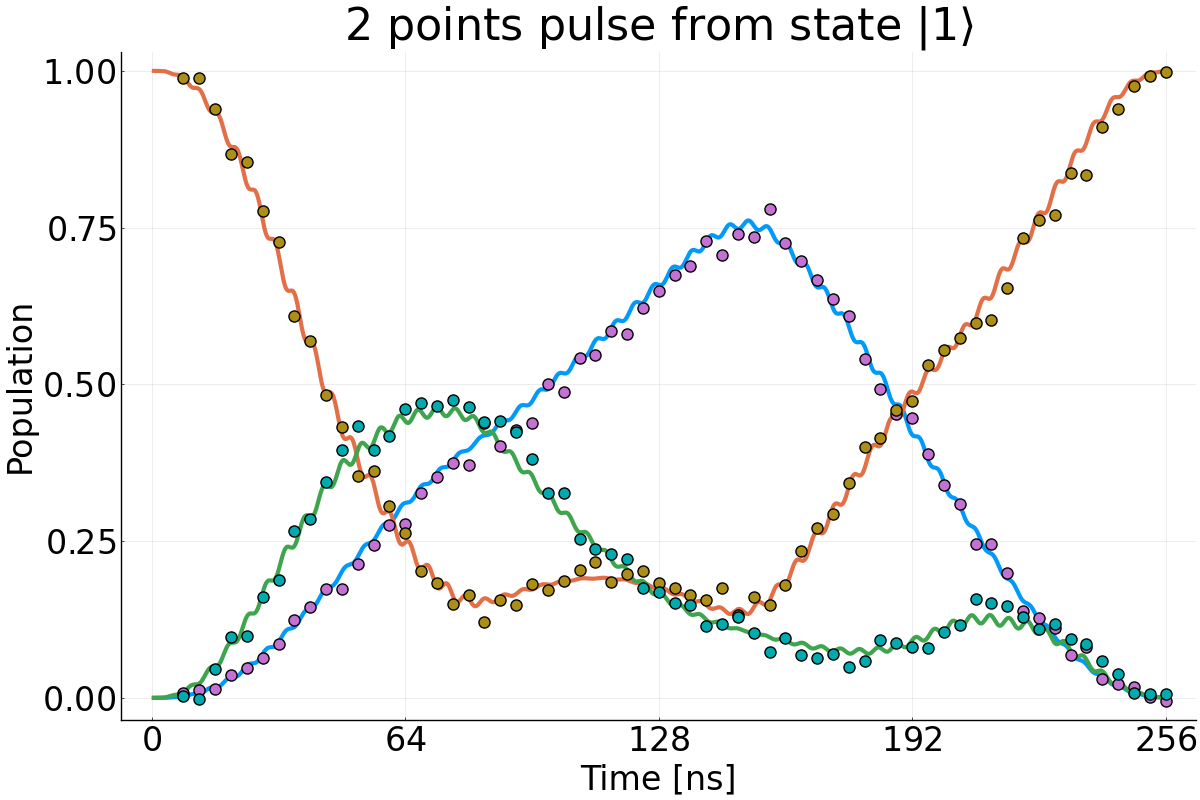}
  \includegraphics[width=0.30\textwidth]{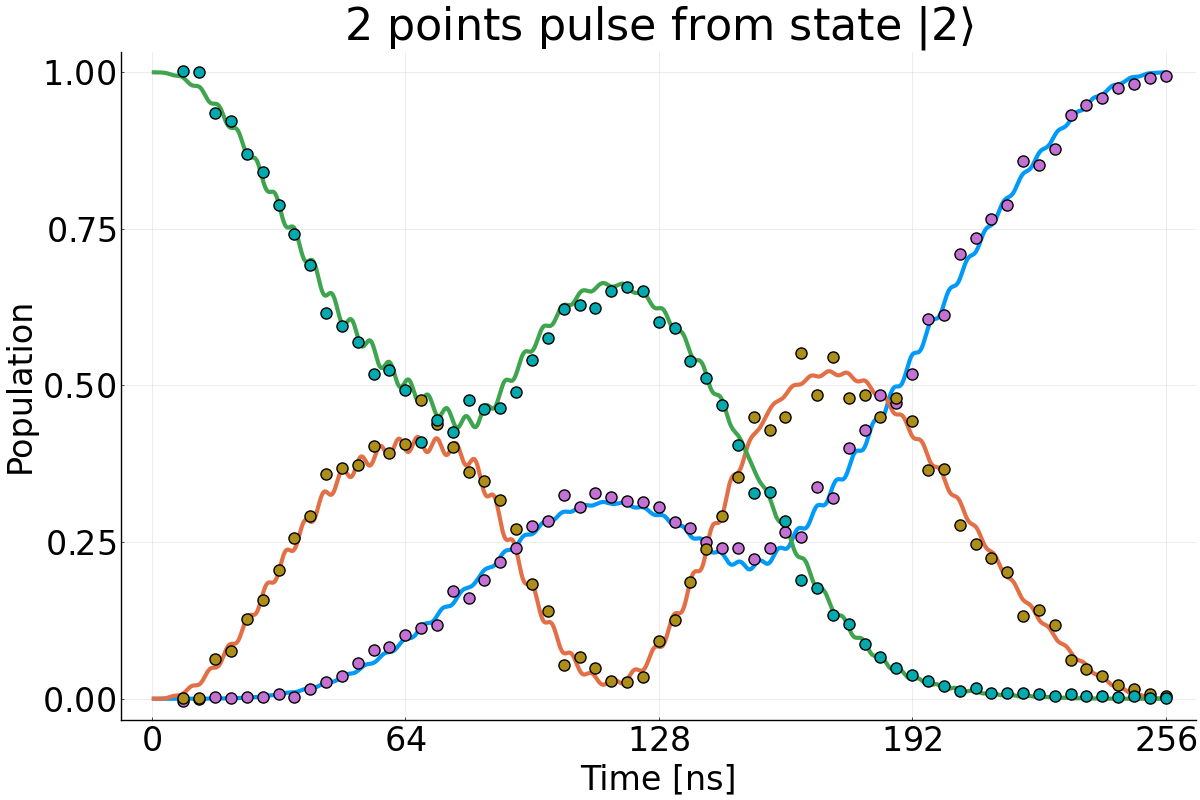}
  \includegraphics[width=0.38\textwidth]{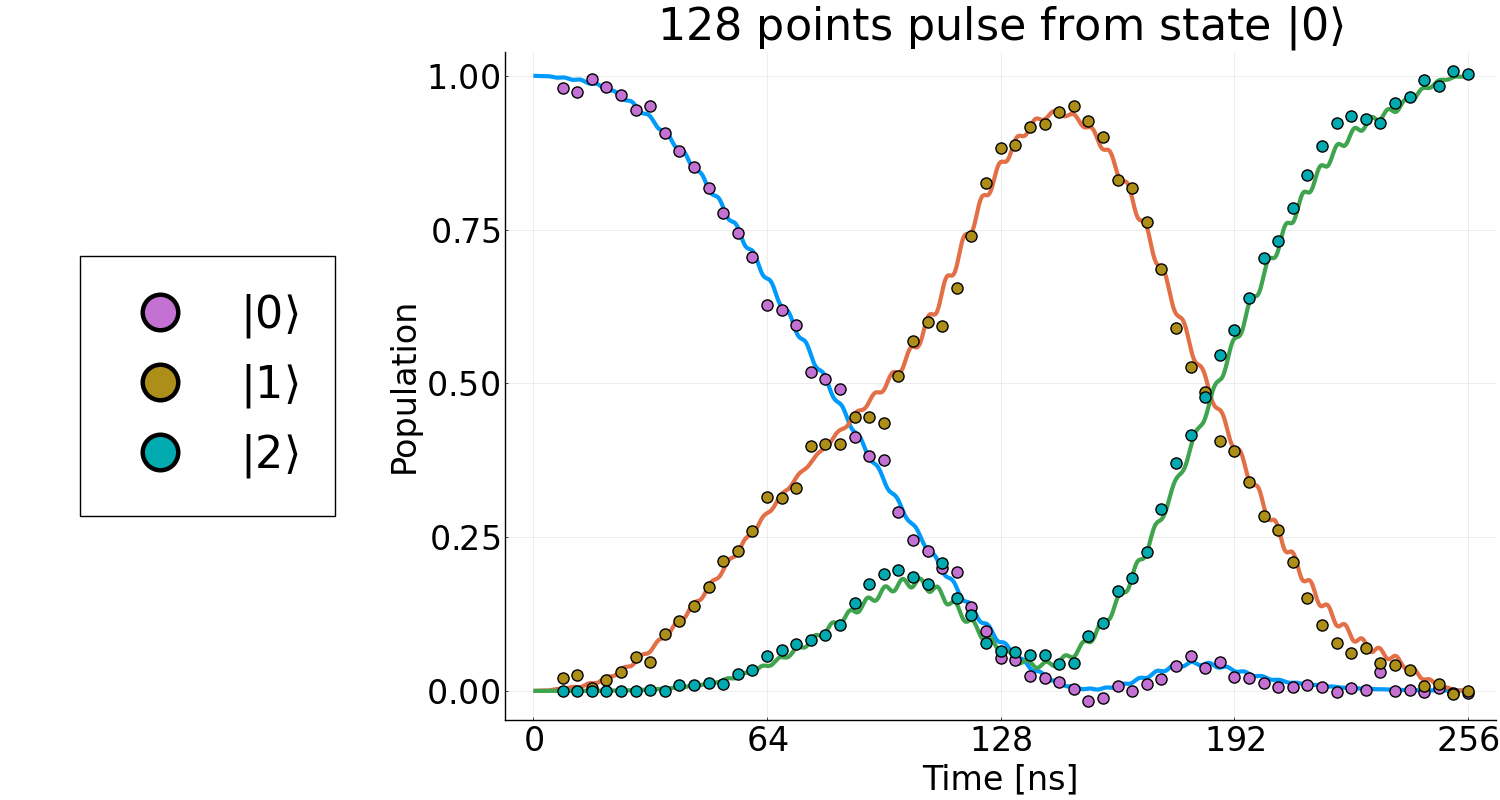}
  \includegraphics[width=0.30\textwidth]{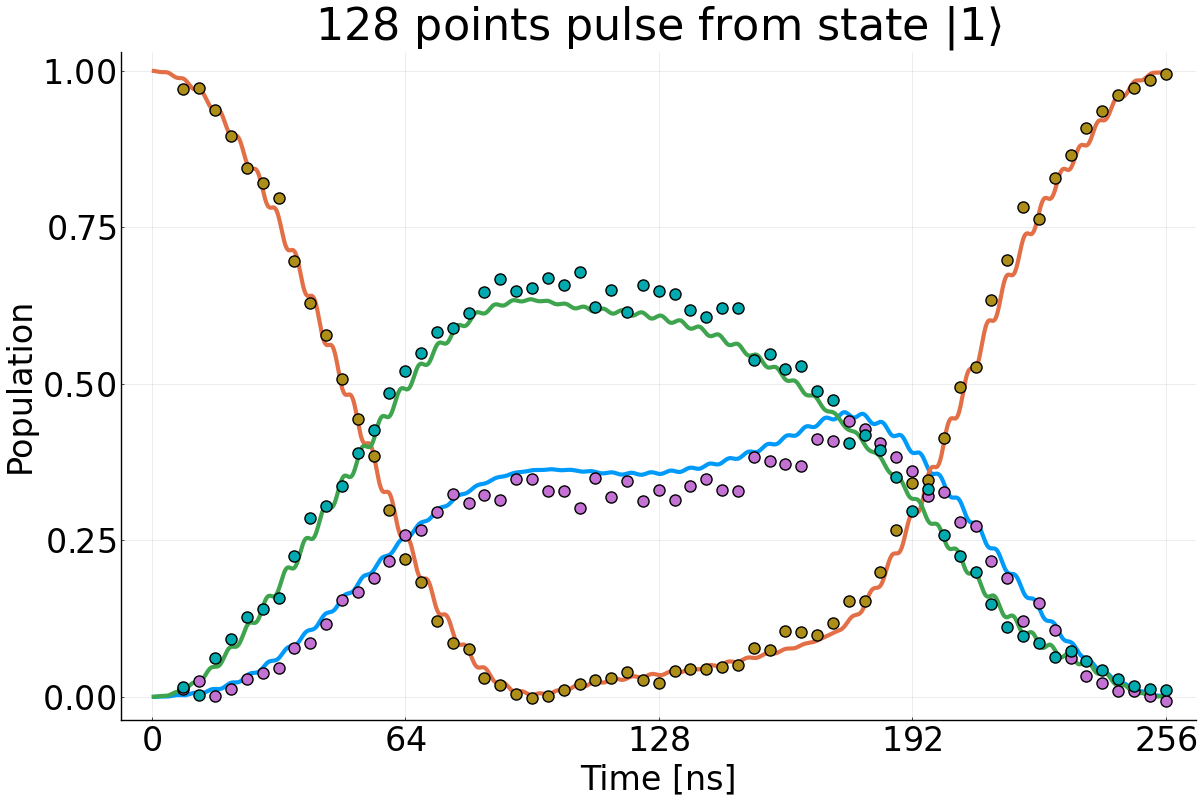}
  \includegraphics[width=0.30\textwidth]{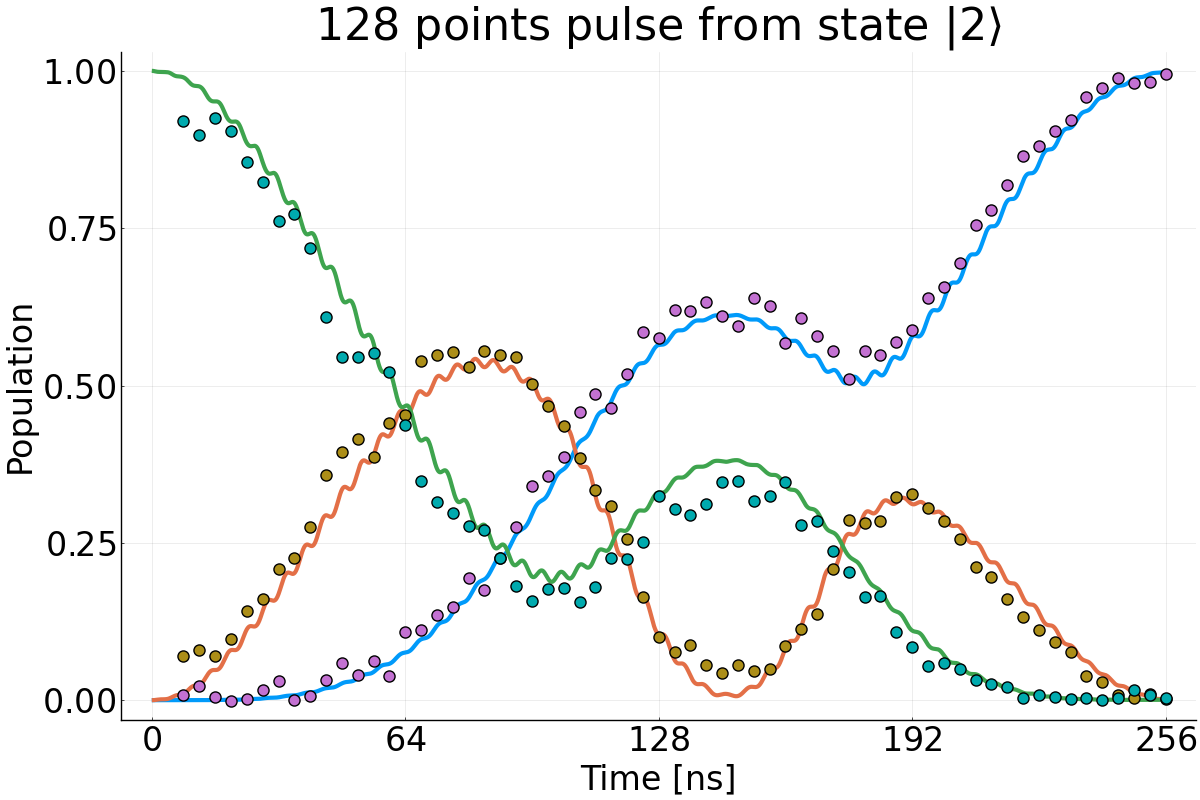}
  \caption{Probability population for different states during a single gate application. Filled circles: experimental data. Solid line: {\tt Juqbox.jl} simulation results. Top row: deterministic pulse. Middle row: 2 points pulse. Bottom row: 128 points pulse.\label{fig:single-gate}}
  \end{center}
\end{figure}

The first test is to measure the evolution of the population for different states during a single gate application. The measurement results are overlaid with {\tt Juqbox.jl} simulation results and displayed in Figure \ref{fig:single-gate}. Each point of the measurement results is the average of 1000 repetitions. For all the three control pulses, experimental results qualitatively match simulation results.

\subsection{Experimental results - gate repetition}
\begin{figure}[ht]
  \begin{center} 
  \includegraphics[height=0.16\textheight]{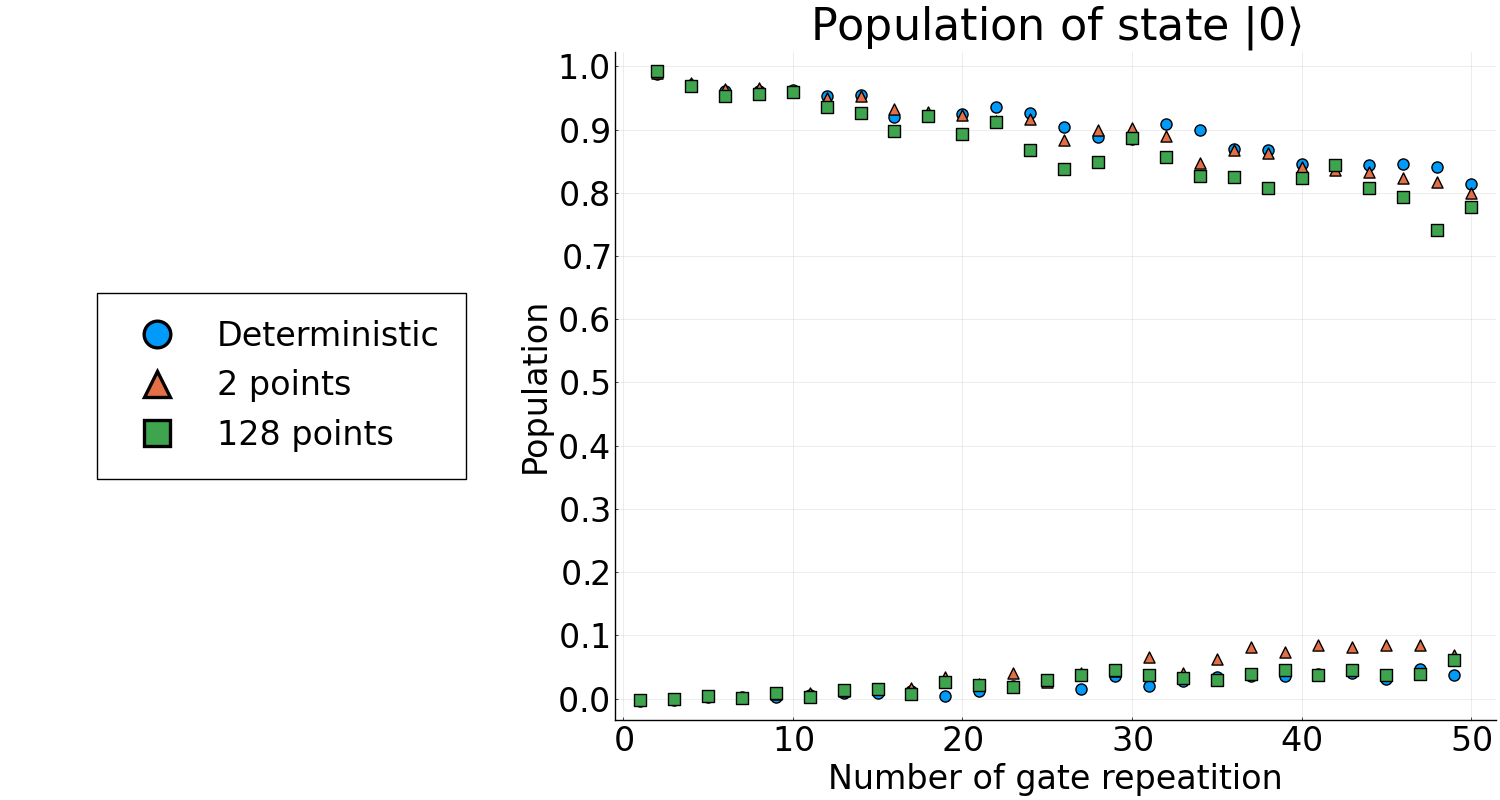}
  \includegraphics[height=0.16\textheight]{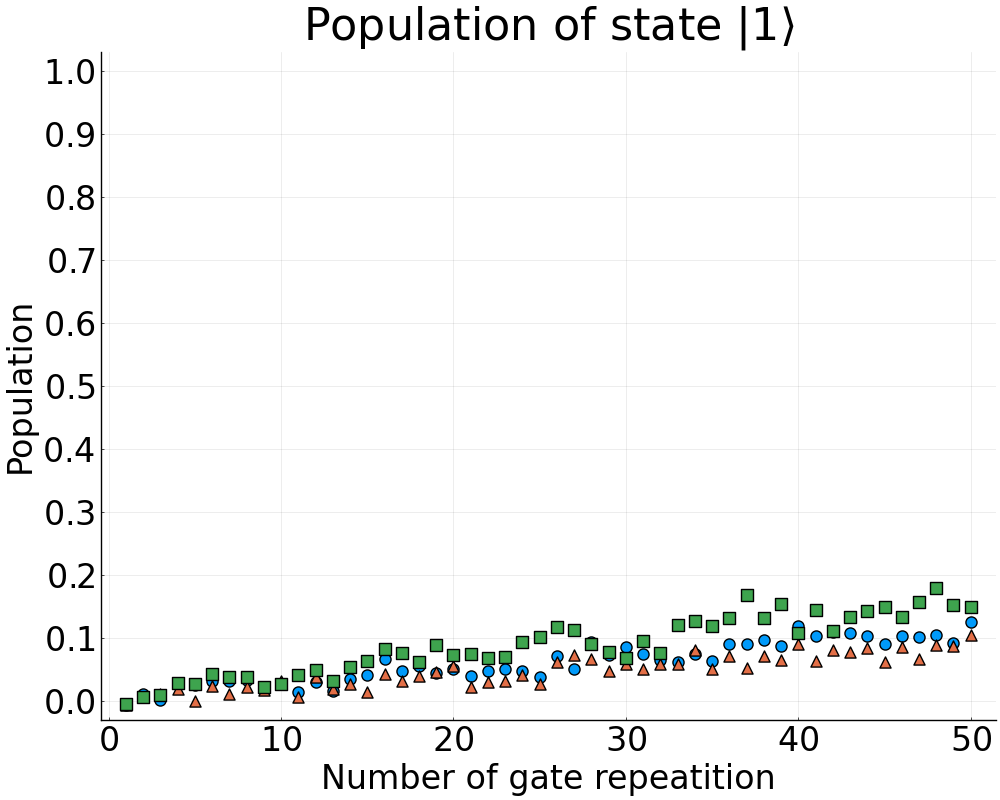}
  \includegraphics[height=0.16\textheight]{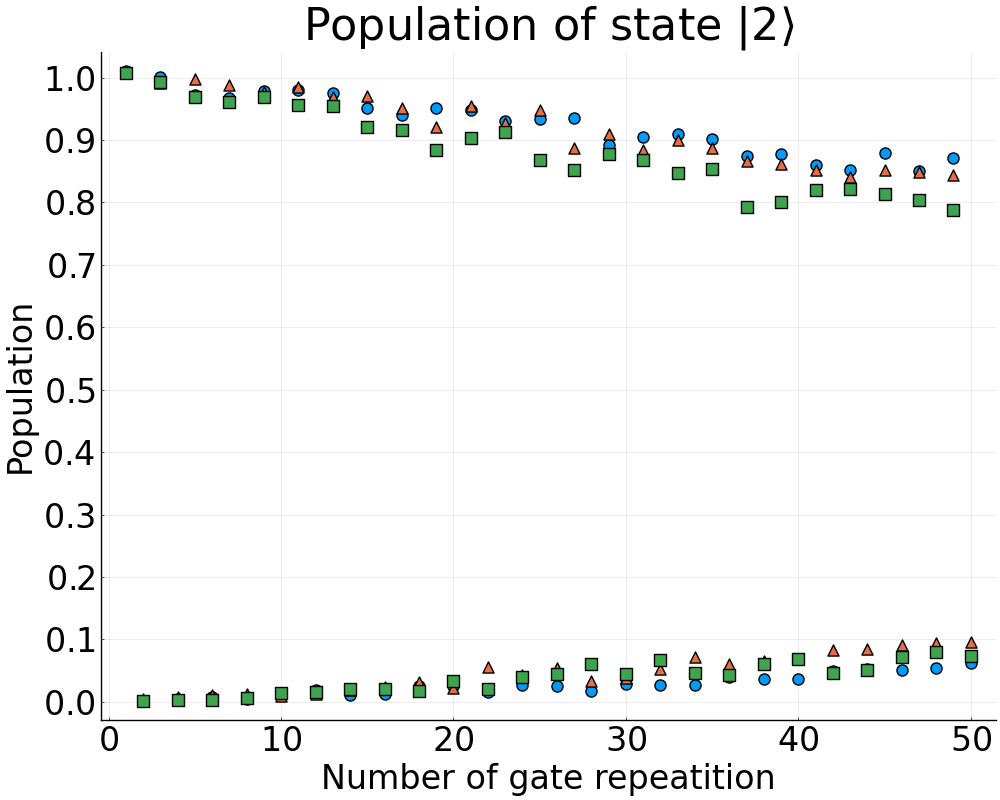}
  \includegraphics[height=0.16\textheight]{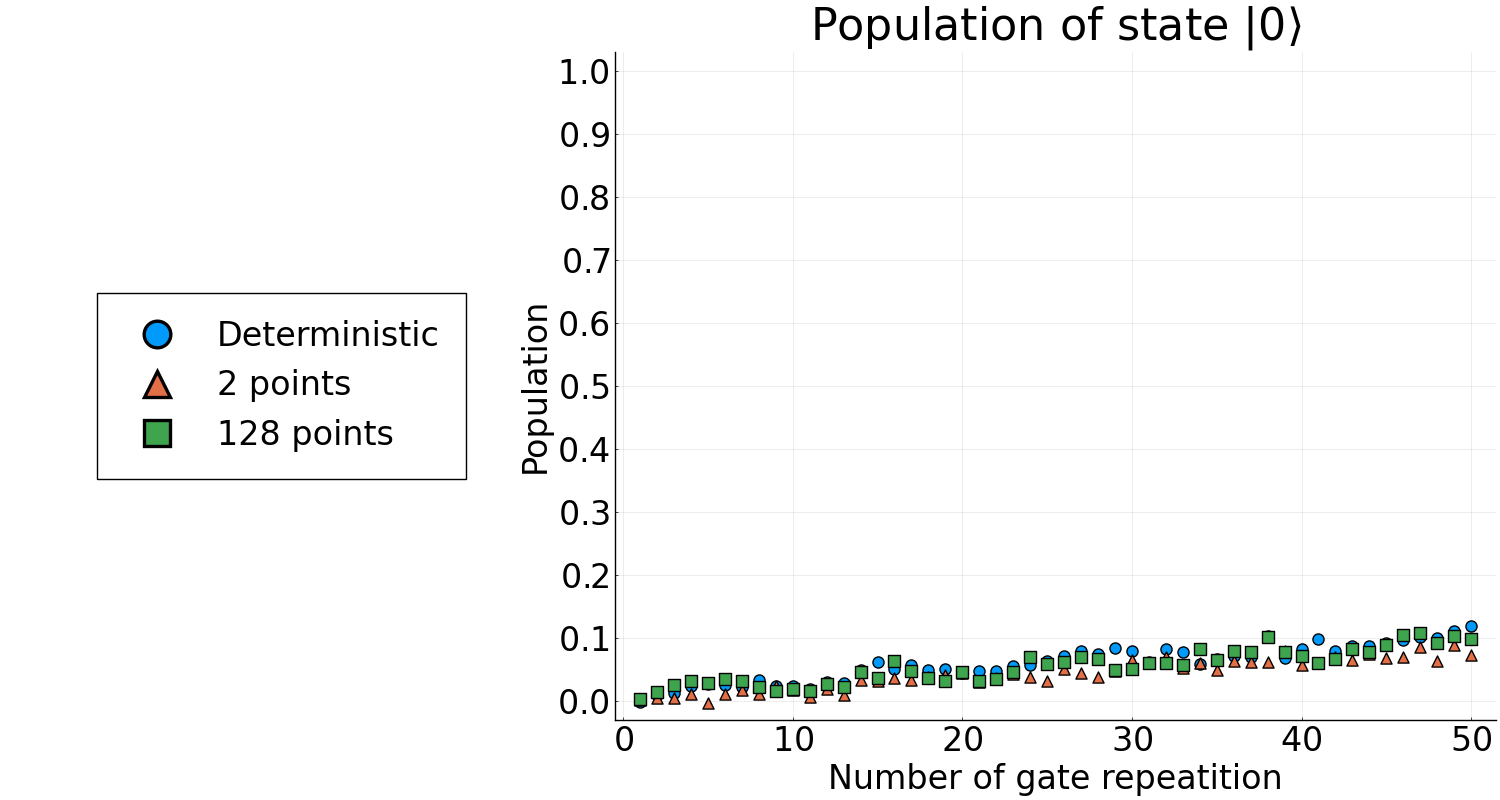}
  \includegraphics[height=0.16\textheight]{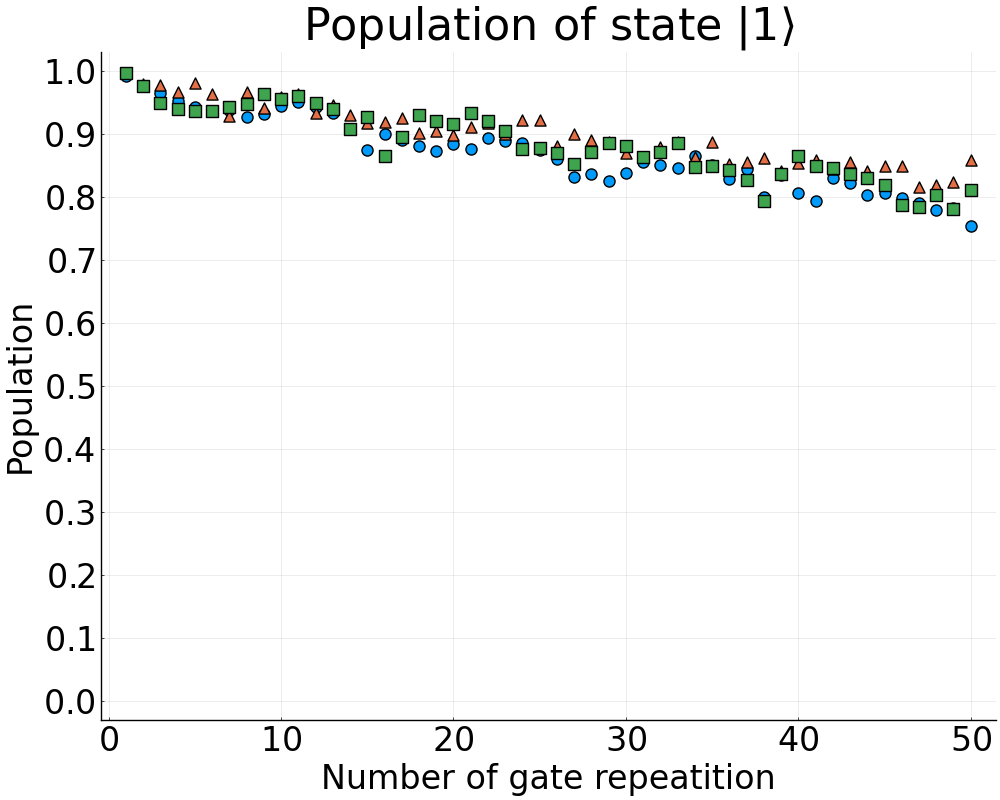}
  \includegraphics[height=0.16\textheight]{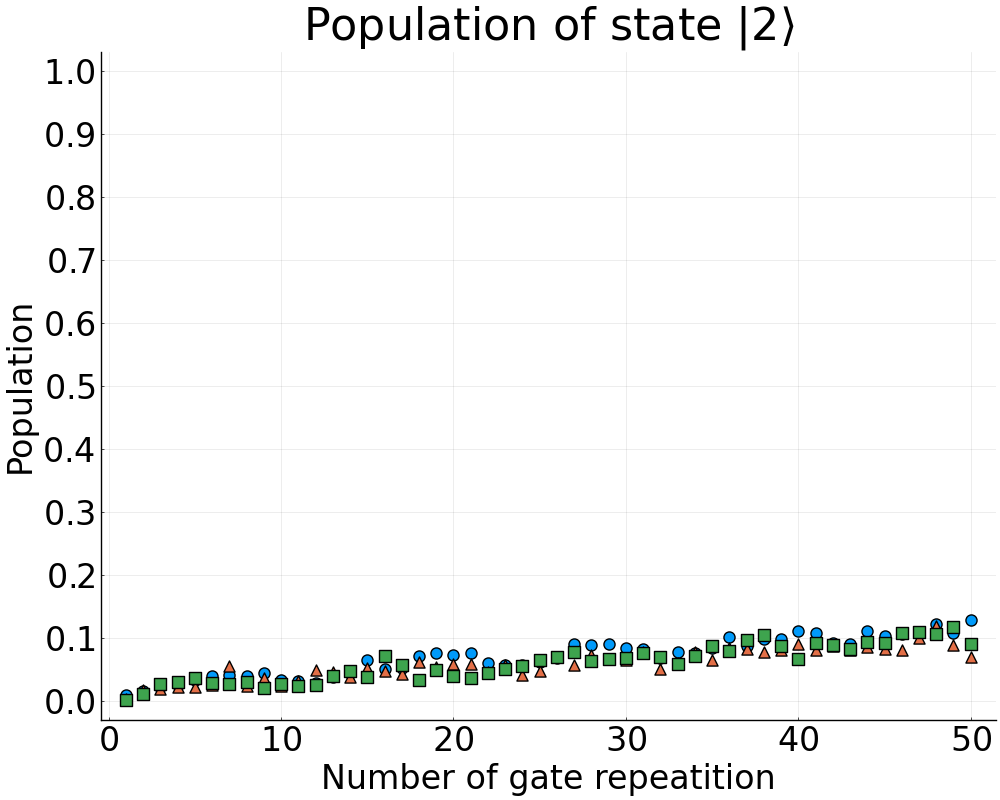}
  \includegraphics[height=0.16\textheight]{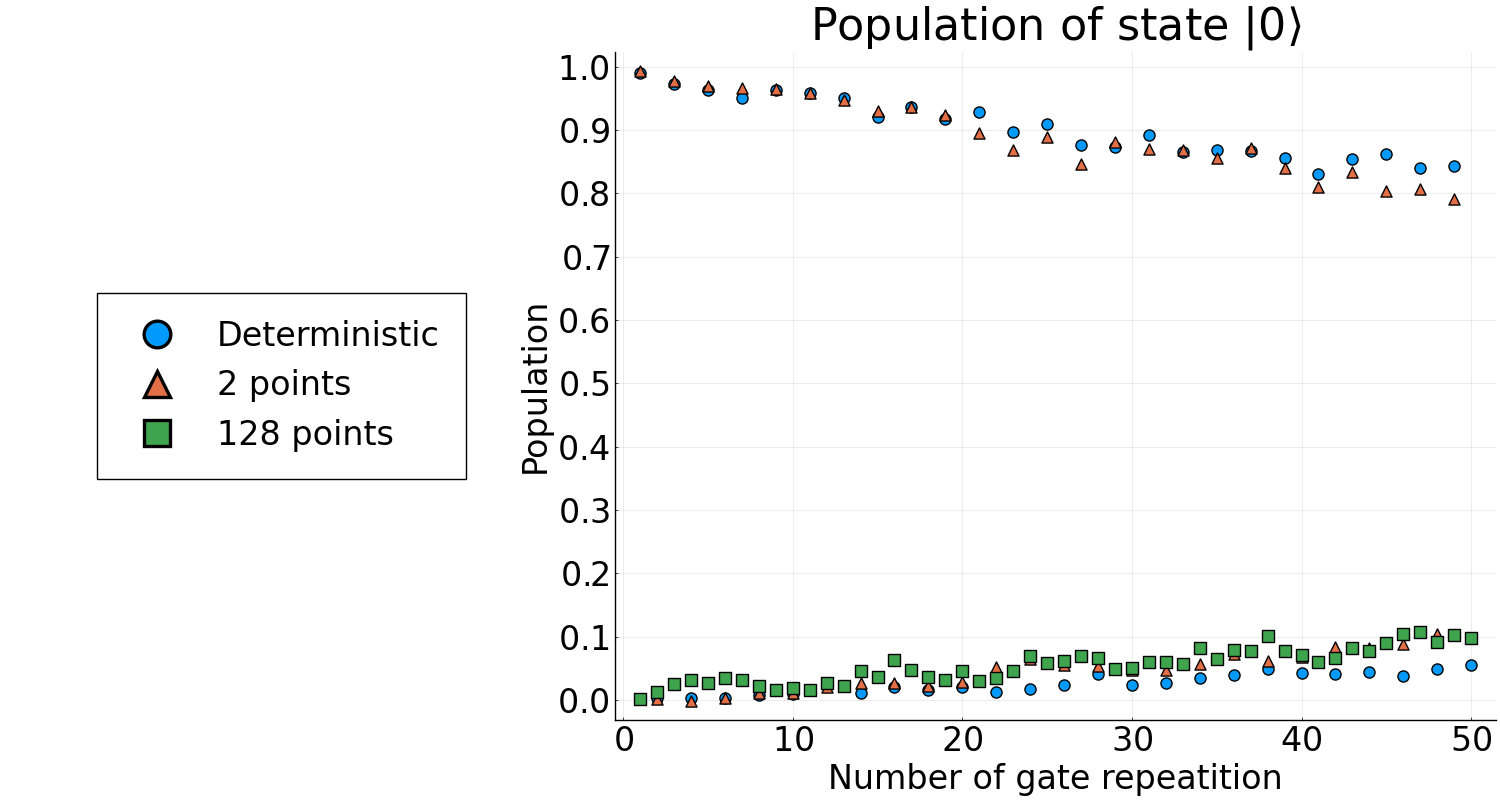}
  \includegraphics[height=0.16\textheight]{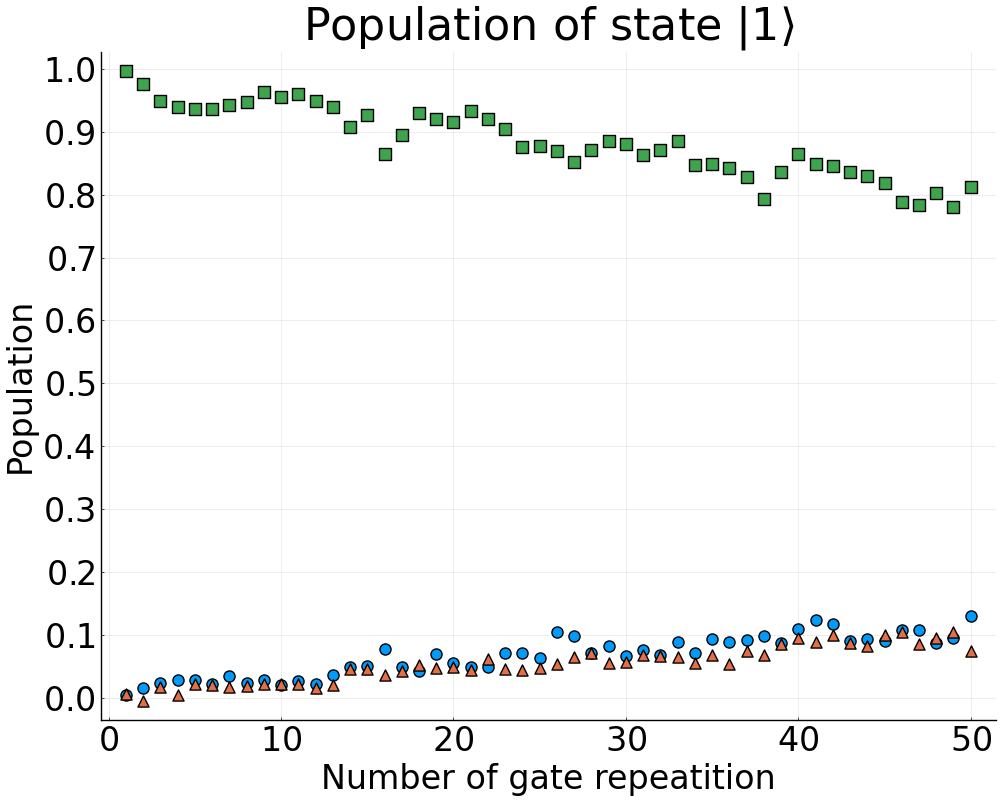}
  \includegraphics[height=0.16\textheight]{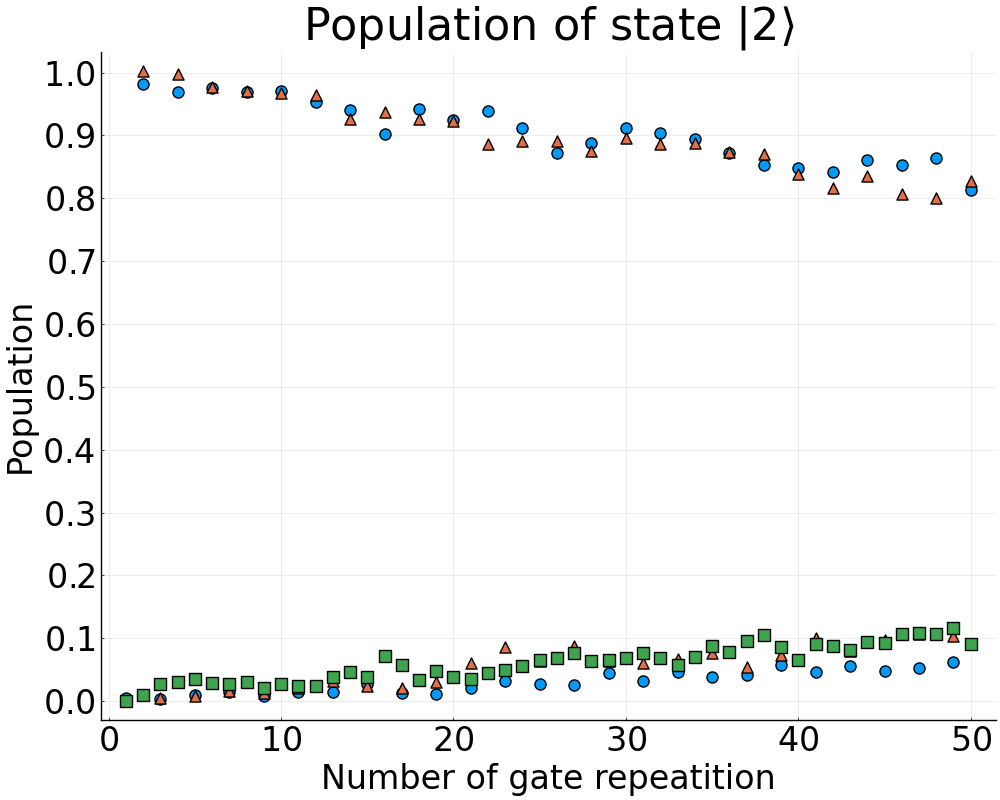}
  \caption{Probability populations for different states of repeatedly applying the $0\leftrightarrow2$ SWAP gate. Left column: initial state is $|0\rgl$. Middle column: initial state is $|1\rgl$. Right column: initial state is $|2\rgl$. \label{fig:repeated-gate}}
  \end{center}
\end{figure}

The second test is to repeat the $0\leftrightarrow2$ SWAP gate $50$ times with the initial state as state $|0\rgl$, $|1\rgl$ and $|2\rgl$. The population is measured after each gate application. The population of different states are presented in Figure \ref{fig:repeated-gate}. For the initial state $|0\rgl$ and $|2\rgl$, we observe the expected exchange of populations between state $|0\rgl$ and state $|2\rgl$. Moderate leakage to state $|1\rgl$ is also observed as the number of gate repetitions increases. For the initial state $|1\rgl$, no obvious exchange is observed, and the leakage to state $|0\rgl$ and $|2\rgl$ is observed as the number of gate repetitions grows. We suspect that the leakage in the gate repetition experiments is due to the decoherence effect in the qudit device, since the amplitude of the envelope for the population of the state $|0\rgl$ and $|2\rgl$ decreases.

\begin{table}[h]
\begin{center}
\begin{tabular}{|c|c|c|c|c|c|c|c||c|c|}
\hline
& Deterministic & $2$ points & $128$ points \\ \hline
Gate fidelity &  99.40\%       & 99.49\%    & 99.37\%     \\ \hline
Entanglement fidelity &  99.06\%       & 99.14\%    & 98.74\%  \\ \hline
\end{tabular}
\caption{Average gate and entanglement fidelities for different pulses. \label{tab:fidelity}}
\end{center}
\end{table}

Based on gate repetition results, we then compute the process matrix $\chi$ and the corresponding process operator $\mathcal{E}_\chi$ of our $0\leftrightarrow2$ SWAP gate:
\begin{subequations}
\label{eq:process}
\begin{align}
&    \mathcal{E}_{\chi}(\rho)=\sum_{m,n=0}^{3^2-1}\chi_{mn}B_m\rho B_n^\dagger,\\
&    B=\{I,Z_{01},Z_{12},X_{01},X_{12},Y_{01},Y_{12},X_{01}X_{12},X_{12}X_{01}\}.
\end{align}
\end{subequations}
Here,  $\chi$ is the process matrix which completely determines the process $\mathcal{E}_\chi$, and $B$ is a complete gate set whose elements form a basis for $3$ by $3$ unitary matrices. We compute $\chi$ by a constrained least squares fit (described in Appendix \ref{sec:app}). Following \cite{chuang1997prescription}, we compute the entanglement and gate fidelity by
\begin{subequations}
\begin{align}
    &F_e(\rho,U,\mathcal{E}_\chi)=\sum_{m,n}\chi_{mn}\textrm{Tr}(U^+B_m\rho)\textrm{Tr}(\rho B_n^+U), \label{eq:gate_fidelity}\\
    &F_g(|\psi\rgl,U,\mathcal{E}_\chi)=\lgl\psi| U^\dagger\mathcal{E}_\chi(|\psi\rgl\lgl\psi|)U|\psi\rgl.
    \label{eq:entanglement_fidelity}
\end{align}
\end{subequations}
Here, $\rho$ is a density matrix, and $|\psi\rgl$ is a state vector. 

Following \cite{wu2020high}, we compute the average gate fidelity of $10000$ randomly sampled state vectors $|\psi\rgl$ according to \eqref{eq:gate_fidelity} and the average entanglement fidelity of $10000$ randomly sampled density matrices $\rho$ (constructed by pure state ensemble) according to \eqref{eq:entanglement_fidelity}. The average gate and entanglement fidelities for different pulses are summarized in Table \ref{tab:fidelity}. 
The coherent control errors, defined as the gate fidelity minus the entanglement fidelity, are around $0.34\%$ for the deterministic pulse, $0.35\%$ for the 2 points pulse and $0.63\%$ for the 128 points pulse. The 2 points control pulse leads to slightly better fidelities, while the overall performance of the three control pulses are comparable. We suspect such comparable performance is due to the fact that the charge dispersion $\epsilon_{1,2}$ and other stochastic uncertainties in the transition frequencies are not large enough to have significant influence over the short time duration of our control pulses.

\subsection{Tuning of control pulses must be done for each pulse} 

We also perform two tests to show that the tuning of control pulses needs to be done in a  pulse-by-pulse manner, and that tuning needs to be done frequently.  

The spectra of the controls, displayed in Figure  \ref{fig:control-pulses}, are similar in shape and amplitude and it is reasonable to see if it will suffice to tune \eqref{eq:calibration_rescale} for one of the three pulses and then use that tuning for the other two. However when we use the deterministic pulse to find $r_c$ and $A_{r_c}$ in \eqref{eq:calibration_rescale}, we find that these values cannot be used for tuning the 2 and 128 point pulses. 

\begin{figure}[]
  \begin{center} 
  \includegraphics[width=0.45\textwidth]{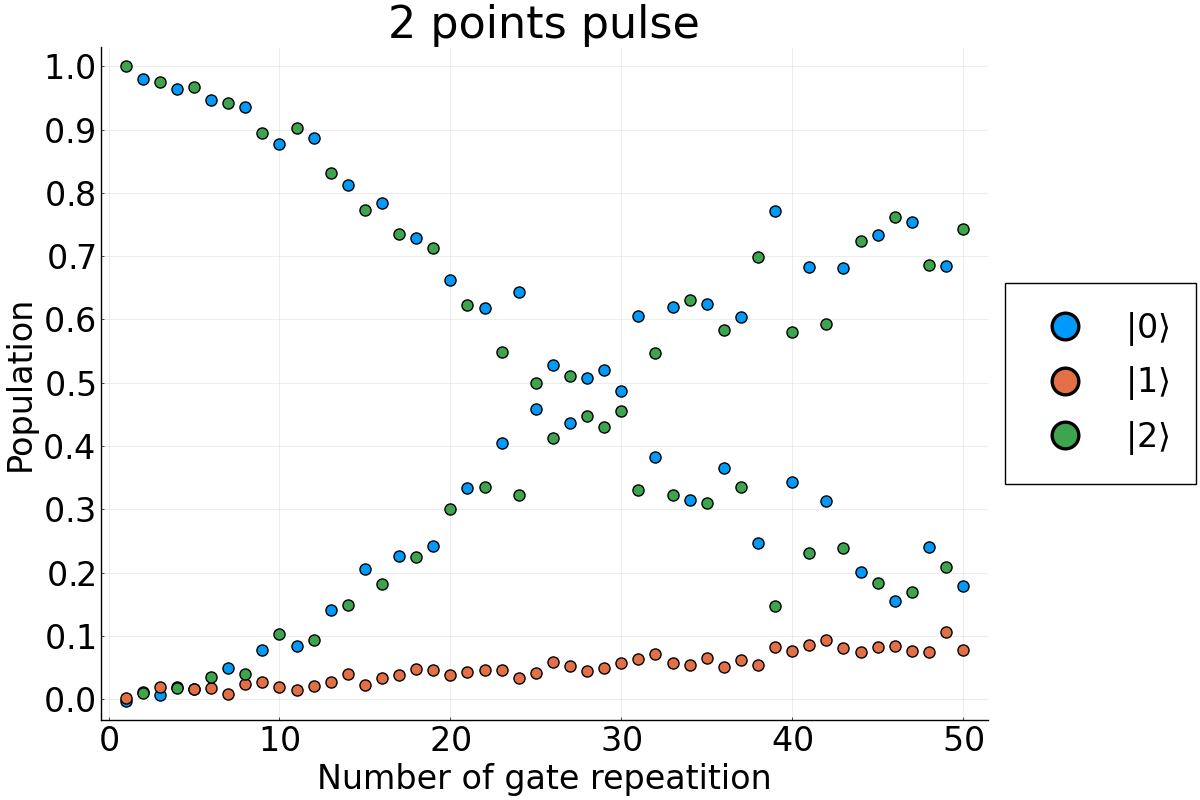}
  \includegraphics[width=0.45\textwidth]{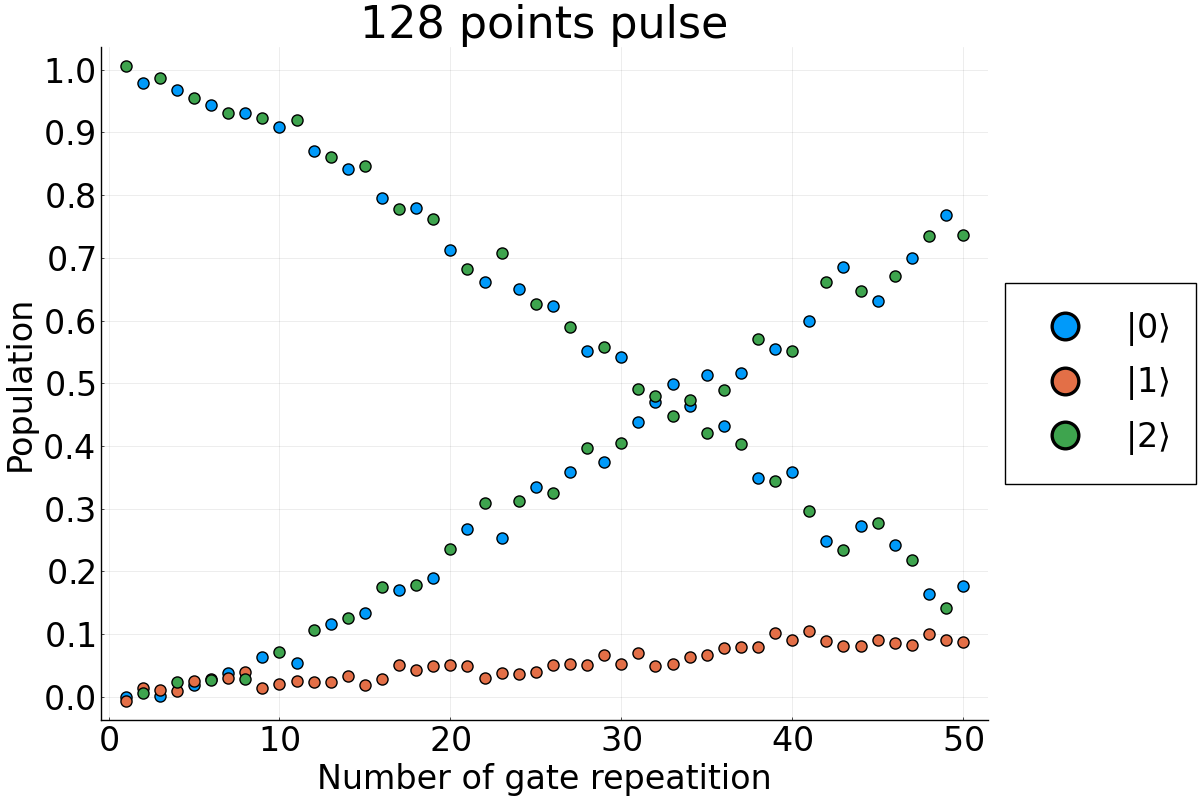}
  \caption{Use $r_c$ and $A_c$ from the calibration of the deterministic pulse to calibrate the $2$ points and $128$ points pulses. The population of repeatedly applying the $0\leftrightarrow2$ SWAP gate with initial state $|0\rgl$. \label{fig:pulse-filter-mismatch}}
  \end{center}
\end{figure}

Specifically, in  Figure \ref{fig:pulse-filter-mismatch} we display the populations (upon 50 gate repetitions) for these two pulses when using the deterministic values for $r_c$ and $A_{r_c}$. As can be seen the fidelity deteriorates rapidly, indicating that a pulse-by-pulse tuning is necessary. However, further investigations are needed to identify the source of this problem.



\subsection{Tuning needs to be performed frequently}
\begin{figure}[]
  \begin{center} 
  \includegraphics[width=0.45\textwidth]{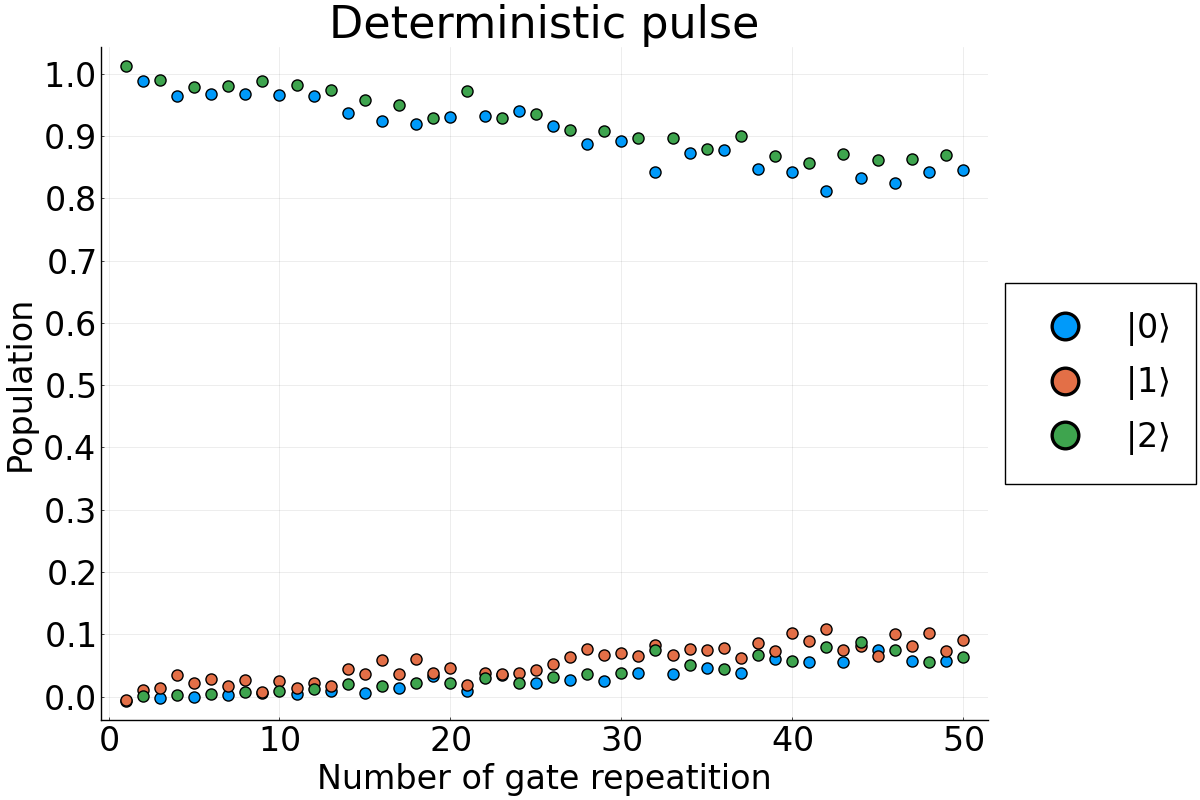}
  \includegraphics[width=0.45\textwidth]{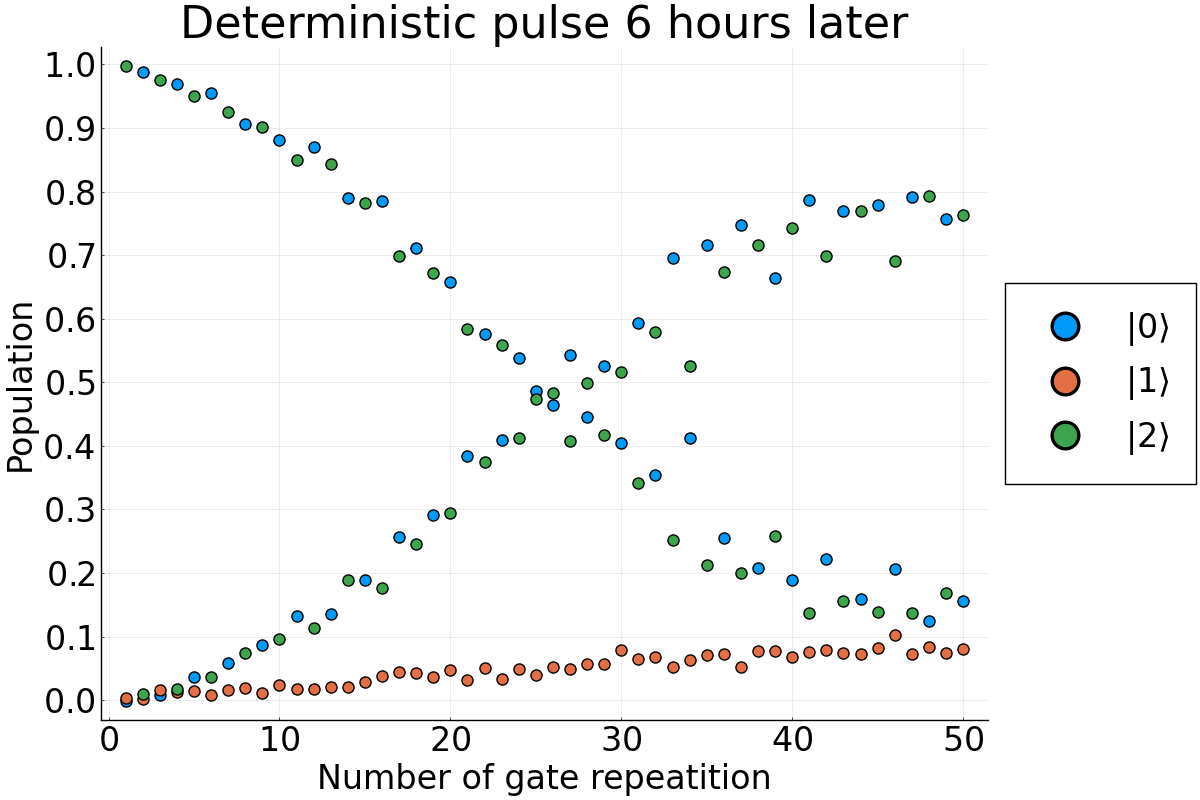}
  \caption{Comparison of the performance of the deterministic pulse with the $r_c$ and $A_{r_c}$ in Table \ref{tab:calibration}. The population of repeatedly applying the $0\leftrightarrow2$ SWAP gate with initial state $|0\rgl$. Left: right after the tuning. Right: 6 hours after the tuning. \label{fig:time-filter-mismatch}}
  \end{center}
\end{figure}

Our second test is to show that a tuning needs to be carried out frequently. For the deterministic pulse (the results are similar for the other pulses) we used the values for  $r_c$ and $A_{r_c}$ from Table \ref{tab:calibration} at the time they were obtained, and then we use them again after $6$ hours. In both cases we repeatedly apply the gate $50$ times with the initial state $|0\rgl$. In Figure \ref{fig:time-filter-mismatch}, which displays the results for the two experiments, it can be observed that the population exchange between state $|0\rgl$ and state $|2\rgl$ right is quite accurate for the experiment that takes place just after then tuning. For the experiment that takes place $6$ hours after the tuning the performance is much worse. 

At the time of writing it is unclear what is the cause of the drift in $r_c$ and $A_{r_c}$ but we note that the drift is relatively small. 

It would be interesting to carry out a more comprehensive study of how the tuning parameters change over time while collecting other external laboratory data such as temperature, time of day etc. It is possible, or even likely, that there is a correlation between one or more external variables. If such a correlation can be detected, say with the room temperature $T_{\rm R}$, then the natural question becomes if we can find a temperature corrected model $(\hat{r}_c(T_{\rm R}),\hat{A}_{r_c}(T_{\rm R}))$ that can be used in place of re-tuning. Readout stability may also play a significant role and needs to be investigated further.


\section{Conclusion}
This report presents a case study in which the processes of calibrating a quantum device, collecting experimental data for characterization, applying deterministic and Bayesian methods for estimating device properties, finding optimal control pulses for unitary gates, tuning the control pulse, and finally, experimental validation of the optimized control pulses.

We have documented a few interesting findings: Parity events are observed in the device we use, leading to beatings in the Ramsey fringes for higher transitions, rendering these to be difficult to use directly in the characterization process. Fortunately, the beating can be captured by a simple mathematical modeling of the parity event, simply taking the average of forward Lindblad solves using different parity. 

Further, we studied the fidelity of a $0\leftrightarrow2$ SWAP gate using different optimal control techniques. We found that a two points risk neutral control pulse, which takes the parity event into account, leads to slightly better experimental gate and entanglement fidelities. In addition we observe a frequency dependency in the transmission of control pulses from the warm to the cold side of the quantum device, which require additional tuning. This tuning is found to be unique for each pulse and it is also found to be changing in time. 

There are many open questions for future studies. For example, can we model the distortion of the control pulse from the warm to the cold side and take these effects into account during the optimal control? If decoherence effects are accounted for during optimal control, will the gate fidelity be improved in practice? Will more advanced Bayesian or deterministic characterization techniques improve the efficiency and accuracy of the quantum device?

\appendix
\begin{appendices}

\section{Rabi oscillation and $\pi$ pulse \label{sec:rabi}}
Consider a two-level Rabi oscillator in the rotating frame:
\begin{align}
    \dot{\psib} = -i(\Omega a+\Omega^\dagger a^\dagger)\psib,\;\Omega=|\Omega|e^{i\theta}\in\mathbb{C}.
\end{align}
The solution operator is 
\begin{align}
    U(t) = \left(
    \begin{array}{cccc}
    \cos(|\Omega|t) & (\sin(\theta)-i\cos(\theta))\sin(|\Omega|t)\\
   -(\sin(\theta)+i\cos(\theta))\sin(|\Omega|t) & \cos(|\Omega|t)
    \end{array}
    \right).
\end{align}

Particularly, with $t_\pi|\Omega|=\frac{\pi}{2}$, we have
\begin{align}
    U(t_\pi) = \left(
    \begin{array}{cccc}
    0 & (\sin(\theta)-i\cos(\theta))\\
   -(\sin(\theta)+i\cos(\theta))\sin(|\Omega|t) & 0
    \end{array}
    \right),
\end{align}
and $U(t_\pi)$ swaps state $|0\rgl$ and state $|1\rgl$, which  realizes a $\pi_{01}$ pulse. 
\section{Curve fitting techniques\label{sec:app-curve-fitting}}
We follow \cite{qiskit_url_calibration} to estimate device parameters and calibrate $\pi$-pulses by curve fitting. 

In the frequency sweep, ideally, the peak or the valley of the measured $I$  ($Q$ or the population) data is at the location for the transition frequency. We fit the frequency sweep data with a Lorentz shape
\begin{align}
   \frac{A}{\pi}\cdot\frac{B}{(\omega_d-\omega_{k,k+1}^{\textrm{est}})^2+B^2}+ C,
\end{align}
where $\omega_d$ is the drive frequency, $A$, $B$ and $C$ are hyper parameters, and $\omega_{k,k+1}^{\textrm{est}}$ estimates the transition frequency.

In the amplitude sweep, ideally, the measured $I$ ($Q$ or the population) data forms a sinusoidal curve with respect to the pulse amplitude. We fit the amplitude sweep data with 
\begin{align}
    B\cos\left(2\pi \frac{A^{\textrm{sweep}}}{A^{\textrm{est}}} - \phi\right) + C,
\end{align}
where $A^{\textrm{sweep}}$ is the pulse amplitude applied in the amplitude sweep, $B$, $C$ are hyper parameters, and $A^{\textrm{est}}$ estimates the amplitude of the ideal $\pi$ pulse.

In the Rasmey experiment, ideally, the curve of measured $I$ ($Q$ or the population) data is a decaying sinusoidal curve oscillating at the detuning frequency $\Delta\omega=\omega_{k,k+1}-\omega_{d}$. We fit the measured data with
\begin{align}
    A\exp(-t_{\textrm{delay}}/T_2^{*,\textrm{est}})\cos\big(2\pi \Delta\omega^{\textrm{est}}t_{\textrm{delay}} - C) + B,
\end{align}
where $t_\textrm{delay}$ is the delay time in the Ramsey experiment, $A$, $B$, $C$ are hyper parameters, $T_2^{*,\textrm{est}}$ estimates the dephasing time and $\Delta\omega^{\textrm{est}}$ estimates the actual detuning. The transition frequency can be estimated as $\omega_{k,k+1}^{\textrm{est}}=\Delta\omega^{\textrm{est}}+\omega_d$.

\section{Computation of the process matrix \label{sec:app}}
We follow the supplementary material of \cite{wu2020high} to compute the process matrix $\chi$. The process matrix $\chi$ satisfies the completion condition
\begin{align}
    \sum_{m,n=0}^{3^2-1}\chi_{mn}B_m^\dagger B_n=I,
    \label{eq:complete_relation}
\end{align}
and it is a Hermitian matrix.

We utilize the Cholesky decomposition $\chi(t)=L^\dagger(\tb)L(\tb)$ to parameter $\chi$. Here, $L(\tb)$ is a lower triangular matrix:
\begin{align}
    L(\tb) = \left(\begin{array}{cccccc}
            t_0 & 0 & 0 & \dots & 0\\
            t_1 & t_2 & 0 & \dots & 0\\
            t_3 & t_4 & t_5 & \dots & 0\\
            \vdots & \vdots & \vdots & \ddots &\vdots\\
            t_{36} & t_{37} & t_{38} & \dots & t_{44}
          \end{array}\right)
          +i\left(\begin{array}{ccccccc}
            0 & 0 & 0 & \dots & 0 \\
            t_{45} & 0 & 0 & \dots & 0 \\
            t_{46} & t_{47} & 0 & \dots &0 \\
            \vdots & \vdots & \ddots & \ddots & \vdots\\
            t_{72} & t_{73} & \dots  & t_{80} & 0
          \end{array}\right),\quad \tb=(t_0,\dots,t_{80})^T.
\end{align}
To infer the process matrix $\chi$, we solve the constrained minimization problem:
\begin{equation}
 \tb=\arg\min_{\tb}\sum_{k=0}^2\sum_{j=0}^2\sum_{n=1}^{N_{\textrm{rep}}} \left(P^{n,j,k}(\tb)-P^{n,j,k}_{\textrm{exp}}\right)^2,\quad\textrm{subject to }  \sum_{m,n=0}^{3^2-1}\chi_{mn}(\tb)B_m^\dagger B_n=I,\label{eq:chi_minimization}
\end{equation}
where $N_{\textrm{rep}}$ is the maximum number of gate repetitions, $P^{n,j,k}_{\textrm{exp}}$ is the population for state $j$ after $n$ gate repetitions with  the initial state $|k\rgl$, and 
$P^{n,j,k}(\tb)$ is the $j$-th diagonal element of the density matrix 
$$\rho^{(n,k)}=\mathcal{E}_{L^\dagger(\tb)L(\tb)}(\rho^{(n-1,k)}), \quad\text{with}\; \rho^{(0,k)}=|k\rgl\lgl k|.$$ The solution of \eqref{eq:chi_minimization} minimizes the mismatch between the populations of different states determined by the process matrix  $L^\dagger(\tb)L(\tb)$ and the measured experimental data. 

\section{Relation between $\gamma_{2,k}$ in \eqref{eq:lindblad_terms} and the pure dephasing time $T_{2,k}$ \label{sec:app:T2-gamma}}
To derive the relation between $\gamma_{2,k}$ and the pure dephasing time $T_{2,k}$, we consider a pure dephaing system:
\begin{equation}
\dot{\rho} = \mathcal{L}_2\rho \mathcal{L}_2^\dagger-\frac{1}{2}\left(\mathcal{L}_2^\dagger\mathcal{L}_2\rho+\rho\mathcal{L}_2^\dagger\mathcal{L}_2\right),
\label{eq:dephasing_equation}
\end{equation}
where $\mathcal{L}_2$ is defined in \eqref{eq:lindblad_terms} and $\rho$ is the density matrix.

With direct calculations,  \eqref{eq:dephasing_equation} can be written as:
\begin{equation}
\left(
\begin{array}{cccc}
\dot{\rho}_{00}  & \dot{\rho}_{01}  & \dot{\rho}_{02}  & \dot{\rho}_{03} \\
\dot{\rho}_{10}  & \dot{\rho}_{11}  & \dot{\rho}_{12}  & \dot{\rho}_{13} \\
\dot{\rho}_{20}  & \dot{\rho}_{21}  & \dot{\rho}_{22}  & \dot{\rho}_{23}  \\
\dot{\rho}_{30}  & \dot{\rho}_{31}  & \dot{\rho}_{32}  & \dot{\rho}_{33} 
\end{array}\right)
=\left(
\begin{array}{cccc}
0 & b_{01}\rho_{01} & b_{02}\rho_{02} & b_{03}\rho_{03}\\
b_{10}\rho_{10} & 0 &  b_{12}\rho_{12} & b_{13}\rho_{13}\\
b_{20}\rho_{20} & b_{21}\rho_{21} & 0 &b_{23}\rho_{23}\\
b_{30}\rho_{30} & b_{31}\rho_{31} & b_{32}\rho_{32} & 0\\
\end{array}\right), 
\label{eq:dephasing_equation_simplified}
\end{equation}
where 
\begin{align}
b_{ij} = b_{ji} = \sqrt{\gamma_{2,i}\gamma_{2,j}}-\frac{1}{2}(\gamma_{2,i}+\gamma_{2,j}),\; i\neq j\quad \text{and}\quad \gamma_{2,0} = 0.
\end{align}
Therefore $\rho_{ij}(t) = \exp(b_{ij}t)\rho_{ij}(0)$. 

With $k\geq1$ the pure dephasing time $T_{2,k}$ satisfies $\rho_{k,k-1}(t)=\exp(-t/T_{2,k})\rho_{k,k-1}(0)$. Hence, $-\frac{1}{T_{2,k}}=b_{k,k-1}$, $k=1,2,3$.
When $k=1$ we have
\begin{equation}
-\frac{1}{T_{2,1}}=b_{10}=-\frac{1}{2}\gamma_{2,1}\Rightarrow \gamma_{2,1}=\frac{2}{T_{2,1}}.
\end{equation}
When $k=2$ we have
\begin{equation}
-\frac{1}{T_{2,2}}=b_{21}=\sqrt{\gamma_{2,1}}\sqrt{\gamma_{2,2}}-\frac{1}{2}\gamma_{2,1}-\frac{1}{2}\gamma_{2,2}.
\end{equation}
Hence $\sqrt{\gamma_{2,2}}$ solves the quadratic equation
\begin{equation}
\gamma_{2,2}-2\sqrt{\gamma_{2,1}}\sqrt{\gamma_{2,2}}+\gamma_{2,1}-\frac{2}{T_{2,2}}=0.\label{eq:quadratic_2}
\end{equation}
The discriminant of this equation is 
\begin{equation}
\Delta = 4\gamma_{2,1}-4(\gamma_{2,1}-\frac{2}{T_{2,2}}) = \frac{8}{T_{2,2}}.
\end{equation}
The solution of the quadratic equation \eqref{eq:quadratic_2} is
\begin{equation}
\sqrt{\gamma_{2,1}}\pm \sqrt{2/T_{2,2}}=\sqrt{2/T_{2,1}}\pm\sqrt{2/T_{2,2}}.
\end{equation}
Physically we know that $T_{2,1}>T_{2,2}$, so 
\begin{equation}
\sqrt{\gamma_{2,2}} = \sqrt{\gamma_{2,1}}+\sqrt{2/T_{2,2}}.
\end{equation}
Similar to the case of $k=2$, we can show that $\sqrt{\gamma_{2,3}}$ satisfy 
\begin{equation}
\sqrt{\gamma_{2,3}} = \sqrt{\gamma_{2,2}}+\sqrt{2/T_{2,3}}.
\end{equation}
by solving the quadratic equation
\begin{equation}
\gamma_{2,3}-2\sqrt{\gamma_{2,2}}\sqrt{\gamma_{2,3}}+\gamma_{2,2}-\frac{2}{T_{2,3}}=0.
\end{equation}
In summary, 
\begin{equation}
\sqrt{\gamma_{2,k}}=\sqrt{\gamma_{2,k-1}}+\sqrt{2/T_{2,k}},\quad \text{with}\quad k=1,2,3,\quad\text{and}\quad \gamma_{2,0}=0.
\end{equation}

\clearpage
\section{Protocols}
\begin{framed}
\begin{protocol}[$\pi$-calibration]\label{prot:pi}
\hfill
\begin{enumerate}
\item Sweep over $k=0,1$, calibrate $\pi_{k,k+1}$ pulse and obtain estimations of $\omega_{k,k+1}$. 
\begin{enumerate}
    \item Frequency sweep to find an estimated value of $\omega_{k,k+1}$ denoted by $\omega_{k,k+1}^{\rm est}$.
    \begin{enumerate}
    \item Sweep over a range of drive frequencies $\omega_d\in[\omega_{L},\omega_U]$. 
    \item If needed, apply calibrated $\pi$ pulses to prepare the device to state $|k\rgl$. 
    \item Apply the $\pi_{k,k+1}$ pulse with the drive frequency $\omega_d$ and collect the corresponding experimental  data. 
    \item Find the rough transition frequency $\omega_{k,k+1}^\textrm{est}$ with curve fitting. For example, in the bottom left picture of Figure \ref{fig:calibration_device_flow}, the location of the minima is taken as $\omega_{1,2}^\textrm{est}$, the estimate for the $1$-$2$ transition frequency. This estimate is then used as the drive frequency for the amplitude. 
    \end{enumerate}
    \item Amplitude sweep to calibrate the amplitude of $\pi_{k,k+1}$ pulse. 
    \begin{enumerate}
        \item Prepare the device to state $|k\rgl$ with the calibrated $\pi$-pulses.  
        \item Sweep over a range of the amplitudes for the $\pi_{k,k+1}$ pulse.
        \item Apply the $\pi_{k,k+1}$ pulse repeatedly for $5$ times with the drive frequency $\omega_{k,k+1}^\textrm{est.}$ and collect data.  
        \item Update the amplitude of the $\pi_{k,k+1}$ based on curve fitting results (see the bottom middle picture of Figure \ref{fig:calibration_device_flow}).
    \end{enumerate}
    \item Perform a short Ramsey $0$-$1$ experiment (Protocol \ref{prot:Ramsey}) to improve the estimation of the transition frequency $\omega_{k,k+1}^\textrm{est}$.
    \item Repeat the amplitude sweep and the short Ramsey experiment to update the amplitude and improve the accuracy of  $\omega_{k,k+1}^\textrm{est}$ until the change of $\omega_{k,k+1}^\textrm{est}$ is small enough.
\end{enumerate}
\end{enumerate}
\end{protocol}
\end{framed}

\begin{framed}
\begin{protocol}[$T_1$-decay]\label{prot:T1}
\hfill
\begin{enumerate}
\item Sweep over different delay times $t_j=j\Delta t$, $j=0,1,\dots,N_{\rm T}$.
    \begin{enumerate}
    \item Repeat the experiment with the delay time $t_j$ for $1000$ shots.
        \begin{enumerate}
        \item Apply a sequence of $\pi$-pluses to prepare the system to the state $|k\rgl$.
        \item Let the system evolve freely for the delay time $t_j=j\Delta t$, and then measure  populations.
        \end{enumerate}
    \end{enumerate}
\end{enumerate}
\end{protocol}
\end{framed}

\begin{framed}
\begin{protocol}[Ramsey experiment]\label{prot:Ramsey}
\hfill
\begin{enumerate}
\item Sweep over different delay times $t_j=j\Delta t$, $j=0,1,\dots,N_{\rm T}$.
    \begin{enumerate}
    \item Repeat the experiment with the delay time $t_j$ for $1000$ shots.
        \begin{enumerate}
        \item If necessary, apply a sequence of $\pi$-pluses to prepare the system to the state $|k\rgl$.
        \item Apply the $\pi_{k,k+1}/2$ pulse at a detuned frequency $\omega_d$ with the nominal detuning $\Delta \omega$ to drive the system halfway to the state $|k+1\rgl$.
        \item Let the system evolve freely for the delay time $t_j=j\Delta t$.
        \item Apply the $\pi_{k,k+1}/2$ pulse at a detuned frequency $\omega_d$ with the  detuning $\Delta \omega$ again, and then measure populations.
        \end{enumerate}
    \end{enumerate}
\end{enumerate}
\end{protocol}
\end{framed}

\begin{framed}
\begin{protocol}[Hahn-Echo experiment]\label{prot:Hahn-Echo}
\hfill
\begin{enumerate}
\item Sweep over different delay times $t_j=j\Delta t$, $j=0,1,\dots,N_{\rm T}$.
    \begin{enumerate}
    \item Repeat the experiment with the delay time $t_j$ for $1000$ shots.
        \begin{enumerate}
        \item If necessary, apply a sequence of $\pi$-pluses to prepare the system to the state $|k\rgl$.
        \item Apply the $\pi_{k,k+1}/2$ pulse to drive the system halfway to the state $|k+1\rgl$.
        \item Let the system evolve freely for the delay time $t_j=j\Delta t$.
        \item Apply the $\pi_{k,k+1}$ pulse.
        \item Let the system evolve freely for the delay time $t_j=j\Delta t$.
        \item Apply the $\pi_{k,k+1}/2$ pulse, and then measure populations.
        \end{enumerate}
    \end{enumerate}
\end{enumerate}
\end{protocol}
\end{framed}

\begin{framed}
\begin{protocol}\label{prot:freq_filter}
\hfill
\begin{enumerate}
\item Sweep over different values of $r$, and multiply the B-spline coefficients of the carrier wave corresponding to the $1$-$2$ transition with $r$.
\begin{enumerate}
    \item Convert the B-spline coefficients of carrier waves to the inputs of the IQ-mixer.
    \item Sweep over different pulse amplitude $A$, and find $A_r$ corresponding to the largest population of state $|2\rgl$ after repeatedly applying the $0\leftrightarrow2$ SWAP gate $15$ times with the initial state $|0\rgl$. 
\end{enumerate}
\item Find the $(r_c,A_{r_c})$ which results in the largest population of state $|2\rgl$ after $15$ gate repetitions with the initial state $|0\rgl$. Return the pulse determined by $\hat{d}(t;r_c,A_{r_c})$ as the tuned pulse.  
\end{enumerate}
\end{protocol}
\end{framed}

\end{appendices}
\bibliographystyle{plain}
\bibliography{ref}
\end{document}